\documentclass[twocolumn]{aa}

\usepackage{graphicx}
\usepackage{txfonts}

\newcommand{\afe}{\ensuremath{[\alpha/{\rm Fe}]}}
\newcommand{\hda}{\ensuremath{{\textrm{H}}\delta_{A}}}
\newcommand{\hdf}{\ensuremath{{\textrm{H}}\delta_{F}}}
\newcommand{\hga}{\ensuremath{{\textrm{H}}\gamma_{A}}}
\newcommand{\hgf}{\ensuremath{{\textrm{H}}\gamma_{F}}}
\newcommand{\lapprox}{\,\rlap{\lower 2.5pt \hbox{$\sim$}}\raise 1.5pt\hbox{$<$}\,} 
\newcommand{\m}{$-$}
\newcommand{\teff}{{T_{\rm eff}}}
\newcommand{\zh}{[{Z/{\rm H}}]}

\newcommand{\Hb}{\ensuremath{{\rm H}\beta}}
\newcommand{\Hg}{\ensuremath{{\rm H}\gamma}}
\newcommand{\HgA}{\ensuremath{{\rm H}\gamma_{\rm A}}}

\newcommand{\Hd}{\ensuremath{{\rm H}\delta}}

\newcommand{\CNone}{\ensuremath{{\rm CN}_1}}
\newcommand{\CNtwo}{\ensuremath{{\rm CN}_2}}
\newcommand{\FeC}{\ensuremath{{\rm C}_2\,}4668}
\newcommand{\Mgone}{\ensuremath{{\rm Mg}_1}}
\newcommand{\Mgtwo}{\ensuremath{{\rm Mg}_2}}
\newcommand{\Mgb}{\ensuremath{{\rm Mg}\, b}}
\newcommand{\TiOone}{\ensuremath{{\rm TiO}_1}}
\newcommand{\TiOtwo}{\ensuremath{{\rm TiO}_2}}

\newcommand{\aFe}{\ensuremath{\alpha/{\rm Fe}}}
\newcommand{\ZH}{\ensuremath{Z/{\rm H}}}
\newcommand{\Zsun}{\ensuremath{Z_\odot}}

\begin{document}
   \title{The sensitivity of Lick indices to abundance variations}

   \author{A.J. Korn
          \thanks{Leopoldina fellow at Uppsala
          Astronomical Observatory, Box 515, 751 20 Uppsala, Sweden}
          \and
          C. Maraston\thanks{current address: University of Oxford, Denys Wilkinson Building, Keble Road, Oxford, OX1 3RH, UK}
          \and
          D. Thomas$^{\star\star}$}

   \offprints{A.J. Korn, \email{akorn@astro.uu.se} }

   \institute{Max-Planck-Institut f\"ur extraterrestrische Physik (MPE),
              Giessenbachstra\ss e, 85748 Garching, Germany
             }

   \date{submitted Oct.~'04; accepted April~'05; {\bf Tables 6\,--\,32 are available electronically at www-astro.physics.ox.ac.uk/$\sim$dthomas}}

   \abstract{We present results of model atmosphere/line formation
calculations which quantitatively test how the 21 classical and four
higher-order Balmer-line Lick/IDS indices (Worthey et al.~1994;
Worthey \& Ottaviani 1997) depend on individual elemental abundances
(of carbon, nitrogen, oxygen, magnesium, iron, calcium, natrium,
silicon, chromium, titanium) and overall metallicity in various
stellar evolutionary stages and at various metallicities. At low
metallicities the effects of an overall enhancement of
$\alpha$-elements are also investigated. The general results obtained
by Tripicco \& Bell (\cite{TB95}) at solar metallicity are confirmed,
while details do differ. Tables are given detailing to which element
every index reacts significantly, as a function of evolutionary stage
and composition.

This work validates a number of assumptions
implicitly made in the stellar population models of Thomas, Maraston
\& Bender (\cite{TMB03}), which utilized the results of Tripicco \& Bell
(\cite{TB95}) to include the effects of element abundance ratios
variations.  In particular, these computations confirm that
fractional changes to index strengths computed at solar metallicity
(and solar age) can be applied over a wide range of abundances and
ages, also to model old stellar populations with non-solar
abundance ratios. The use of metallicity-dependent response functions
not only leads to a higher degree of self-consistency in the
stellar population models, but is even required for the proper
modelling of the Balmer-line indices. We find that the latter
become increasingly sensitive to element abundances with increasing
metallicity and decreasing wavelength. While \Hb\ still responds only
moderately to abundance ratio variations, the higher-order Balmer
indices \Hg\ and \Hd\ display very strong dependencies at high
metallicities. As shown in Thomas, Maraston \& Korn (2004), this
result allows to remove systematic effects in age determinations based
on different Balmer-line indices.

   \keywords{Radiative transfer -- Line: formation -- Stars: late-type -- Galaxies: abundances -- Galaxies: evolution -- Galaxies: fundamental parameters}
   }

   \maketitle
%

\section{Introduction}
In order to constrain galaxy formation models, and with it the
cosmological evolution of the baryonic Universe, the analysis of
stellar populations is of primary importance, because it provides the
unique way of measuring the metal enrichment and the epoch(s) of star
formation. Furthermore, if it was possible to measure the relative
abundances of individual elements in integrated spectra, in the same
way one does for single stars (e.g. magnesium, iron, calcium, oxygen),
this would provide further important information on the detailed
chemical enrichment. In massive galaxies the situation is, however,
more complicated. The dispersion of the velocity distribution
significantly reduces the {\em intrinsic} spectral resolution, in
other words single absorption lines are not resolved. Typical galaxy
spectra exhibit relatively broad absorption features, that contain a
huge number of lines from a large variety of chemical elements. This
obviously complicates the chemical abundance analysis.

Substantial progress in exploiting galaxy spectra has been made by the
Lick group who defined a set of so-called absorption-line indices in
the visual region ($4000\,\AA \lapprox \lambda \lapprox 6400\,\AA$)
with relatively wide band-passes on the feature ($\Delta\lambda\sim
20\,-\,40\,$\AA) and two windows blue- and redward of the band-pass
defining pseudo-continua (Burstein et al. \cite{BFGK84}). The various
indices were measured on a large sample of Milky Way stars, mostly
located in the solar vicinity, and were related with the stellar
parameters, effective temperature ($\teff$), surface gravity ($g$),
total metallicity ($Z$), by analytical relations, known as the Lick
fitting functions (Gorgas et al. \cite{Goretal93}; Worthey et
al. \cite{WFGB94}). The latter are then plugged into an evolutionary
synthesis code which produces integrated Lick indices for stellar
population models (Worthey~1994). The comparison of galaxy data with
these models allows one in principle to quantify the metallicity of
the galaxy.

It had been realized from the very start, that the interpretation of
the Lick indices would not be as simple as the naming of the indices
suggests. The modelled Lick indices fail to match the data of
elliptical galaxies (Worthey et al. \cite{WFG92}; Davies, Sadler \&
Peletier \cite{DSP93}; Carollo \& Danziger \cite{CD94}), more
specifically magnesium indices were found to be much stronger than
predicted by the models at a given iron index. The calibration of the
models with globular cluster (GC) data having abundance patterns known
from high-resolution spectroscopy have secured the conclusion that the
mismatch is indeed caused by an elemental abundance effect, in
particular an enhanced \afe~ratio (Maraston et al. \cite{Maretal03}).
There are several astrophysical environments that have
non-solar elemental proportions, which urges the computation of models
for a variety of chemical mixtures. This task is difficult to
accomplish empirically by using observed spectra, because these are
bound to trace the specific chemical history of the portion of
Universe the stars were born in (cf. efforts of Borges et
al. \cite{Boretal95}, Idiart et al. \cite{ITF97}, Lejeune et
al. \cite{LCB97}, and LeBorgne et al. \cite{LRPLFS04}).

The theoretical approach has the undoubted advantage of allowing a
much larger variety of chemical mixtures and ages, and was the way
embarked on by the seminal study of Tripicco \& Bell~(1995, hereafter TB95). The
authors investigated how the 21 original indices depend upon
individual abundance variations for representative positions along a
solar-metallicity 5\,Gyr isochrone. On the whole, their calculations
succeeded in reproducing many of the observed index strengths, while
some deficits were uncovered.

Thomas, Maraston \& Bender (2003a, hereafter TMB03) have implemented
the results by TB95 for the individual stars in a theoretical scheme,
using an extension of the method introduced by Trager et
al.~(2000). This allows the computation of a complete stellar
population model, thereby providing integrated Lick indices for a
variety of chemical mixtures. In first applications, $\alpha$/Fe
and Ca/Fe ratios of early-type galaxies have been derived (Thomas,
Maraston, Bender \cite{TMB03b}; Thomas et al. \cite{Thoetal04}).

Although the models have been carefully calibrated to reproduce the
absorption lines of globular clusters with known element abundance
ratios, a number of approximations have entered their modelling, due
to the limited range of results published by TB95.

Firstly, TB95 measured the index variations for solar metallicity
only, while the corresponding partial derivatives have been applied by
TMB03 to a large range of chemical compositions. Although the
calibrating GCs span a wide range in metallicities, super-solar
metallicities are not covered, while they are very important for many
applications. Also, the very low metallicities encountered in Halo GCs
($\zh\sim-2$) remain to be calibrated.\\ Secondly, TB95 selected the
three representative stars, a main-sequence dwarf, a turnoff and a
giant star on a 5 Gyr isochrone (shown to match the colour-magnitude
diagram of the open cluster M67), while TMB03 used the results for
these stars to compute models for a wide range of ages from 1 to 15
Gyr (and thus a range of stellar parameters as well).

Thirdly, TB95 analysed the classical Lick indices, while absorption
indices in other wavelength ranges are nowadays in wide use, and
especially for high-redshift galaxies bluer wavelength settings are
required.

The principle goal of the present paper is to verify and
extend the work of TB95, the prior by using an independent model
atmosphere code, the latter by investigating how all 25 Lick indices
react to abundance variations at different metallicities. That is, we
extend the computations to include the four higher-order Balmer-line
indices \hga, \hgf, \hda, \hdf\ defined by Worthey \&
Ottaviani~(1997) and the modification of the Fe5270 index by the
SAURON collaboration (Davies et al. \cite{Davetal01}; Falcon-Barroso et al. \cite{F-Betal04}; H.~Kuntschner,
private communication).

Section 2 describes the computations we have performed and confronts
the synthetic spectra with high-resolution spectra of the Sun and
Arcturus. In Section 3 the derived index values are compared with the
computations by TB95 and the empirical fitting functions
(relating line index strengths and stellar parameters, hereafter
referred to as FFs) of Worthey et al. (\cite{WFGB94}). The behaviour of
every index to abundance variations is subsequently discussed
individually. Section 4 puts the new computations in the context of
stellar population models, while Section 5 summarizes our results.


\section{Computations}

\subsection{Model Atmospheres}
All spectra and indices presented below are based on the model
atmosphere code originally written by T. Gehren (Gehren \cite{G75a},
b) and later named MAFAGS. Just like e.g. ATLAS (Kurucz \cite{ATLAS})
or MARCS (Gustafsson et al. \cite{GBEN75}) , this code produces model
temperature structures for the photospheres of cool stars assuming a
plane-parallel geometry in hydrostatic equilibrium and LTE. The models
are flux-conserving and line blanketed by means of Opacity Distribution Functions
(ODFs, Kurucz \cite{ODF}). As these ODFs were computed
assuming\footnote{$\log$\,$\varepsilon$(Fe)\,=\,$\log\,(n_{\rm
Fe}/n_{\rm H})\,+\,12$} $\log\,\varepsilon$(Fe)\,=\,7.67, we rescale
them by \m0.16\,dex to account for the low solar iron abundance of
$\log\,\varepsilon$(Fe)\,=\,7.50 $\pm$ 0.05 as given by Grevesse \&
Sauval (\cite{GS98}). For all models of all metallicities ODFs with a
microturbulence of 1\,km/s were used, while the microturbulence
entering the line formation calculation was allowed to vary between
1\,km/s and 2.5\,km/s.

Convection is approximated by mixing-length theory (B{\"o}hm-Vitense
\cite{bohm-vitense}). No overshooting is considered. Unlike in the
other codes mentioned above, a mixing length $\alpha_{\rm
conv}$\,=\,$l/H_{\rm p}$\,=\,0.5 is used in order to simultaneously
model H$\alpha$ and the higher Balmer lines (cf. Fuhrmann et
al. \cite{FAG93} and van't Veer-Menneret \& M\'{e}gessier
\cite{vant}). This choice has little effect on metal lines, as they, in contrast to H$\beta$ and higher members of the Balmer series,
form at optical depths in which the flux is predominantly carried by
radiation. A low convective efficiency leads to a steeper temperature gradient at high optical depths which alters the line formation of H$\beta$ and the higher-order Balmer lines (see below and Section 3).

It is worth noticing that in the framework of a refined theory of
convection (Canuto \& Mazzitelli \cite{canutomazzitelli}) a single
value of $\alpha_{\rm conv}$ is capable of fulfilling the constraints
of stellar evolution (i.e., reproducing the solar radius at the solar
age) and Balmer profile fitting (Bernkopf \cite{bernkopf}).

Following TB95, we have calculated model atmospheres with solar
abundance ratios and ones with the abundances of carbon, nitrogen,
oxygen, magnesium, iron, calcium, sodium, silicon, chromium and
titanium each doubled in turn. In addition, we have also increased the
overall metallicity [$Z$/H] of the model by +0.3\,dex from each base
metallicity\footnote{[Fe/H]\,=\,$\log$\,$\varepsilon$(Fe)$_\star$\,\m\,$\log$\,$\varepsilon$(Fe)$_\odot$}
[Fe/H] of \{+0.67, +0.35, 0.0, \m0.35, \m1.35, \m2.25\}. Only in the
latter case the ODF was changed according to the modified metallicity
isolating the abundance effect whenever a single elemental abundance
was varied. This assumes that the element whose abundance is varied
behaves like a trace element and its variation has no effect on the
temperature structure of the atmosphere. For most of the elements
considered here, this is a valid assumption. It is not in two cases: iron and carbon.
As iron is responsible for more than half of all line blanketing
effects, the true index change due to varying the iron abundance will
be somewhere between what is tabulated as `Fe' and `[$Z$/H]'.

Carbon is an even more complicated case. Enhancing C by 0.3\,dex brings the C/O ratio so close to unity that a carbon star is produced. While the according changes to the molecular equilibria of C$_2$, CH, CO and CN are properly accounted for in the line formation, the feedback of the significantly modified line blanketing of carbon-star spectra on the underlying atmospheric structure cannot be modelled in an ODF approach to calculating model atmospheres. This can lead to dramatically overestimated (or underestimated) index strengths. Worthey (\cite{W04}) argues that the excess of carbon (the fraction not locked up in CO) in the C-enhanced models of TB95 overestimates the true sensitivity of certain indices to carbon, as the Swan bands of C$_2$ react quadratically to the carbon abundance. This is correct, but not the main cause for a potential overestimation of the sensitivity. Rather, we believe it is caused by the shortcomings inherent to the ODF approach (also adopted by TB95) when applied to carbon stars.

To guide the reader in an attempt to assess the reliability of the index variations due to carbon, we have computed additional sequences for the solar-metallicity stars tabulated in Tables 12\,--\,14: carbon was enhanced by 0.15\,dex only (column 15). This smaller increase will not produce a carbon star and reflects the approach taken by Houdashelt et al. (\cite{Hetal02}). Another attempt was performed by increasing both carbon and oxygen by 0.3\,dex, thus preventing the formation of a carbon star. The ``C0.15'' sequence ought to correctly predict the sensitivity to an increase of carbon by 40\,\%. ``C$+$O0.3'' $-$ ``O'' (column 16 $-$ column 6) is deemed a better proxy to the actual C variation that what is tabulated in column 4. As carbon was not varied in TMB03, this whole problem has no impact on the scientific results published so far.

For solar metallicity, three representative models along a 1\,Gyr
isochrone were also computed. They allow an assessment of the overall
influence the parameter `age' has on the index strengths, mainly
via increased turnoff temperatures.

For the two lowest metallicities additional sequences were computed in
which a global enhancement of the $\alpha$ elements by a factor of two
was assumed. This change in abundance was assumed not to influence the
stellar parameters, to be able to isolate the effect of $\alpha$
enhancement. To be consistent with previous assumptions (TMB03),
$\alpha$ elements were chosen to include O, Ne, Mg, Si, S, Ar, Ca, Ti (particles that are build up with
$\alpha$-particle nuclei) plus the elements N and Na. The abundance of C was kept fixed at the base metallicity, such that the $\alpha$ enhancement is compensated for by a deficit of
iron-group elements. For more details on this point, see TMB03.

\subsection{Synthetic Spectra}
For computing the emergent spectral energy distribution we utilize the
line formation code LINFOR which is derived from MAFAGS. It has to be
realized that the traditional two-step approach to calculating spectra
(step 1: determining the temperature structure and partial pressures,
step 2: calculating the spectrum) is a convenient, but artificial,
separation of the one task of solving the radiative transfer equation
with boundary conditions. It is convenient, as (in codes like ATLAS or
MAFAGS) a resolution of 20\,\AA\ (the resolution of the ODFs) is
enough to compute realistically blanketed model temperature structures
in step 1, while the use of a line formation code (like SYNTHE for
ATLAS, SSG for MARCS or LINFOR for MAFAGS) in principle allows one to
compute spectra at arbitrary resolutions.

LINFOR considers the formation of lines for 37 elements and 14
diatomic molecules: H$_2$, CH, NH, OH, C$_2$, CN, CO, N$_2$, NO,
O$_2$, MgH, SiH, CaH and TiO. Triatomic molecules are currently not
included. All important external broadening mechanisms are considered,
i.e. Doppler, van-der-Waals, resonance and Stark broadening.  In the
specific case of Fe\,{\sc i}, $\log$\,C$_6$ values were determined to
be 0.15\,dex smaller than those proposed by Anstee \& O'Mara
(\cite{AOM95}) and changed accordingly (see Gehren et al. \cite{GKS02}
for more details). For most of the other elements, $\log$\,C$_6$
values according to Uns\"{o}ld (\cite{unsoeld}) or Kurucz (\cite{K92})
are used.

The line list serving as input to LINFOR is not, strictly speaking,
fully consistent with the ODFs used in the model atmosphere
computation. This is because the atomic data of individual lines has
been modified over the years based on constraints arising from the
reproduction of the solar spectrum. However, the vast majority of
lines still has unaltered atomic data most of which taken from Kurucz
\& Peytremann (\cite{kupey}), Kurucz (\cite{K92}) and Kurucz \& Bell
(\cite{kubell}). Our line list does not contain TiO (Kurucz
(\cite{K02}) quotes his list to contain close to 40 million lines due
to this molecule alone), therefore we refrain from discussing the two
TiO indices in Sect. \ref{results} at great length.

\subsection{The Sun and Arcturus}
\begin{figure}
   \centering
   \includegraphics[bb=205 53 558 738,angle=90,width=.5\textwidth,clip]{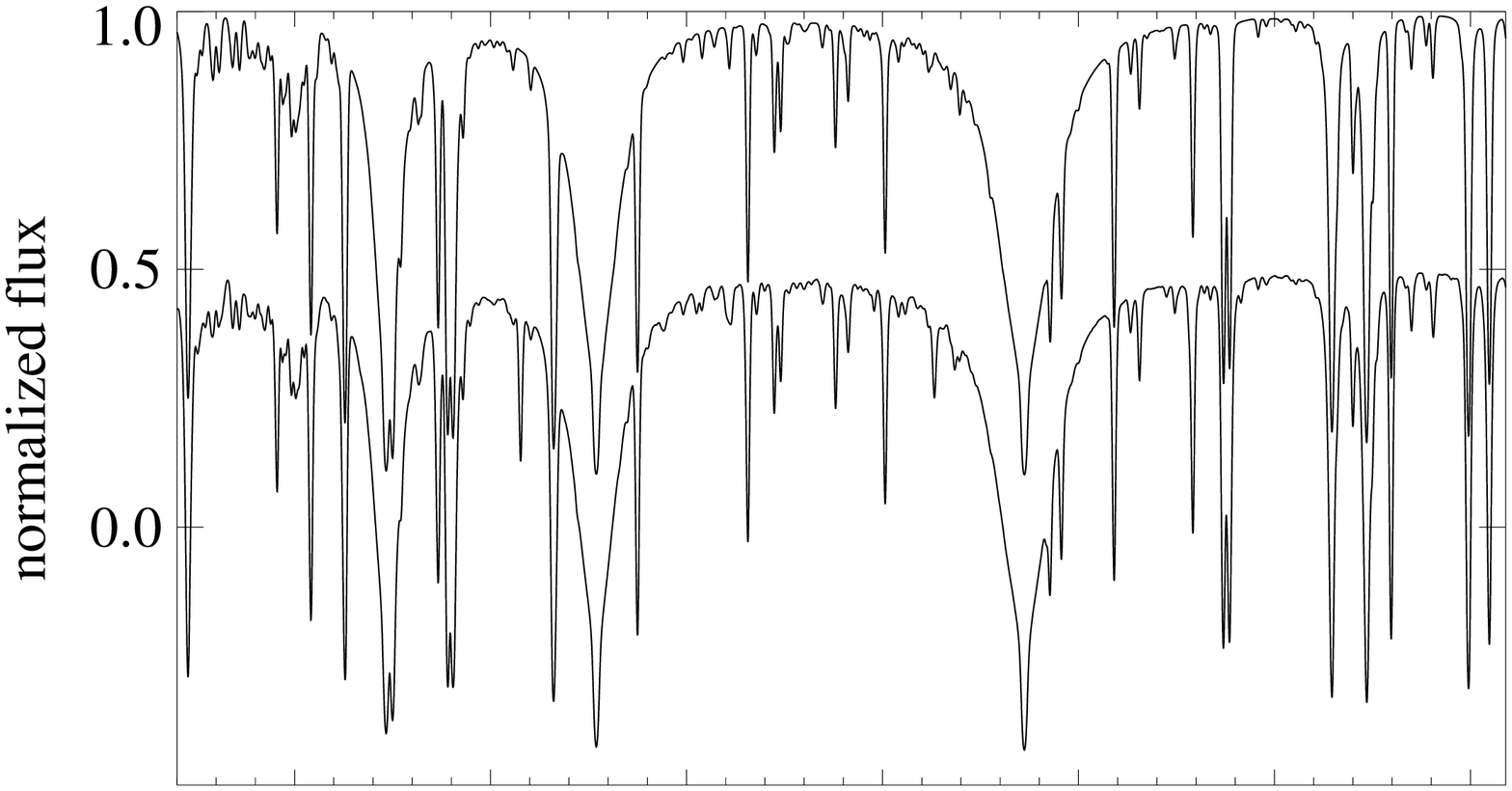}
   \includegraphics[bb=130 53 530 738,angle=90,width=.5\textwidth,clip]{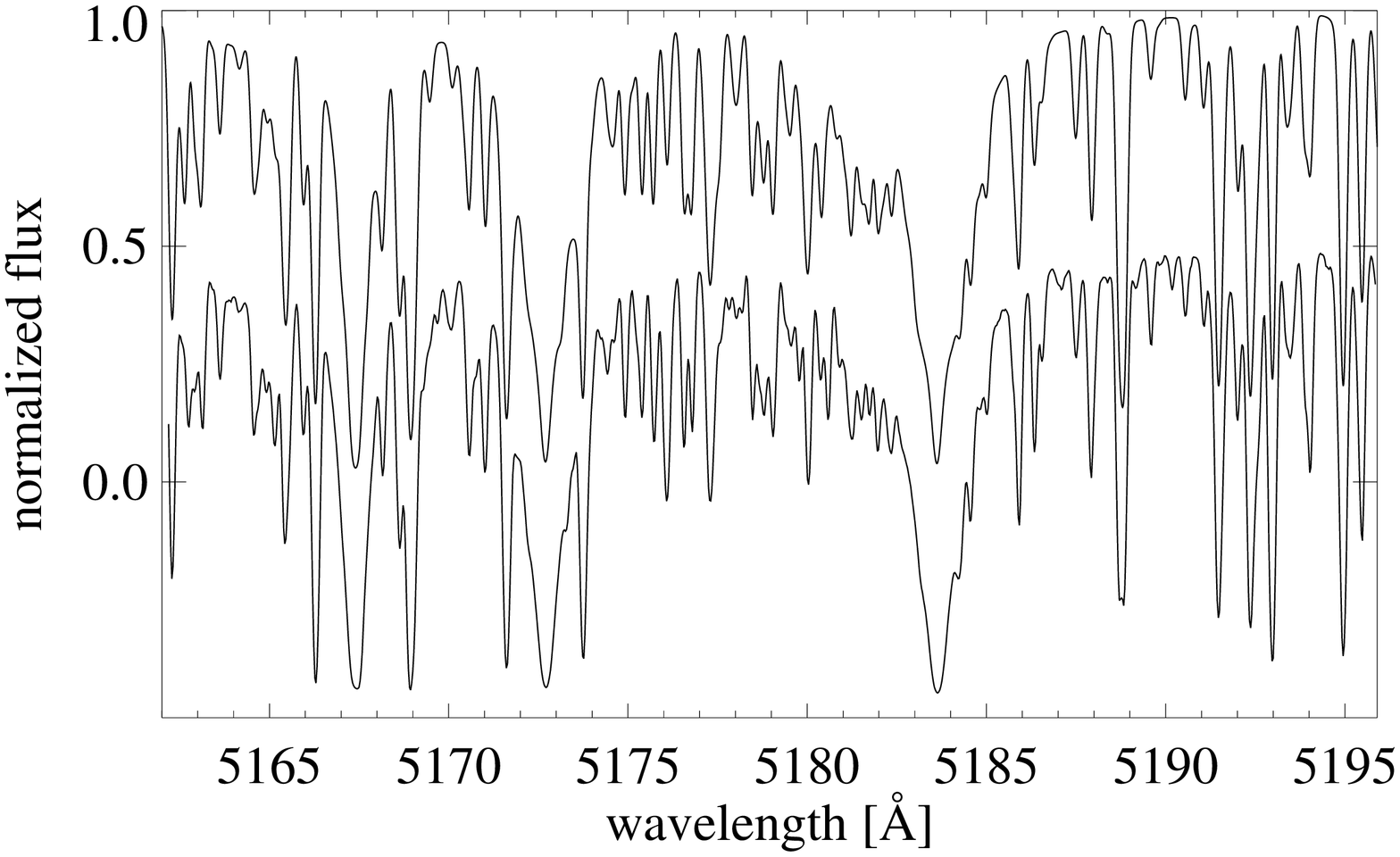}
\caption{A comparison between observed and computed spectra around the
Mg\,{\sc i}{\em b} lines. Upper panel: the Sun, lower panel:
Arcturus. In both panels the observations are offset by \m0.5 units}
         \label{MgIb}
\end{figure}

\begin{figure}
   \centering
   \includegraphics[bb=205 53 558 738,angle=90,width=.5\textwidth,clip]{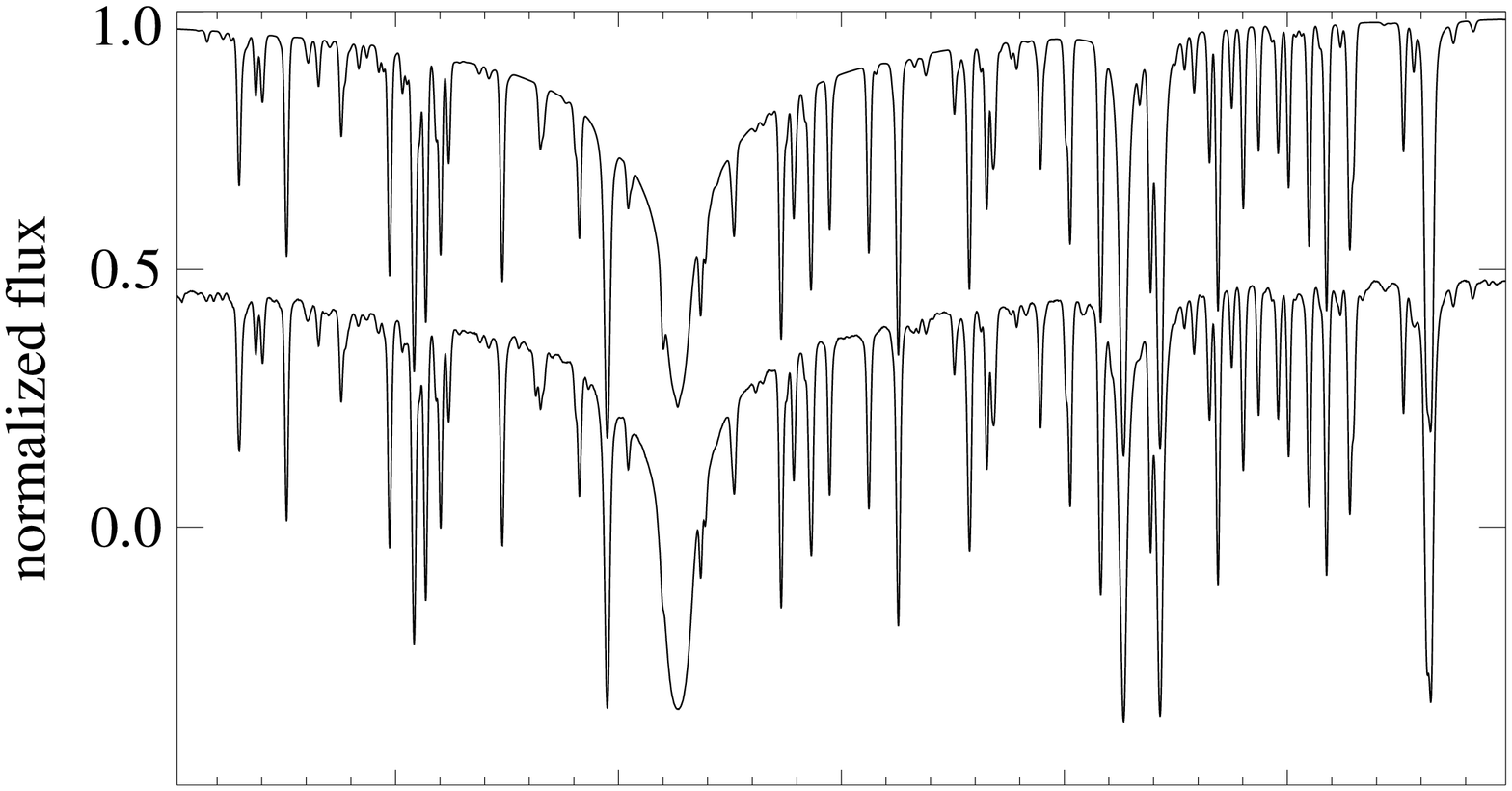}
   \includegraphics[bb=130 53 530 738,angle=90,width=.5\textwidth,clip]{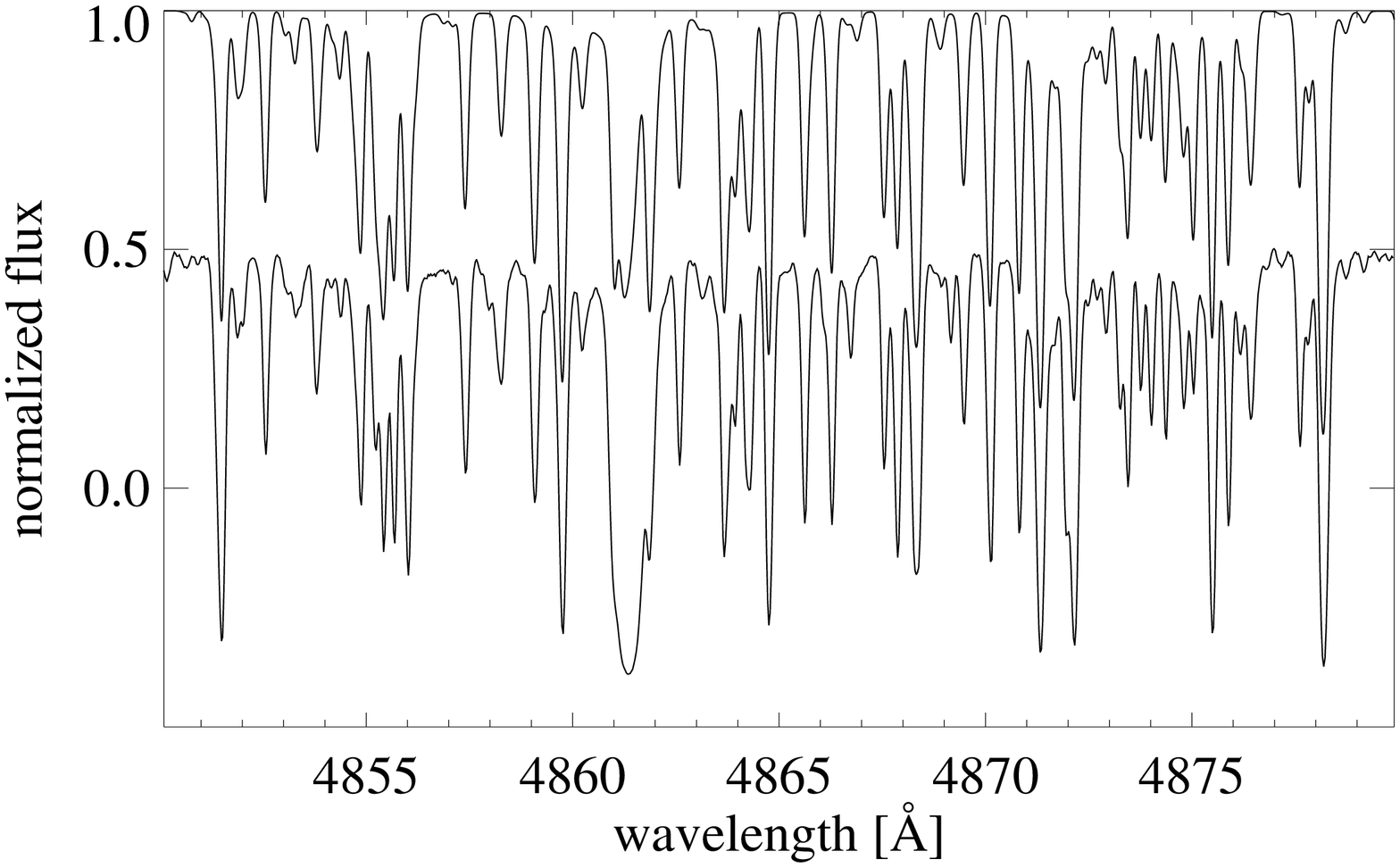}
\caption{A comparison between observed and computed spectra around
H$\beta$. Upper panel: the Sun, lower panel: Arcturus. In both panels
the observations are offset by \m0.5 units. Due to unmodelled physics,
the line strength of H$\beta$ comes out too low in Arcturus}
         \label{Hbetacomp}
\end{figure}

The Sun clearly constitutes the most important point of reference for
our calculations. We have therefore compared our synthetic spectra
with the most commonly used solar flux atlas, the Kitt Peak atlas
(Kurucz et al. \cite{KFBT84}), in two ways. Firstly, we can compare
the spectra by eye. This way individual lines obviously missing from
our line list were identified and added utilizing online compilations
for atomic data like those of the National (American) Institute of
Standards and Technology, NIST. We, however,
refrained from adding artificial lines as substitutes for unidentified
lines in the Sun of which there are still thousands in the optical (a
good example is the line at 5170.76\,\AA,
cf. Fig. \ref{MgIb}). Figures \ref{MgIb} and \ref{Hbetacomp} give a
qualitative indication of how well our modelling performs. These
Figures can be directly compared to Fig. 1 \& 2 of TB95 and show that
the performance of the two independent codes is quite comparable.

Secondly, since the position of the continuum is known in both the
observed and the theoretical spectrum of the Sun, we can directly
compare the equivalent widths $W_\lambda$ of the band-passes of
the 25 Lick indices without the use of the pseudo-continua. This
comparison was done after folding the theoretical spectra with a
projected rotational velocity $v \sin i$ of 1.8\,km/s and a
macroturbulence $\Xi_{\rm rt}$ of 3.5\,km/s in the radial-tangential
approximation (Gray \cite{Gray77}) to account for the observed solar line broadening. The results
are given in Table \ref{Wlamcomp} with the last column indicating that
there is a general absorption deficit in our modelling of
(15\,$\pm$\,5)\,\%. Some indices deviate markedly from this mean
value: \FeC, NaD and the two TiO indices have significantly higher
$\Delta W_\lambda$\footnotetext{The spectral regions around Fe5709 and
Fe5782 were re-rectified by +1.0 respectively +1.8\,\% to account for
an unmodelled global suppression of the observed continuum in the Kitt Peak atlas. This
procedure reduces the $\Delta W_\lambda$ values for these two
indices.}, G4300, H$\beta$, Mg$_2$, Mg$b$ and H$\gamma$\,F perform
better than average. There is no clear trend with wavelength. We note
that using $\alpha_{\rm conv}$\,=\,1.5 would result in a $\Delta
W_\lambda$ twice as large for H$\beta$ (14.1\,\%) and 4\,--\,8\,\%
larger for the higher-order Balmer-line indices.

\begin{table}
\caption{Direct integration of the Lick index band-passes, both on the
observed spectrum of the Sun (Kitt Peak National Observatory, KPNO)
and on the theoretical spectrum computed using the MAFAGS atmosphere
code ($T_{\rm eff}$\,=\,5780, $\log\,g$\,=\,4.44, [Fe/H]\,=\,0,
$\xi_{\rm mic}$\,=\,1.0\,km/s, $\alpha_{\rm conv}$\,=\,0.5). The final
column gives the absorption deficit $\Delta
W_\lambda$\,=\,100\,($W_\lambda$(KPNO)\,$-$\,$W_\lambda$(MAFAGS))/$W_\lambda$(KPNO). All
numbers are rounded to the first decimal place.}
\label{Wlamcomp}
\begin{center}
\begin{tabular}{rrrr}
\hline
\hline\\
Lick index & $W_\lambda$(KPNO) & $W_\lambda$(MAFAGS) & $\Delta W_\lambda$ \\[1mm]
 & [\AA] & [\AA]\hspace{0.6cm} & [\%] \\[1mm]
\noalign{\smallskip}\hline \noalign{\smallskip}\\[-3mm]
CN$_1$, CN$_2$ & 9.1 & 7.4\hspace{0.6cm} & 18.0 \\
Ca4227 & 4.0 & 3.7\hspace{0.6cm} & 9.4 \\
G4300 & 14.1 & 13.2\hspace{0.6cm} & 6.5 \\
Fe4383 & 12.7 & 11.3\hspace{0.6cm} & 11.2 \\
Ca4455 & 4.4 & 3.9\hspace{0.6cm} & 11.5 \\
Fe4531 & 6.8 & 5.6\hspace{0.6cm} & 17.9 \\
\FeC & 11.2 & 7.6\hspace{0.6cm} & 31.9 \\
H$\beta$ & 5.9 & 5.5\hspace{0.6cm} & 7.1 \\
Fe5015 & 10.0 & 8.6\hspace{0.6cm} & 14.2 \\
Mg$_1$ & 6.7 & 5.7\hspace{0.6cm} & 13.8 \\
Mg$_2$ & 7.7 & 7.3\hspace{0.6cm} & 5.0 \\
Mg$b$ & 6.7 & 6.4\hspace{0.6cm} & 4.5 \\
Fe5270 & 5.0 & 4.3\hspace{0.6cm} & 13.9 \\
Fe5335 & 3.5 & 3.0\hspace{0.6cm} & 13.6 \\
Fe5406 & 2.8 & 2.4\hspace{0.6cm} & 12.7 \\
Fe5709 & 1.5 & 1.3\hspace{0.6cm} & 13.5 \\
Fe5782 & 1.0 & 0.8\hspace{0.6cm} & 21.0 \\
NaD & 2.5 & 1.9\hspace{0.6cm} & 25.6 \\
TiO$_1$ & 1.7 & 0.7\hspace{0.6cm} & 57.4 \\
TiO$_2$ & 3.0 & 1.9\hspace{0.6cm} & 37.2 \\
H$\gamma$\,A & 11.9 & 10.5\hspace{0.6cm} & 11.9 \\
H$\gamma$\,F & 6.1 & 5.7\hspace{0.6cm} & 6.8 \\
H$\delta$\,A & 10.3 & 9.3\hspace{0.6cm} & 10.3 \\
H$\delta$\,F & 6.5 & 5.9\hspace{0.6cm} & 9.3 \\ [1mm]
\noalign{\smallskip}\hline \noalign{\smallskip}
\end{tabular}
\end{center}
\end{table}

The cause of this absorption deficit is not to be found primarily in
individual strong lines that are missing in our line list; rather is
has to be attributed to many unmodelled lines of low oscillator
strength and small equivalent width. To make progress in this context
significantly larger line lists would have to be implemented
(cf. efforts by Kurucz \cite{K02}) which is beyond the scope of
this project.

The lower panels of Figures \ref{MgIb} and \ref{Hbetacomp} compare
theory and observation for the thick-disk standard star Arcturus, a
first-ascent red giant. The spectrum was obtained with the FOCES
spectrograph (Pfeiffer et al. \cite{pfeiffer}) on the Calar Alto 2.2m
telescope in May 2000 (kindly made available by K. Fuhrmann). The
spectrum covers 4200\,\AA\,$<$\,$\lambda$\,$<$\,9000\,\AA\ and has a
resolving power of $R$\,=\,60\,000. The signal-to-noise ratio (S/N)
varies between 70 in the bluest part of the spectrum and 430 in the
near-IR; at the wavelengths presented here it is around 230 (Mg\,{\sc
i}b lines) and 200 (H$\beta$), respectively.

For the modelling we take stellar parameters from the literature:
$T_{\rm eff}$\,=\,4300\,K (Peterson et al. \cite{PDK93}),
$\log\,g$\,=\,1.6 (between Peterson et al. \cite{PDK93} and Decin et
al. \cite{D97}), [Fe/H]\,=\,$-$0.5, $\xi_{\rm mic}$\,=\,1.7\,km/s. The
abundances of O, Mg, Si and Ca were enhanced by 0.3\,dex to account
for Arcturus' $\alpha$ enhancement.

As can be appreciated from the lower panels of Figures \ref{MgIb} and
\ref{Hbetacomp}, our model for Arcturus is, within its limitations, an
appropriate representation of the conditions in this cool red
giant. One marked exception is the H$\beta$ line itself with obvious
modelling deficits that have to be attributed to unmodelled physics
(non-LTE and/or chromospheric contributions, see Przybilla \& Butler
\cite{nob04}). The overall good correspondence between theory and
observation shows that the choices for transition probabilities and
damping constants made for the Sun are essentially correct. Absorption
deficits of similar magnitude as in the solar case are, however, also
encountered for Arcturus.

\subsection{Spectral Indices}
The output spectra are normalized and have a flux point spacing of
0.1\,\AA, just like those of TB95.  To put them onto the Lick/IDS
system, they were convolved with Gaussians of variable FWHM
interpolated among the values given in Table 8 of Worthey \& Ottaviani
(\cite{WO97}). This ought to not have a significant influence on the
comparison with TB95 (see below). The final flux point spacing is
typically 0.5\,\AA, a factor of two higher than that of the IDS
spectrograph.

We did not run calculations testing the influence of microturbulence,
offsets, FWHM, continuum and resolution on the index values. The
reader is referred to Tables 1\,--\,3 of TB95 for a thorough
evaluation of these aspects. Note, however, that the dependence on
FWHM, continuum and resolution was found to be below the 2$\sigma$
(twice the observed standard error) level in all cases. This leaves
microturbulence and potential wavelengths distortions/offsets as
additional sources of systematic error.\\ 936 Lick indices were
computed at each base metallicity making it 8424 spectral indices in
total.


\section{Results}\label{results}
\subsection{Comparison with TB95}
Tables 12\,--\,14 give the Lick index name (column
1), the computed index strength (column 2), the standard error taken
from TB95 (column 3) and the absolute variation of the index strengths
upon increasing the abundance of the element given at the top of the
column by 0.3\,dex. [$Z$/H] (column 14) refers to an overall increase
in metallicity by 0.3\,dex.  The Tables can be compared with Tables
4\,--\,6 of TB95. Note, however, that we chose to tabulate the
absolute index changes, i.e., not the ones normalized to the assumed
standard error.

Common and disparate features of the two data sets
are discussed below, first globally, then index by index.

In comparing the computed index values given in column 2 to those of
TB95, one notices agreement at the 2$\sigma$ level in 2/3 of all cases
(42 out of 63).

For the {\em turnoff star} (6200/4.1, Table 13), the
agreement is always within 2$\sigma$, in 2/3 of all cases (14 out of
21) even within 1$\sigma$. It was expected that the turnoff star would
show the highest degree of consistency as {\em a)} the temperature
structure and electron pressure is robust to small changes in the
assumed composition and {\em b)} molecules play a significantly
reduces r\^{o}le in computing the spectra. Therefore, it does not come
as a surprise either that the fractional changes show a similar
overall behaviour: in 17 out of the 21 cases the sensitivity is found
to be highest for the same particular element. The seemingly
discrepant cases are Ca4227, Fe5335, NaD an TiO$_2$. We find Ca4227
and NaD to depend more strongly on the name-giving element than on
metallicity (TB95 vice versa), in the case of Fe5335 it is the other
way around. The two TiO indices do not significantly depend on any of
the 12 elements, neither in this study nor in TB95 (in the M-star
regime, they would naturally depend on Ti and O).

Turning to the {\em main sequence star} (4575/4.6, Table 12), there are two
indices which turn out more than 2$\sigma$ stronger than their TB95
counterparts: H$\beta$ and Fe5015. In both cases, this strengthening
turns out to bring these indices into much better agreement with both
the observational data on M67 and the Worthey et al. FFs. The index
strength of Fe5015 is still 1\,\AA\ short of the FF for the dwarf, but
the scatter among the M67 dwarfs is of the same order (cf. Fig.~13 of
TB95).

Seven indices have I$_0$ values smaller by more than 2$\sigma$:
CN$_1$, CN$_2$, Ca4227, G4300, Ca4455, \FeC\ and NaD. In two cases
(Ca4227 and NaD), the new values lie closer to the corresponding FF
and the M67 data. Two other cases (the two CN indices) are now in
better agreement with M67, but move away from the FF. The remaining
three cases (G4300, Ca4455 and \FeC) perform worse with respect to
both references.

Comparing the sensitivity to particular abundance variations,
agreement is found on the element causing the largest index change in
16 cases. In two cases (Fe5015 and Mg$_1$) the element in first and
second place trade places, and we find Ca4455 to (insignificantly) react more strongly
to Ca itself than to metallicity, Cr or Fe. Furthermore, no
sensitivity to C is found in the two TiO indices (as claimed by
TB95). In fact, no significant sensitivity is found at all for any of these
indices, neither for Ca4455 and H$\beta$ (cf.~the red giant).

In the {\em red giant} (4255/1.9, Table 14), four indices are significantly
stronger: Fe4531, H$\beta$, Fe5015 and Fe5335. Just like in the case
of the main sequence star, H$\beta$ and Fe5015 are now in excellent
agreement with the M67 fiducials of TB95. Fe4531 and Fe5335 are 2\,\AA\ respectively
1\,\AA\ too large to be in accord with M67 and the FF.

The same seven indices that were found to be more than 2$\sigma$ weaker in the
dwarf are also weaker in the giant; TiO$_2$ has to be added to this
list. Four indices (CN$_1$, Ca4455, \FeC\ and TiO$_2$) clearly perform
worse, the new strength of NaD places it on the opposite side of the
distribution of giants in M67 from TB95 and about equally close to the
FF, and three indices (CN$_2$, Ca4227 and G4300) fare better using
MAFAGS.

Turning to the abundance variations, the dominant r\^{o}le of a
particular element found by TB95 is recovered for 11 indices. In a
further 5 cases (Fe5015, Fe5270, Fe5335, Fe5406 and Fe5782) the order
of the first two element is reversed (sometimes making them most
sensitive to Fe, sometimes to [$Z$/H]), in one case (Ca4227) that of
the dominant three. In four cases (Ca4455, H$\beta$, TiO$_1$ and
TiO$_2$) we find no significant sensitivity to any of the elements
studied. For H$\beta$ this is confirmation of the TB95 result.

Overall, the one-by-one comparison shows that the computations
presented here are on par with those published by TB95. Irrespective
of whether improvements were found or not, understanding the cause of
disparate results would bring on progress. Since we do not have access
to R. Bell's version of the MARCS code nor to the line-formation code
SSG and the line list used, this is unfortunately not possible. Taken
at face value, the differences certainly tell us something about the
uncertainties involved in the modelling of Lick indices. As a rule of
thumb, we conclude that the bluest classical indices except
Fe4383\footnote{Fe4383 can be considered an exception. We agree with
TB95 that this index is best suited for the determination of the iron
abundance.} (that is, CN$_1$, CN$_2$, Ca4227, G4300, Ca4455, Fe4531 and
\FeC) bear higher modelling uncertainties than the other 13
indices. Together with the wavelength independence of $\Delta
W_\lambda$ (see Table~\ref{Wlamcomp}), this indicates that line-list
differences are largest in regions of high line density, as one might
have expected.

\subsection{Comparison with Fitting Functions at various metallicities}
\begin{table}
\caption{Stellar parameters used in the computation of the synthetic Lick indices. }
\label{stellparam}
\begin{center}
\begin{tabular}{rrcrrr}
\hline
\hline\\
$\zh$ & age [Gyr] & stage &$\teff$ & log\,$g$ & \afe \\[1mm]
\noalign{\smallskip}\hline \noalign{\smallskip}\\[-2mm]
+0.67   & 5\hspace*{0.6cm}     & MS & 4667 & 4.59 & 0.0 \\
        &       & TO & 5694 & 4.16 &  \\
        &       & RG & 4590 & 3.42 &  \\[1mm]
+0.35   & 5\hspace*{0.6cm}     & MS & 4543 & 4.58 & 0.0 \\
        &       & TO & 5969 & 4.18 &  \\
        &       & RG & 4236 & 2.00 &  \\[1mm]
+0.00   & 5\hspace*{0.6cm}     & MS & 4575 & 4.60 & 0.0 \\
        &       & TO & 6200 & 4.10 &  \\
        &       & RG & 4255 & 1.90 &  \\[1mm]
+0.00   & 1\hspace*{0.6cm}     & MS & 5297 & 4.56 & 0.0 \\
        &       & TO & 8048 & 3.91 &  \\
        &       & RG & 4336 & 1.83 &  \\[1mm]
\m0.35  & 13\hspace*{0.6cm}    & MS & 4466 & 4.60 & 0.0 \\
        &       & TO & 5822 & 4.22 &  \\
        &       & RG & 4414 & 1.77 &  \\[1mm]
\m1.35  & 13\hspace*{0.6cm}   & MS & 4385 & 4.83 & 0.0/+0.3 \\
        &       & TO & 6383 & 4.16 &  \\
        &       & RG & 4662 & 1.83 &  \\[1mm]
\m2.25  & 13\hspace*{0.6cm}    & MS & 5124 & 4.70 & 0.0/+0.3 \\
        &       & TO & 6724 & 4.15 &  \\
        &       & RG & 4822 & 1.83 &  \\ [1mm]
\noalign{\smallskip}\hline \noalign{\smallskip}
\end{tabular}
\end{center}
\end{table}

Table \ref{stellparam} gives details of the stellar parameters used
for the computation of the synthetic Lick indices. The data pairs
\{$\teff$, log\,$g$\} were read off from isochrones of the given
metallicity and age (from Cassisi, Castellani, Castellani
\cite{CCC97} and Salasnich et al. \cite{Saletal00} for the highest
metallicity, see Maraston \cite{Maraston98,Maraston04} for details)
for dwarf and turnoff stars. For the giants, they represent average
locations in which the fuel consumption on the Red Giant Branch is
maximum (see Maraston \cite{Maraston04}).

Tables 6\,--\,32 give the index values and
index variations for representative dwarfs (MS), turnoff stars (TO)
and red giants (RG) for metallicities ranging from +0.67 to \m2.25. At
solar metallicity Tables 15\,--\,17 give
additional values for objects of age 1\,Gyr. At metallicities \m1.35
and \m2.25 Tables 24\,--\,26 and
30\,--\,32 contain additional data sets
computed under the assumption of a general enhancement of $\alpha$
ele\-ments. Tables \ref{sensitivity_MS}\,--\,\ref{sensitivity_RG} give
an overview of the sensitivity of Lick indices to abundance variations
(including sensitivity to [\ZH]).

Figures \ref{Fe4383}\,--\,\ref{HdeltaF} in the Appendix are a
graphical representation of the metallicity dependence of the index
and the corresponding FF values. They may serve as an at-a-glance
source of information on where modelling deficits are practically
absent or dominating. Modelled index strengths are denoted by the
(coloured) boxes, the bullets are the corresponding FF
values with IDS standard errors of Worthey et al. (\cite{WFGB94}). The diamonds indicate the
absolute index change that results from a 0.3\,dex increase of the
element producing the highest index change. This dominant element was
sometimes found to vary with evolutionary phase, see e.g. H$\beta$.

Offsets between our results and the FFs beyond 1$\sigma$ are found in
2/3 of all cases (cf. the Figures in the Appendix). Note that it is
practically impossible to name the cause for a mismatch between theory
and observation. This is because a high index strength can be caused
by too few absorbers in the continuum band-passes or too many in the
band-pass itself. A mismatch can also result from ill-determined
stellar parameters in the sample of stars defining the FFs such that
one is not comparing like with like. There certainly is room for the
latter at the low-metallicity end where absolute effective
temperatures are still uncertain at the 200\,K level (see e.g. Barklem
et al. \cite{barklem}) and where the FFs are poorly constrained
owing to the lack of stars in the Lick library.

In what follows the 25 indices are discussed individually.

\begin{table*}
\caption{Sensitivity of individual indices to abundance variations in
the {\bf dwarf phase} of evolution. Only the significant ($>$ one
standard error) top three elements are given. For Fe5270(Sauron), the
standard error of Fe5270(Lick) was assumed to be applicable.}
\label{sensitivity_MS}
\begin{center}
{\small
\begin{tabular}{rcccccccc}
\hline
\hline\\
Index & +0.67 & +0.35 & 0.00 & \m0.35 & \m1.35 & \m1.35 & \m2.25 & \m2.25 \\
 & & & & & & $\alpha$-enh. & & $\alpha$-enh. \\[1mm]
\noalign{\smallskip}\hline \noalign{\smallskip}\\[-2mm]
CN$_1$ & C,O,N & C,O,N & C,N,O & C,N & & & & \\
CN$_2$ & C,O,N & C,O,N & C,N,O & C,N & & & & \\
Ca4227 & Ca,[$Z$/H],C & Ca,[$Z$/H],C & Ca,C,[$Z$/H] & Ca,[$Z$/H],C & Ca,[$Z$/H],C & Ca,[$Z$/H],C & & \\
G4300 & C,O,Ti & C,O,Ti & C,O,Fe & C,O,Ti & C,O & C,O & C,[$Z$/H],O & C,[$Z$/H],O \\
Fe4383 & Fe,Mg,[$Z$/H] & Fe,Mg,[$Z$/H] & Fe,Mg,[$Z$/H] & Fe,Mg,[$Z$/H] & Fe,[$Z$/H] & Fe,[$Z$/H] & [$Z$/H],Fe & \\
Ca4455 & & & & & & & & \\
Fe4531 & Ti,[$Z$/H],Mg & Ti,[$Z$/H],Mg & Ti,[$Z$/H] & Ti,[$Z$/H] & [$Z$/H],Ti & [$Z$/H],Ti & & \\
\FeC & C,O & C & C & C & & & & \\
H$\beta$ & & & & & & & & \\
Fe5015 & Ti,[$Z$/H],Mg & Ti,[$Z$/H],Mg & Ti,[$Z$/H],Mg & Ti,[$Z$/H],Mg & Ti,Mg,[$Z$/H] & Ti,Mg,[$Z$/H] & & \\
Mg$_1$ & C,Mg,Fe & C,Mg,[$Z$/H] & C,Mg,Fe & Mg,C,Fe & Mg,[$Z$/H],Fe & Mg,[$Z$/H] & & \\
Mg$_2$ & Mg,[$Z$/H],Fe & Mg,[$Z$/H],Fe & Mg,[$Z$/H],Fe & Mg,[$Z$/H],Fe & Mg,[$Z$/H],Fe & Mg,[$Z$/H],Fe & Mg,[$Z$/H] & Mg,[$Z$/H]\\
Mg $b$ & Mg,Fe,[$Z$/H] & Mg,Fe,Cr & Mg,Fe,Cr & Mg,Fe,Cr & Mg,[$Z$/H],Cr & Mg,[$Z$/H],Cr & Mg,[$Z$/H] & Mg,[$Z$/H]\\
Fe5270 & Fe,[$Z$/H],Mg & Fe,[$Z$/H],Mg & Fe,[$Z$/H],Mg & Fe,[$Z$/H],Mg & [$Z$/H],Fe & [$Z$/H],Fe & &\\
Fe5270(Sauron) & Fe,[$Z$/H],Mg & Fe,[$Z$/H],Mg & Fe,[$Z$/H],Mg & Fe,[$Z$/H],Mg & [$Z$/H],Fe & [$Z$/H],Fe & & \\
Fe5335 & Fe,[$Z$/H],Mg & Fe,[$Z$/H],Mg & Fe,[$Z$/H],Mg & Fe,[$Z$/H],Mg & Fe,[$Z$/H] & Fe,[$Z$/H] & & \\
Fe5406 & Fe,[$Z$/H],Mg & Fe,[$Z$/H],Mg & Fe,[$Z$/H],Mg & Fe,[$Z$/H],Mg & Fe,[$Z$/H] & Fe,[$Z$/H] & & \\
Fe5709 & [$Z$/H] & [$Z$/H] & [$Z$/H] & [$Z$/H] & & & & \\
Fe5782 & Cr & Cr & Cr & Cr & & & & \\
Na D & Na,[$Z$/H],Mg & Na,[$Z$/H],Mg & Na,[$Z$/H],Mg & Na,[$Z$/H],Mg & Na,[$Z$/H] & Na,[$Z$/H] & & \\
TiO$_1$ & & & & & & & & \\
TiO$_2$ & & & & & & & & \\
H$\gamma$\,A & Fe,Mg,[$Z$/H] & Fe,Mg,[$Z$/H] & Fe,Mg,[$Z$/H] & Fe,Mg,C & C,[$Z$/H],Fe & C,[$Z$/H],Fe & C,[$Z$/H] & C,[$Z$/H] \\
H$\gamma$\,F & Fe,C,Mg & Fe,C,Mg & Fe,C,Mg & Fe,C,Mg & C,Fe,O & C,O,Fe & C,[$Z$/H] & C \\
H$\delta$\,A & Fe,Mg,[$Z$/H] & Fe,Mg,[$Z$/H] & Fe,Mg,[$Z$/H] & Fe,C,Mg & Fe,[$Z$/H],C & Fe,[$Z$/H],C & & \\
H$\delta$\,F & Fe,[$Z$/H],Mg & Fe,[$Z$/H],Mg & Fe,[$Z$/H],Mg & Fe,[$Z$/H],Mg & Fe,[$Z$/H] & Fe,[$Z$/H] & & \\
\noalign{\smallskip}\hline \noalign{\smallskip}
\end{tabular}
}
\end{center}
\end{table*}

\begin{table*}
\caption{Sensitivity of individual indices to abundance variations in
the {\bf turnoff phase} of evolution. Only the significant ($>$ one
standard error) top three elements are given. For Fe5270(Sauron), the
standard error of Fe5270(Lick) was assumed to be applicable.}
\label{sensitivity_TO}
\begin{center}
{\small
\begin{tabular}{rcccccccc}
\hline
\hline\\
Index & +0.67 & +0.35 & 0.00 & \m0.35 & \m1.35 & \m1.35 & \m2.25 & \m2.25 \\
 & & & & & & $\alpha$-enh. & & $\alpha$-enh. \\[1mm]
\noalign{\smallskip}\hline \noalign{\smallskip}\\[-2mm]
CN$_1$ & N,C,[$Z$/H] & N,[$Z$/H],C & & & & & & \\
CN$_2$ & N,C,[$Z$/H] & N,[$Z$/H],C & & & & & & \\
Ca4227 & C,Ca,N & Ca & & & & & & \\ G4300 & C,Fe & C,Fe & C,[$Z$/H] & C & & [$Z$/H] & & \\
Fe4383 & Fe,[$Z$/H] & Fe & & [$Z$/H] & & & & \\
Ca4455 & & & & & & & & \\
Fe4531 & [$Z$/H] & [$Z$/H] & [$Z$/H] & [$Z$/H] & & & & \\
C$_2$\,4668 & C,[$Z$/H],O & C,[$Z$/H] & C & C & & & & \\
H$\beta$ & [$Z$/H] & [$Z$/H] & [$Z$/H] & & & & & \\
Fe5015 & [$Z$/H] & [$Z$/H] & [$Z$/H] & [$Z$/H] & & & & \\
Mg$_1$ & C,[$Z$/H],O & C,[$Z$/H],Fe & C & C & & & & \\
Mg$_2$ & Mg,[$Z$/H],C & [$Z$/H],Mg,C & Mg,[$Z$/H] & Mg,[$Z$/H],C & & & & \\
Mg $b$ & Mg,[$Z$/H],C & Mg,[$Z$/H] & Mg,[$Z$/H] & Mg,[$Z$/H] & & & & \\
Fe5270 & [$Z$/H],Fe & [$Z$/H],Fe & [$Z$/H] & [$Z$/H] & & & & \\
Fe5270(Sauron) & [$Z$/H],Fe,Ca & [$Z$/H],Fe,Ca & [$Z$/H],Fe,Ca & [$Z$/H],Fe,Ca & [$Z$/H],Ca,Fe & [$Z$/H],Ca,Fe & [$Z$/H],Fe,Ca & [$Z$/H],Fe,Ca \\
Fe5335 & Fe,[$Z$/H] & [$Z$/H],Fe & [$Z$/H] & [$Z$/H] & & & & \\
Fe5406 & Fe,[$Z$/H] & & & & & & & \\
Fe5709 & [$Z$/H] & [$Z$/H] & & & & & & \\
Fe5782 & & & & & & & & \\ Na D & Na,[$Z$/H] & Na,[$Z$/H] & Na & Na & & & & \\
TiO$_1$ & & & & & & & & \\
TiO$_2$ & & & & & & & & \\
H$\gamma$\,A & C,Fe,Mg & C & C & C,[$Z$/H] & & & & \\
H$\gamma$\,F & C & C & C & C & & & & \\
H$\delta$\,A & Fe & & & & & & & \\
H$\delta$\,F & Fe & & & & & & & \\
\noalign{\smallskip}\hline \noalign{\smallskip}
\end{tabular}
}
\end{center}
\end{table*}

\begin{table*}
\caption{Sensitivity of individual indices to abundance variations in
the {\bf giant phase} of evolution. Only the significant ($>$ one
standard error) top three elements are given. For Fe5270(Sauron), the
standard error of Fe5270(Lick) was assumed to be applicable.}
\label{sensitivity_RG}
\begin{center}
{\small
\begin{tabular}{rcccccccc}
\hline
\hline\\
Index & +0.67 & +0.35 & 0.00 & \m0.35 & \m1.35 & \m1.35 & \m2.25 & \m2.25 \\
 & & & & & & $\alpha$-enh. & & $\alpha$-enh. \\[1mm]
\noalign{\smallskip}\hline \noalign{\smallskip}\\[-2mm]
CN$_1$ & C,O,N & C,O,N & C,O,N & C,N,O & N,C & N,C,[$Z$/H] & & \\
CN$_2$ & C,O,N & C,O,N & C,O,N & C,N,O & N,[$Z$/H],C & N,[$Z$/H],C & & \\
Ca4227 & Ca,[$Z$/H],C & Ca,[$Z$/H],C & Ca,[$Z$/H],C & C,Ca,[$Z$/H] & C & C & & \\
G4300 & C,O,Fe & C,O,Fe & C,O & C,O & C & & C,[$Z$/H] & C,[$Z$/H] \\
Fe4383 & Fe,Mg,[$Z$/H] & Fe,[$Z$/H],Mg & Fe,[$Z$/H],Mg & Fe,[$Z$/H] & [$Z$/H],C & C,[$Z$/H] & & \\
Ca4455 & & & & & & & & \\
Fe4531 & [$Z$/H],Ti & [$Z$/H],Ti & [$Z$/H],Ti & [$Z$/H] & [$Z$/H] & [$Z$/H] & [$Z$/H] & \\
C$_2$\,4668 & C,O,[$Z$/H] & C,O,[$Z$/H] & C,O,[$Z$/H] & C,[$Z$/H],O & C & C & & \\
H$\beta$ & & & & & & & & \\
Fe5015 & [$Z$/H],Ti,Mg & [$Z$/H],Ti,Mg & [$Z$/H],Ti,Mg & [$Z$/H],Fe,Mg & [$Z$/H] & [$Z$/H] & [$Z$/H] & [$Z$/H] \\
Mg$_1$ & C,Mg,[$Z$/H] & C,Mg,[$Z$/H] & C,[$Z$/H],Mg & C,[$Z$/H],Mg & C,[$Z$/H] & C,[$Z$/H] & & \\
Mg$_2$ & Mg,[$Z$/H],C & Mg,[$Z$/H],C & Mg,[$Z$/H],C & Mg,[$Z$/H],C & [$Z$/H],C & [$Z$/H],Mg,C & & \\
Mg $b$ & Mg,Fe,[$Z$/H] & Mg,[$Z$/H],Fe & Mg,[$Z$/H],Fe & Mg,[$Z$/H],C & & Mg & & \\
Fe5270 & Fe,[$Z$/H] & Fe,[$Z$/H] & Fe,[$Z$/H] & [$Z$/H],Fe & [$Z$/H] & [$Z$/H] & & \\
Fe5270(Sauron) & Fe,[$Z$/H],Mg & Fe,[$Z$/H],Mg & Fe,[$Z$/H],Mg & [$Z$/H],Fe,Mg & [$Z$/H],Fe,Ca & [$Z$/H],Fe,Ca & [$Z$/H],Fe,Ca & [$Z$/H],Ca,Fe \\
Fe5335 & [$Z$/H],Fe,Mg & [$Z$/H],Fe & [$Z$/H],Fe & [$Z$/H],Fe & [$Z$/H] & & & \\
Fe5406 & [$Z$/H],Fe & [$Z$/H],Fe & [$Z$/H],Fe & [$Z$/H],Fe & [$Z$/H] & & & \\
Fe5709 & [$Z$/H] & [$Z$/H] & [$Z$/H] & [$Z$/H] & [$Z$/H] & & & \\
Fe5782 & Cr,[$Z$/H] & Cr & Cr,[$Z$/H] & Cr,[$Z$/H] & & & & \\
Na D & Na,[$Z$/H],Mg & Na,[$Z$/H] & Na,[$Z$/H] & Na,[$Z$/H] & & & & \\
TiO$_1$ & & & & & & & & \\
TiO$_2$ & & & & & & & & \\
H$\gamma$\,A & Fe,[$Z$/H],Mg & Fe,[$Z$/H],Mg & Fe,[$Z$/H],Mg & [$Z$/H],Fe & [$Z$/H] & [$Z$/H] & [$Z$/H],C & [$Z$/H],C \\
H$\gamma$\,F & Fe,C & [$Z$/H],Fe,C & Fe,[$Z$/H] & & & & C,[$Z$/H] & C,[$Z$/H] \\
H$\delta$\,A & Fe,[$Z$/H],Mg & Fe,[$Z$/H] & Fe,[$Z$/H] & Fe,[$Z$/H] & C & C & & \\
H$\delta$\,F & Fe,[$Z$/H],Mg & [$Z$/H],Fe,Mg & [$Z$/H],Fe & [$Z$/H],Fe & & & & \\
\noalign{\smallskip}\hline \noalign{\smallskip}
\end{tabular}
}
\end{center}
\end{table*}

\subsubsection*{CN$_1$ \& CN$_2$ (Fig. \ref{CN1})}
The index is most sensitive to the abundances of C and N. It is also
sensitive to O via the preferred formation of CO. Metallicity plays some r\^{o}le in the turnoff phase.

Contrary to TB95, our computations underpredict the index strengths at
high metallicity, in particular for the dwarfs and the
giants. Interestingly, this behaviour is reversed in the
low-metallicity dwarfs making the dependence on metallicity rather
weak. The FFs do show such a weak trend for the turnoff stars where
our modelling is in good agreement with them. The metallicity trend
for the giants is more or less recovered with an offset of
0.1\,mag. An increase of 0.2\,dex in C would be sufficient to account
for this offset at high metallicities. However, as mixing with
CN-cycled material on the giant branch would lower the C abundance,
this can not serve as an explanation for the low index values
predicted here.

\subsubsection*{Ca4227 (Fig. \ref{Ca4227})}
Ca, C and [$Z$/H] dominate the overall index strength,
$\alpha$-enhancement has no effect on it.\\ The index strengths for
the dwarfs rises steeper than the FFs and do not saturate. At solar
metallicity, the agreement is acceptable (within the 0.3\,dex
variation of Ca), somewhat better than in TB95. In the giants, a
steeper rise with metallicity is also predicted which starts at higher
metallicities only.

\subsubsection*{G4300 (Fig. \ref{G4300})}
This index is very sensitive to the C abundance, O and Fe are
important contributors as well. The sensitivity to
$\alpha$-enhancement is generally low.

The turnoff stars and dwarfs are compatible with the FFs within a
0.3\,dex variation of the C abundance. The high-metallicity giants are
also consistent and are in fact in better agreement with the FFs than
in the study of TB95. Yet the index strengths are significantly
overestimated at low metallicities. The saturation in index strength
takes place at much lower metallicities than what is deduced from the
FFs.

\subsubsection*{Fe4383 (Fig. \ref{Fe4383})}
This index is particularly sensitive to the iron abundance itself
which is no surprise given the intrinsic strength of Fe\,{\sc i} 4383
and Fe\,{\sc i} 4404 (both multiplet 41). Metallicity [$Z$/H] and the
Mg abundance also influence the index strength, the latter one via its
influence on the continuous opacity arising from H$^-$.

There is a very slight tendency for the index strength to be too high
in the turnoff stars and giants, a clear one in the case of the
dwarfs. Here, the dynamic response to $\alpha$-enhancement is also
overestimated, whereas it is well reproduced in the evolved stages of
evolution.

\subsubsection*{Ca4455 (Fig. \ref{Ca4455})}
This index is not very sensitive to any of the elements studied, at all times
below the level of 1$\sigma$. The highest variations are due to Ca
(dwarf phase), [$Z$/H] (turnoff phase) and Fe and Cr (giant phase).

In the absence of a significant response to varying elemental
abundances it is particularly troublesome to see the index perform
increasingly badly towards high metallicities, in all stages of
evolution. TB95 note the strong dependence on band-pass placement, the
highest found for any of the 21 classical Lick indices. The size of the
variation (between 4 and 8\,$\sigma$ for a \m3\,\AA\ wavelength shift
in the giant and dwarf, respectively) is in principle fully sufficient
to account for the offsets found. This might or might not be the sole
source of the discrepancy; in any case, it tells us that one has to be
careful in using this index, as also noted from the comparison
with globular cluster data (Maraston et al. \cite{Maretal03}, TMB03).

\subsubsection*{Fe4531 (Fig. \ref{Fe4531})}
For this index, variations of Ti and [$Z$/H] are important. All other
elements cause an index response at or below the 1\,$\sigma$ level.

The agreement between our model predictions and the FFs is good in the
case of the dwarfs, excellent for the turnoff stars, but mediocre for
the giants. Here, the index strengths are overpredicted at high
metallicities, which goes into the opposite direction from TB95.

\subsubsection*{C$_2$\,4668 (formerly Fe4668) (Fig. \ref{Fe4668})}
This index is almost exclusively a measure of the carbon abundance. It
shows practically no dependence on $\alpha$-enhancement.

The dwarfs do not follow the FFs at all: values too high at low
metallicity, values too low at high metallicity. At [$Z$/H]\,=\,0,
C$_2$\,4668(TB95) is at 2\,\AA\ rather than 1\,\AA\ as here, whereas the FF
is close to 5\,\AA. The turnoff stars, on the other hand, are
well-matched. A variation of C by 0.2\,dex is enough to bring the
giants into concordance with the FF values. Yet, it is quite obvious
that the metallicity dependence is not properly accounted for. TB95
succeed remarkably well in reproducing the FFs and the data on giants
in M67. Just like in the case of Fe4531, our respective baseline
values fall on opposite sides of the FFs.

\subsubsection*{H$\beta$ (Fig. \ref{Hbeta})}
This index is hardly changed by any of the studied elemental
variations, $\alpha$-enhancement has no effect, either (see Thomas et
al. 2004). The dominant variations presented in Fig.~\ref{Hbeta} are
at or below 1/3 $\sigma$ and therefore hard to trace observationally.

The expected behaviour of the index is reproduced well for all
evolutionary stages, even at low temperatures where TB95 found
systematic offsets of around 1\,\AA. We have no explanation for the
1\,\AA\ offset encountered in the $Z$\,=\,0.05\,Z$_\odot$ dwarf.

The seemingly non-monotonic behaviour of the H$\beta$ index (most
apparent in the turnoff stars) is a direct consequence of our choice
of representative stars: they are non-monotonic in $T_{\rm eff}$
causing the temperature sensitivity of the H$\beta$ line to carry
through to the index strength.

\subsubsection*{Fe5015 (Fig. \ref{Fe5015})}
In the turnoff stars, this index is only sensitive to [$Z$/H],
in the cooler stars also to Ti and Mg. This supports the findings of
TB95. $\alpha$-enhancement has some effect on the overall index
strength, in good agreement with the FFs.

This index was considered to be the overall poorest fit by TB95. On
the contrary, the FFs of all stages of evolution are reasonably
well-matched by our computations. All discrepancies can be removed by
a 0.3\,dex increase in Ti alone. This is, however, not to say that all
stars defining the FFs show super-solar [Ti/Fe] abundance ratios
(which is unlikely at or above solar metallicity). We note that the
high sensitivity to offsets in wavelength found in the TB95 dwarf (5
standard errors per \m3\,\AA) cannot explain the residual
discrepancies.

\subsubsection*{Mg$_1$ (Fig. \ref{Mg1})}
C, Mg, [$Z$/H] and Fe dominate the behaviour of this index in
decreasing order of importance. It is interesting to see how the
relative sensitivity to Mg (MgH) wins over that to C (C$_2$) towards
lower metallicities in the dwarf phase.

The overall correspondence between FFs and theoretical index strengths
is good, except for two areas: the I$_0$ values are systematically too
low in the metal-poor dwarfs and systematically too high in the
metal-rich giants. The latter was also found by TB95.

\subsubsection*{Mg$_2$ (Fig. \ref{Mg2})}
As the Mg$_2$ band-pass covers the Mg$b$ lines, the sensitivity to Mg
is higher than in the case of Mg$_1$. It is also influenced by
[$Z$/H], C and Fe.

The behaviour of Mg$_2$ as a function of metallicity is very similar
to Mg$_1$. Among the dwarfs, only the most metal-poor star shows a
significant discrepancy, among the giants it is once more the solar
and super-solar ones. In this case, no offset was reported by TB95 at
solar metallicity.

\subsubsection*{Mg$b$ (Fig. \ref{Mgb})}
The most dominant species are Mg, [$Z$/H] and Fe (plus Cr in the
dwarfs). $\alpha$ enhancement has an effect on the
$Z$\,=\,0.005\,Z$_\odot$ dwarfs and giants which is well-modelled.

The turnoff and red giant stars are well accounted for, in this index
the dwarfs are off the FFs.  This is then a feature common to all
three Mg indices. At the lowest metallicity, our calculation
predict values far below the FFs, which is due to the FFs largely
overestimating the index strengths with respect to real stars (see
Fig. 3 in Maraston et al. \cite{Maretal03}).  At high metallicities,
the behaviour of all three Mg indices is also similar (Mg$b$ performs best) and -- in contrast to the low-metallicity cases in the dwarf phase -- stays within a 0.3\,dex variation of Mg.

\subsubsection*{Fe5270 (Fig. \ref{Fe5270})}
This index is most sensitive to Fe, [$Z$/H] and Mg. A response to
$\alpha$-enhancement is practically only visible in the dwarfs where
its dynamic range is predicted correctly.

The fit is overall acceptable, displaying the same shortcomings as in
TB95: the index strengths is predicted to be somewhat too high at low
effective temperatures, both in the dwarfs and giants (the TB95 values
for giants come closer to the FFs).

\subsubsection*{Fe5335 (Fig. \ref{Fe5335})}
This index behaves very much like Fe5270. Its response is dominated by
[$Z$/H], Fe and Mg. The effect of $\alpha$-enhancement is more
pronounced in the computations than what is encoded in the FFs.

Just like in Fe5270, the theoretical indices for high-metallicity
dwarfs and giants are too strong by typically 1\,\AA. Again, the
results of TB95 are closer to the FFs.

\subsubsection*{Fe5406 (Fig. \ref{Fe5406})}
The response of this index is practically identical to that of Fe5335
(see above). For a direct comparison, Fe5270, Fe5335 and Fe5406 are
all plotted on the same scale in Figures \ref{Fe5270} and
\ref{Fe5406}.

This index performs best of the three above-mentioned iron
indices. Deficits only exist in the high-metallicity giants where the
index strengths of the FFs is overpredicted by 0.5\,\AA. Considering
that the overall index strength is smaller than in the case of Fe5270
and Fe5335, the fractional deficits are all of the same size. It is
then observationally easier to use Fe5270 or Fe5335.

\subsubsection*{Fe5709 (Fig. \ref{Fe5709})}
Another iron index quite similar to Fe5406, but with lower sensitivity to Fe. Interestingly, in the
high-metallicity giants it is more sensitive to Ti than to Fe itself. All variations cease to be significant at
$Z$\,=\,0.005\,Z$_\odot$. The dependence on $\alpha$-enhancement is well-modelled.

The high-metallicity giants are once again predicted to have index
strengths some 0.5\,\AA\ larger than what is observed. This finding is
not shared by TB95, although their prediction keeps rising towards
cooler temperatures while the FF level off and decline at around
4200\,K.

\subsubsection*{Fe5782 (Fig. \ref{Fe5782})}
As already noted by TB95, this index measures Cr rather than Fe,
[$Z$/H] plays some r\^{o}le as well.

The predictions are too low at high metallicity and vice versa making
the metallicity trends generally flatter than in the FFs. Except for
the lowest-metallicity dwarfs and giants, all indices are compatible
with the FF within the dynamic range of a 0.3\,dex variation in Cr.

\subsubsection*{Na\,D (Fig. \ref{NaD})}
This index depends strongly on Na, other than that most significantly
on [$Z$/H]. $\alpha$-enhancement has little effect whatsoever.

The reproduction of the FFs presented here is quite good overall, in
the case of the dwarfs (where deficits are clearly present) better
than by TB95. Except for the lowest-metallicity point, all deficits
are confined to a $\pm$0.3\,dex variation in Na.

\subsubsection*{TiO$_1$ \& TiO$_2$  (Fig. \ref{TiO1})}
In the absence of TiO lines from our line list, the results we get are
quite comparable to those of TB95. However, we do not find a
pronounced sensitivity to C.

While TiO$_2$ is well-reproduced in the turnoff stars, there is a
general deficit of $\approx$ 0.01\,mag in TiO$1$ for all evolutionary
phases rising to $\approx$ 0.03\,mag in the super-solar
giant. Presumably, this rise is entirely due to the increased
importance of TiO at low effective temperatures.

\subsubsection*{H$\gamma$\,A (Fig. \ref{HgammaA})}
The most crucial elements for this index are Fe, C, Mg and metallicity
[$Z$/H]. Some dependence on $\alpha$-enhancement is predicted
(e.g. 1.3\,\AA\ difference between the two $Z$\,=\,0.005\,Z$_\odot$
dwarfs).

When it comes to stellar populations, this index is dominated by the
contribution of the turnoff (by the horizontal branch where
applicable). It is therefore reassuring to see that indices for the
turnoff stars are predicted correctly (within the 0.3\,dex variation
of the element dominating the index strength). Both the dwarfs and
giants fall short of the FFs by several \AA.

\subsubsection*{H$\gamma$\,F (Fig. \ref{HgammaF})}
Overall, H$\gamma$\,F behaves quite similar to H$\gamma$\,A. The
dependence on C is, however, more pronounced.

The turnoff stars are well-matched (on par with H$\gamma$\,A), the
deficits in the dwarfs and giants are smaller here (in relative
terms). On the other hand, the difference in index strengths between
dwarfs/giants and turnoff stars (as read off from the FFs) is not as
large as in H$\gamma$\,A. Taking these two competing properties into
account, it is not easy to decide which index is to be preferred.

\subsubsection*{H$\delta$\,A (Fig. \ref{HdeltaA})}
The three elements to which this pair of indices is most sensitive are
Fe, [$Z$/H] and Mg.

What was said about the behaviour of dwarfs and giants in H$\gamma$
also holds true here. The theoretical sequence for the giants is in
better agreement with the FF at high metallicities, but does not come
close to the FFs for the metal-poor objects. At the lowest
metallicities, the FFs predict the indices of the giants to
compete with those of the turnoff stars, a finding not supported by
our calculations.

\subsubsection*{H$\delta$\,F (Fig. \ref{HdeltaF})}
This index is less sensitive to Fe than H$\delta$\,A. Additionally,
the difference in index strengths between dwarfs/giants and turnoff
stars is not as large as in H$\delta$\,A. Therefore, H$\delta$\,A is
to be preferred, if one wants to use this band-pass.

\section{Stellar population models}
The index responses presented a here are the key ingredient for the
stellar population models with variable element abundance ratios
presented in TMB03. In these models the index response functions of
TB95 have been used, which have, as mentioned in the Introduction, the
major shortcoming to be calculated only for a $5\,$Gyr isochrone with
solar metallicity. The main motivation for this work was to improve on
these simplifications. In this section, we discuss the impact of the
present paper on stellar population models, and compare stellar
population models using the new response functions of this work with
the models presented in TMB03 based on TB95.

\subsection{Inclusion of the response functions}
Before, we briefly summarize, how the calculations of this paper are
included in stellar population models. For details we refer to TMB03.

\subsubsection{The evolutionary phases}
Like TB95, we have measured on each model star -- dwarfs, turnoff,
giants -- the absolute Lick index value $I_0$. Doubling in turn the
abundances $X_i$ of the dominant elements C, N, O, Mg, Fe, Ca, Na, Si,
Cr, and Ti, we determine the index changes $\delta I(i)$.  The
abundance effects are therefore isolated at a given temperature and
surface gravity.  In this way, we obtain the first partial derivative
$\partial I/\partial [X_i]$ of the index $I_0$ for the logarithmic
element abundance increment $\delta [X_i]\equiv\log X_i^1/X_i^0=\log
2=0.3\ {\rm dex}$.

As discussed in the previous sections, the $I_0$ values of our
calculations (like in TB95) do not always match the ones derived from
the (purely empirical) FFs. In TMB03, it was therefore decided to rely
on the values in a differential sense, and adopt only the index
variations ($(\partial I/\partial [X_i])\times 0.3$.  The absolute
$I_0$ values for the three evolutionary phases, instead, are taken
from our underlying 5~Gyr, \Zsun\ SSP model.

The model index is computed by splitting the basic SSP model in the
three evolutionary phases, dwarfs, turnoff stars and giants. We
compute the Lick indices of the base model for each phase separately,
and modify them using the fractional responses $\delta I/I_0$.

\subsubsection{Negative indices}
As extensively discussed in TMB03, this approach assumes that the
index approaches asymptotically the value zero for very low element
abundances, i.e.\ $I\rightarrow 0$ for $X_i\rightarrow 0$ or
$[X_i]\rightarrow -\infty$. This condition, however, is not generally
fulfilled for Lick indices. They can become negative, depending on the
definition of line and pseudo-continuum windows. This typically
happens at young ages and/or low abundances, and must be corrected
before applying fractional index responses.  In TMB03 this problem is
solved by shifting negative index values such that they approach zero
at zero element abundances.  After applying the response functions,
the index is scaled back (see TMB03 for details).  Indices with
positive values were assumed to reach the value zero at zero
abundances and therefore did not require any correction. This approach
seems a bit approximative, but turns out to work very well.

The fact that we have response functions at every metallicity allows
us now to improve on this method. We now apply the fractional
responses to the flux in the absorption line directly, as the latter
is always positive. The absorption index $I$ (measured in \AA) is
linked with the fluxes in line $F_l$ and continuum $F_c$ through the
following equation ($\Delta$ is the bandwidth, see Maraston et
al. (\cite{Maretal03}) for more details and for the equivalent
definition for indices measured in magnitudes):
\begin{equation}
I=\Delta\cdot (1-F_l/F_c)
\label{eqn:indexdef}
\end{equation}
The index variations tabulated in Tables~6\,--\,32 are converted to
variations in $F_l$ with the following equation which can be derived from
Eqn.~\ref{eqn:indexdef}:
\begin{equation}
\frac{\delta F_l}{F^0_l}=\frac{\delta I}{I_0-\Delta}\ .
\end{equation}
Equivalent to Eqn.~7 in TMB03, the fractional responses of the flux in
the line are then multiplied as follows:
\begin{equation}
F_l^{\rm new} = F_l\ \prod_{i=1}^{n}\:\exp
\left\{\frac{1}{F^0_l}\frac{\partial F_l}{\partial
[X_i]}\; 0.3\right\}^{\left(\Delta [X_i]/0.3\right)}\nonumber
\end{equation}
The new index is then computed from $F_l^{\rm new}$ with
Eqn.~\ref{eqn:indexdef}.  We have verified that the resulting stellar
population models are not significantly affected when compared to
models based on the method used in TMB03.  This consistency can be
taken as further evidence that the approximation originally used in
TMB03 has been successful at dealing with negative absorption line
indices.

\subsection{Metallicity dependence}
In TMB03, it had to be assumed that the fractional responses $\delta
I/I_0$ are independent of metallicity. This approximation is sensible
in the linear part of the growth curve, in which equivalent width of
the absorption line and element abundance are linearly related. In
particular at metallicities well above solar, it is not clear whether
this assumption is still valid. The calculations of this paper allow
to test the validity of this simplification, as we have now at hand
response function in the whole metallicity range $-2.25\leq [\ZH]\leq
0.67$.

\subsection{Age dependence}
TB95 calculated the index response functions for stellar parameter
pairs that correspond to an isochrone of 5\,Gyr, and in TMB03 it is
assumed that this age restriction does not significantly impact on the
resulting SSP model. In this paper we have now the tools available to
check the validity of this assumption. For this purpose, we have
additionally computed index response functions for a 1\,Gyr isochrone
at solar metallicity (Tables~15\,--\,17). The resulting \aFe\ enhanced
SSP model (age 1\,Gyr) is in perfect agreement with the 1\,Gyr model
based on the response functions from the 5\,Gyr isochrone
(Tables~12\,--\,14). Deviations are about 1 per cent for G4300 and
Fe4383, significantly below that value for all other indices.

\subsection{Symmetry}
TB95 computed index variations upon enhancing the abundance of an
individual element by 0.3\,dex. We also performed test calculations in
which the abundance was diminished instead. No significant changes
were found in the absolute index changes. This means that the index
variations can be used to trace element abundances both above and
below the base abundance.

\subsection{Classical Lick indices}
\begin{figure*}
\includegraphics[width=0.9\linewidth]{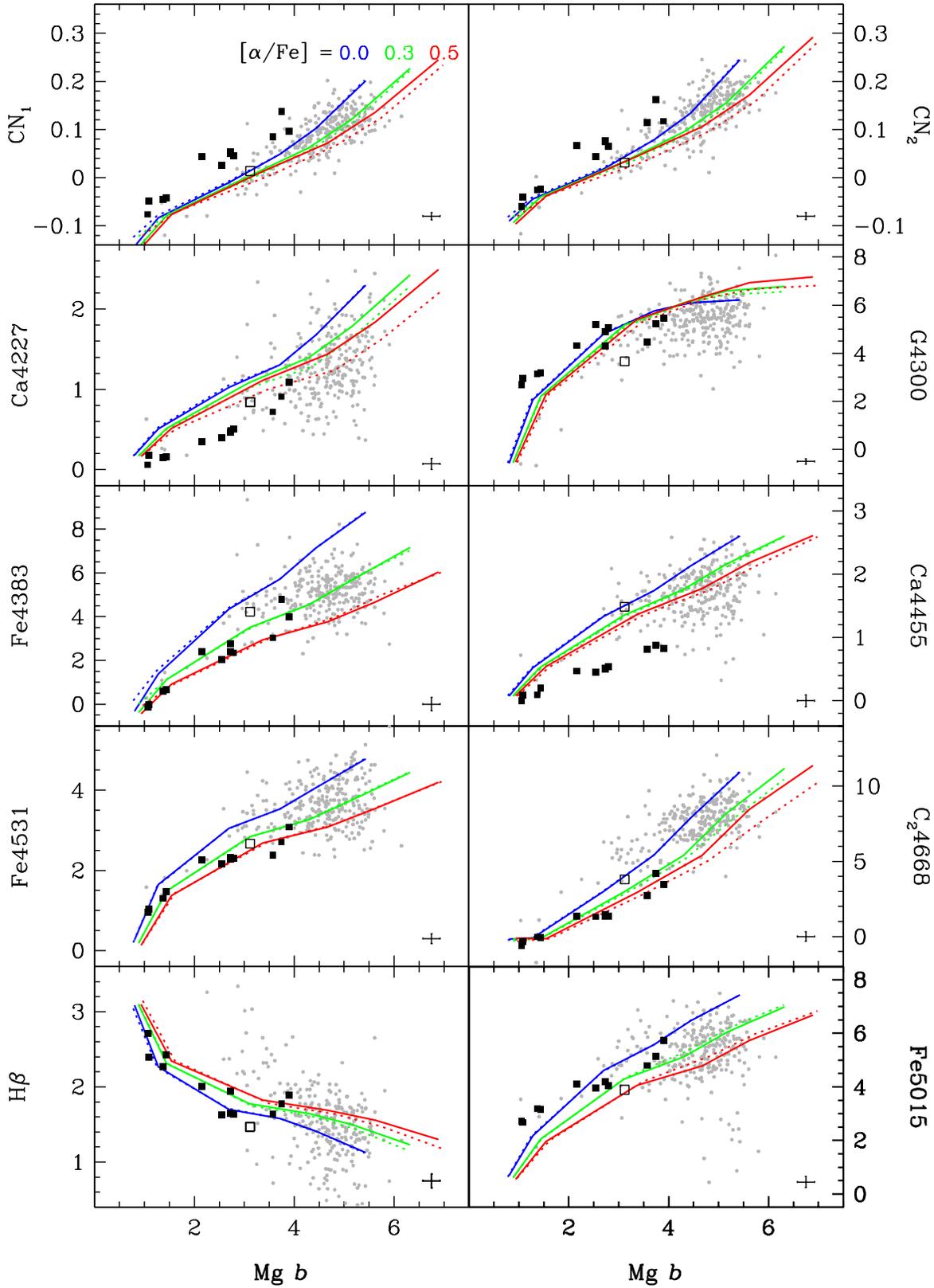}
\caption{\Mgb\ index versus the other 20 Lick indices. Solid lines are
the TMB03 models using the new response functions of this paper,
dotted lines are the models of TMB03 based on the TB95 response
functions. They are plotted for constant age (12~Gyr) and constant
\aFe\ ratios in the metallicity range $-2.25\leq [\ZH]\leq 0.67$. The
three models with $[\aFe]=0.0,\ 0.3,\ 0.5$ are shown in blue, green,
and red, respectively (see labels). Models with solar abundance ratios
($[\aFe]=0.0$) and models with $[\aFe]=0.5$ are those with the lowest
and highest \Mgb\ indices, respectively.  Filled squares are globular
cluster data, the open square is the integrated Bulge light from Puzia
et al. (2002), small grey dots are the Lick data of giant elliptical
galaxies from Trager et al. (1998). Error bars indicate typical errors
of the globular cluster data. The figure is the equivalent of Fig.~2
in TMB03 and continues on the next page}
\label{fig:ssp}
\end{figure*}
\begin{figure*}
\includegraphics[width=0.9\linewidth]{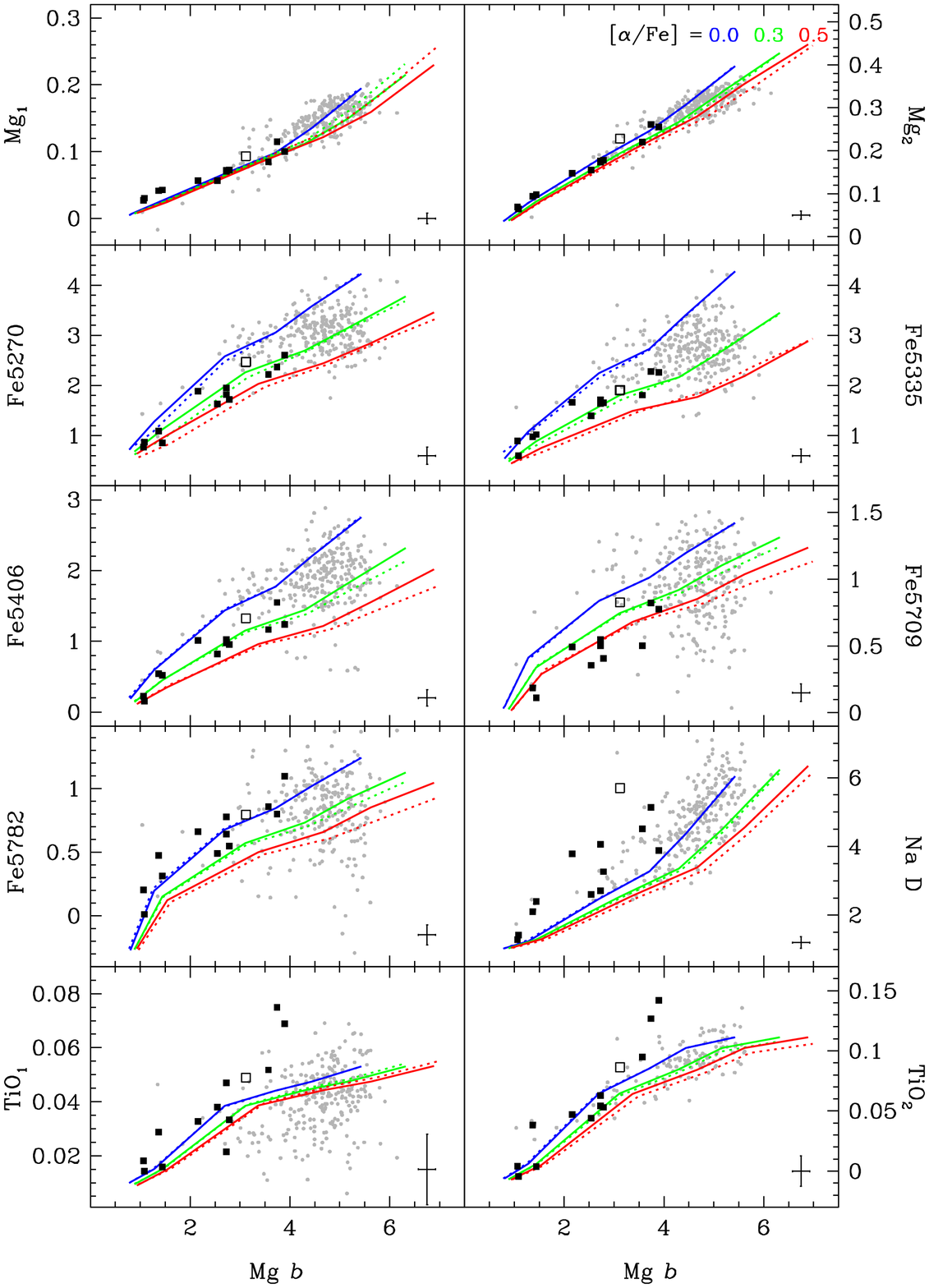}
\end{figure*}
The resulting stellar population models calibrated with galactic
globular clusters are shown in Fig.~\ref{fig:ssp}, which is the
equivalent of Fig.~2 in TMB03. We plot models for constant age
(12~Gyr) and constant \aFe\ ratios in the metallicity range $-2.25\leq
[\ZH]\leq 0.67$. Three models with $[\aFe]=0.0,\ 0.3,\ 0.5$ are shown
in blue, green, and red, respectively (see labels). Models with solar
abundance ratios ($[\aFe]=0.0$) and models with $[\aFe]=0.5$ are those
with the lowest and highest \Mgb\ indices, respectively.  Solid lines
are the TMB03 models using the new response functions of this paper,
dotted lines are the models of TMB03 based on the TB95 response
functions.  Filled squares are globular cluster data, the open square
is the integrated Bulge light (both from Puzia et al. 2002), small
grey dots are the Lick data of giant elliptical galaxies from Trager
et al. (1998).

We recall here that the Galactic globular clusters are \aFe\ enhanced
by about a factor 2, hence the model can be considered well calibrated
when the model with $[\aFe]=0.3$ (middle green line) fits the
observational data (squares). As discussed in detail in TMB03, not all
of the Lick absorption line indices show a satisfactory calibration
and can be considered adequate for stellar population studies. For the
aim of this paper, we are more interested in the direct comparison
with the TMB03 models based on the TB95 response functions.

The main conclusion is the following: the assumption of TMB03 that the
fractional response is independent of metallicity is generally
confirmed for the whole metallicity range considered. The convergence
to minimal index variations at the lowest metallicities and the
corresponding expansion of the models with different \aFe\ ratios at
high metallicities is present also in the models with the
metallicity-dependent response functions presented here. For most
indices, the two models are well consistent within the calibration
uncertainty introduced by observational errors. If discrepancies are
present, they do not originate from the neglect of the metallicity
dependence, but are caused by differences between the response
functions presented here and those of TB95.

The indices \CNone, \CNtwo, \Hb, \Mgb, \Mgtwo, Fe5270, NaD, and
\TiOtwo\ exhibit only relatively small offsets, while the indices
G4300, Fe4383, Ca4455, Fe4531, Fe5015, Fe5335, and \TiOone\ basically
did not change. Largest differences are found for Ca4227, \FeC,
\Mgone, Fe5406, Fe5709, and Fe5782.

Ca4227 and \FeC\ have become more sensitive to \aFe\ ratio, and increase
now slightly with increasing \aFe\ ratios. In both cases, the main
reason is a more pronounced negative response to the element Cr, the
abundance of which decreases relative to the $\alpha$-elements in the
enhanced models. As mentioned earlier, Ca4227 and \FeC\ are most
sensitive to their name-giving elements, Ca and C, respectively, which
do not vary significantly in the \aFe\ enhanced mixtures of the TMB03
models.

\Mgone\ is now somewhat less sensitive to \aFe\ because of slightly
smaller responses to both Mg and Fe in the response function of this
work. Fe5406 and Fe5709 decrease less strongly with increasing \aFe\
because the sensitivity to the element Fe is less pronounced. Fe5782
is interesting, as its sensitivity to \aFe\ ratio comes only from the
positive response to abundance changes of the element Cr. In the
response functions of this paper, Fe5782 even decreases with
increasing Fe abundances, hence counterbalances the positive response
to Cr, which leads to the lower sensitivity to \aFe\ ratios. For all
the three Fe indices, this effect slightly increases with increasing
metallicity.

\Hb\ is the only case, in which the metallicity dependence of the
response functions leaves a small, but measurable trace. The higher
the metallicity, the larger is the \Hb\ strength of the \aFe\ enhanced
models with the new metallicity-dependent response functions relative
to the ones based on TB95. The reason is the negative response to Cr,
due to a Cr line at $4885\,$\AA\ in the red pseudo-continuum of the
index definition. It makes indeed sense that the influence of metallic
lines on a Balmer-line index increases with increasing
metallicity. Still, the effect is small. In the highest metallicity
model ($[\ZH]=0.67$), where the effect is maximized, the doubling of
the \aFe\ ratio leads to an increase of the \Hb\ index by less than 10 per cent.

This is significantly less than the increase of about 100 per cent
recently claimed by Tantalo \& Chiosi (\cite{tc04}) on the basis of
the TB95 response functions (in contrast to the results in
TMB03). According to Tantalo \& Chiosi (\cite{tc04}) this dramatic
increase of \Hb\ results from a strong sensitivity of \Hb\ to Ti
abundance in the dwarf evolutionary phase. Inspecting the response
function of TB95, however, reveals that \Hb\ in the dwarfs actually
decreases with increasing Ti, which leads to the conclusion that the
result of Tantalo \& Chiosi (\cite{tc04}) is likely a numerical
artefact. It should also be noted that the models of Tantalo \& Chiosi
(\cite{tc04}) predict such strong \Hb\ indices (of the order of
$2\,$\AA) that the vast majority of early-type galaxies would be about
$30\,$Gyr old, which is significantly older than the Universe. It is
very important to note that in the response functions of this work,
the sensitivity of \Hb\ to Ti abundance does not exceed the few per
cent level in all evolutionary phases.

\subsection{Higher-order Balmer-line indices}
As mentioned earlier, \Hb\ shows little sensitivity to element ratio
variations as well as to total metallicity, because of the lack of
strong metallic lines in the wavelength range of the index
definition. At bluer wavelengths, this situation naturally
changes. The higher-order Balmer-line indices \Hg\ and \Hd\ around
4340 and $4100\,$\AA\ as defined in Worthey \& Ottaviani (1997) are
significantly more affected by metallic lines. A well-known direct
consequence is that these indices are more sensitive to metallicity
than \Hb. But they are also significantly more sensitive to the \aFe\
ratio as shown here. With increasing metallicity, the influence of
metallic lines rises. As a consequence, unlike for \Hb, the fractional
index responses increase considerably with metallicity as can be
appreciated by comparing Figs.~5 and 8. The impact on the stellar
population models is dramatic as shown in Thomas, Maraston, \& Korn
(2004).  This effect cannot be neglected when these line indices are
used to derive the ages of metal-rich, unresolved stellar populations
like early-type galaxies. In Thomas et al.\ (2004) we show that
consistent age estimates from \Hb\ and \HgA\ are obtained, only if the
effect of \aFe\ enhancement on \HgA\ is taken into account in the
models.  This result rectifies a problem currently present in the
literature, namely that \Hg\ and \Hd\ have up to now led to
significantly younger ages for early-type galaxies than \Hb\ (see
Thomas et al. 2004 for more details).

\section{Conclusions}
We have computed line indices for the 21 classical and 4 higher-order
Balmer-line Lick/IDS indices at six different metallicities ranging
from $-2.25$ to +0.67 in [\ZH]. At the lowest two metallicities
($-$2.25 and $-$1.35) we also computed indices with a general
enhancement of $\alpha$ elements. For all of these, we have varied
individual elemental abundances of ten elements (plus overall
metallicity) and tabulated the index changes. At solar metallicity, we
compare our results with the work of Tripicco \& Bell
(\cite{TB95}). In the majority of cases, we confirm the findings to
TB95. By tabulating index changes for all 25 Lick indices for a wide
range of metallicities, we significantly extend the work of TB95.

The major assumptions of TMB03 were checked and verified by
these calculations: the use of fractional changes produces accurate
results over a wide range of abundances and ages. They can also be
used to compute response functions with enhanced or diminished
abundances of individual elements (notably $\alpha$ elements).
Still, the metallicity-dependent response functions
presented here do lead to a higher degree of self-consistency in the
stellar population models. Furthermore, their use turns out to be of particular
importance for the Balmer-line indices. We find that, different from
the metallic indices, the Balmer-line indices become increasingly sensitive
to element abundances with increasing metallicity and decreasing
wavelength.

Hence, while the \Hb\ index only responds moderately (below 10 per cent)
to abundance ratio variations, the higher-order Balmer-line indices \Hg\
and \Hd\ display very strong dependencies at high metallicities. More
specifically, \Hg\ and \Hd\ significantly increase with increasing
alpha/Fe ratio. This effect must not be neglected when these
indices are used as age indicators of unresolved stellar systems.  As
shown in Thomas, Maraston, \& Korn \cite{TMK04}, the response
functions of this work allow for a proper modelling of these indices,
which helped to remove systematic effects in age determinations based
on different Balmer-line indices previously present in the
literature.

With the general availability of huge, homogeneous sets of galaxy
spectra (e.g. Madgwick et al. \cite{MCCetal03}), direct synthesis of
large wavelength ranges will become a more wide-spread ana\-lysis
technique. To extract useful information from the spectra, the
behaviour of different regions to individual abundance changes will
have to be established. As a starting point and reference, the Lick
indices serve an important purpose in this sense. By defining how they
are expected to react to abundance changes under a variety of
circumstances, we hope to shed light on the chemical evolution of
galaxies throughout the Hubble sequence.

\begin{acknowledgements}
We would like to thank Thomas Gehren (Universit\"{a}ts-Sternwarte M\"{u}nchen) for continuous support with
his atmospheric code MAFAGS. Laura Greggio, Alvio Renzini and Scott
Trager are thanked for repeated fruitful discussions. The referee, Guy Worthey, is thanked for stressing the carbon-star difficulty which prompted us to perform some extra computations. This work was in part supported by the Deutsche Akademie der Naturforscher Leopoldina (Halle, Germany) under grant
BMBF-LPD 9901/8-87.
\end{acknowledgements}

\clearpage

\begin{figure*}
   \centering
   \includegraphics[bb=388 53 570 738,angle=90,width=\textwidth,clip]{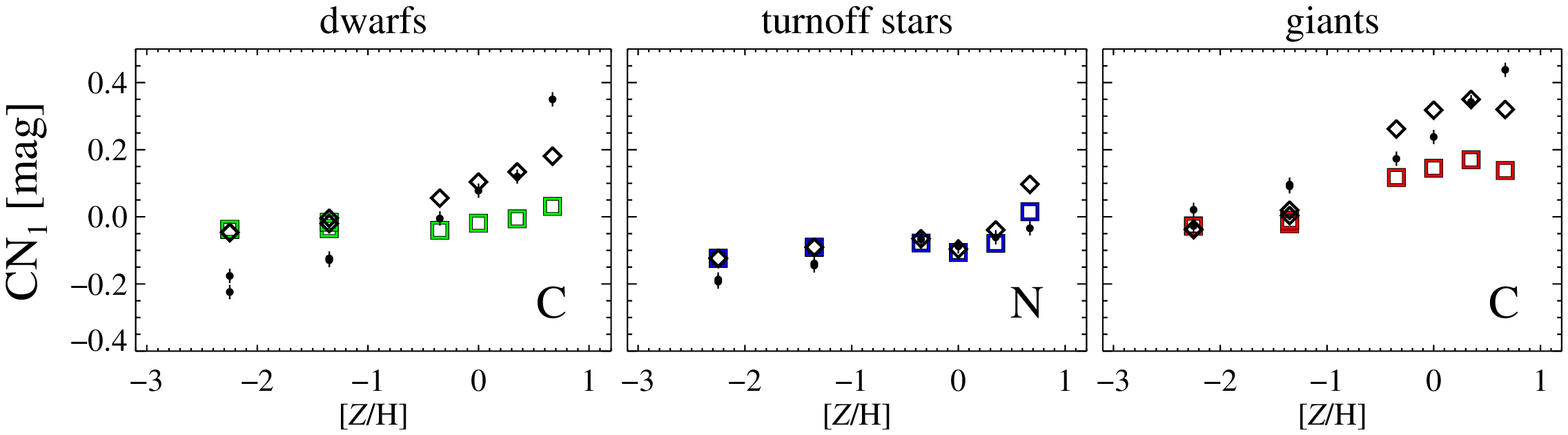}
   \centering
   \includegraphics[bb=388 53 534 738,angle=90,width=\textwidth,clip]{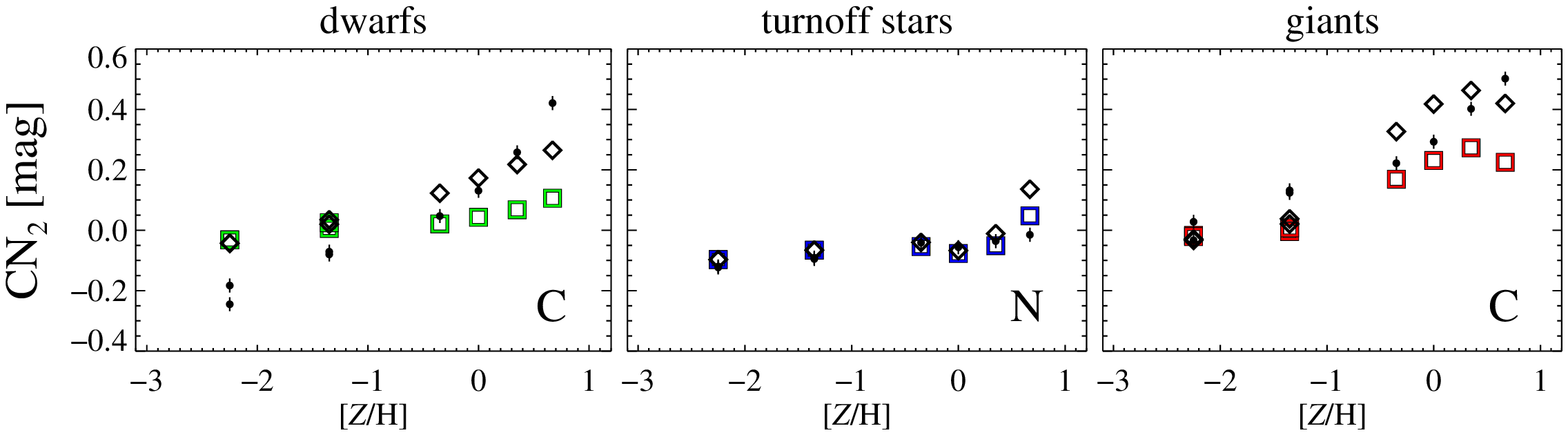}
   \centering
   \includegraphics[bb=388 53 534 738,angle=90,width=\textwidth,clip]{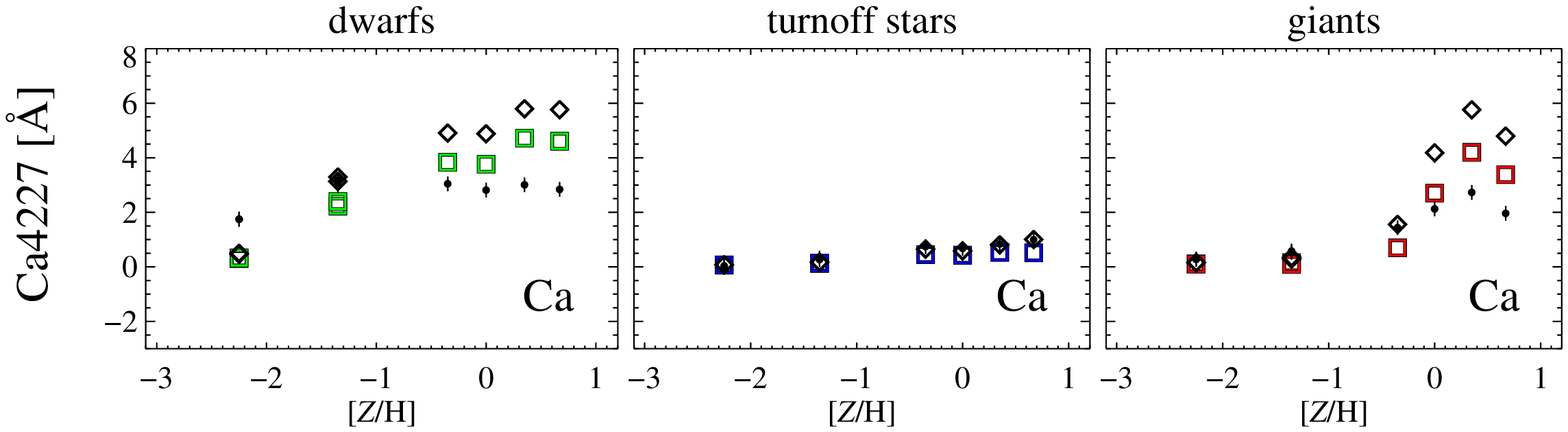}
   \centering
   \includegraphics[bb=388 53 534 738,angle=90,width=\textwidth,clip]{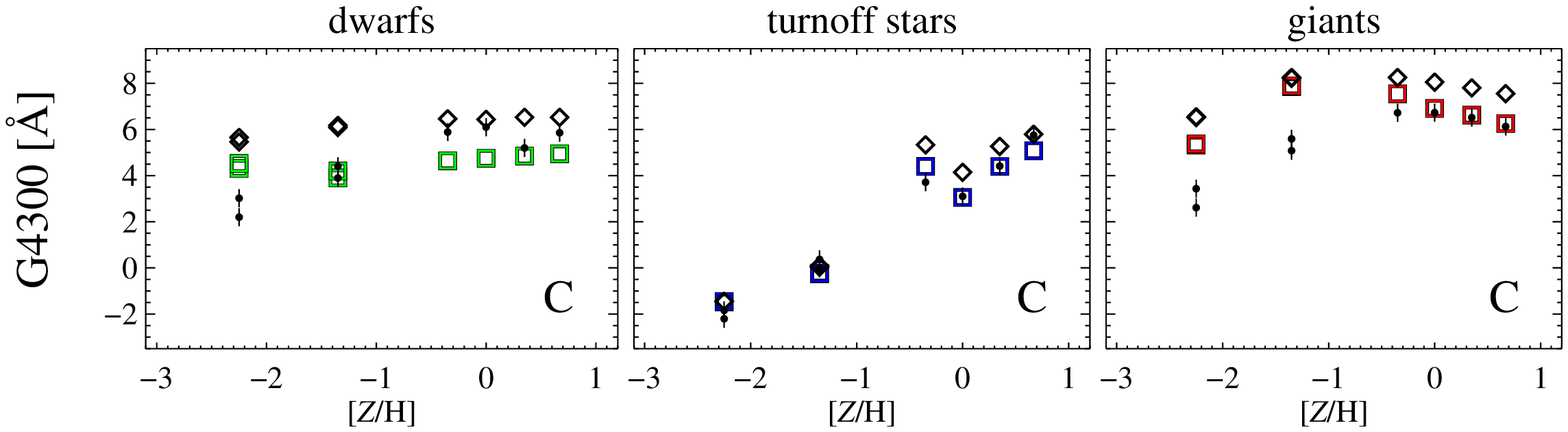}
   \centering
   \includegraphics[bb=350 53 534 738,angle=90,width=\textwidth,clip]{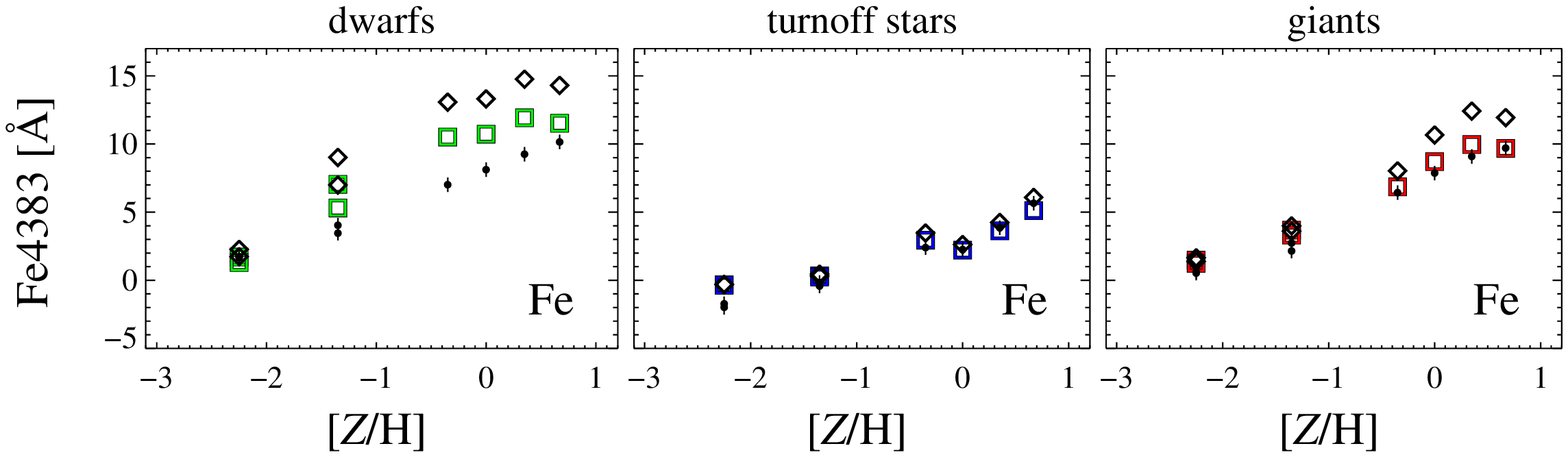}
  \caption{Theoretical index-strength variations as a function of metallicity for CN1, CN2, Ca4227, G4300 and Fe4383 in the three evolutionary stages dwarf (left), turnoff (middle) and giant phase (right). Squares denote index strengths from this work, diamonds the respective index strength obtained by enhancing the index-dominating element (given in the lower right corner) by 0.3\,dex. Bullets are the FFs {\bf with IDS standard errors} of Worthey et al. (\cite{WFGB94})}
         \label{Fe4383}\label{CN1}\label{CN2}\label{Ca4227}\label{G4300}
         \vspace*{-7mm}
\end{figure*}

\clearpage

\begin{figure*}
   \centering
   \includegraphics[bb=388 53 570 738,angle=90,width=\textwidth,clip]{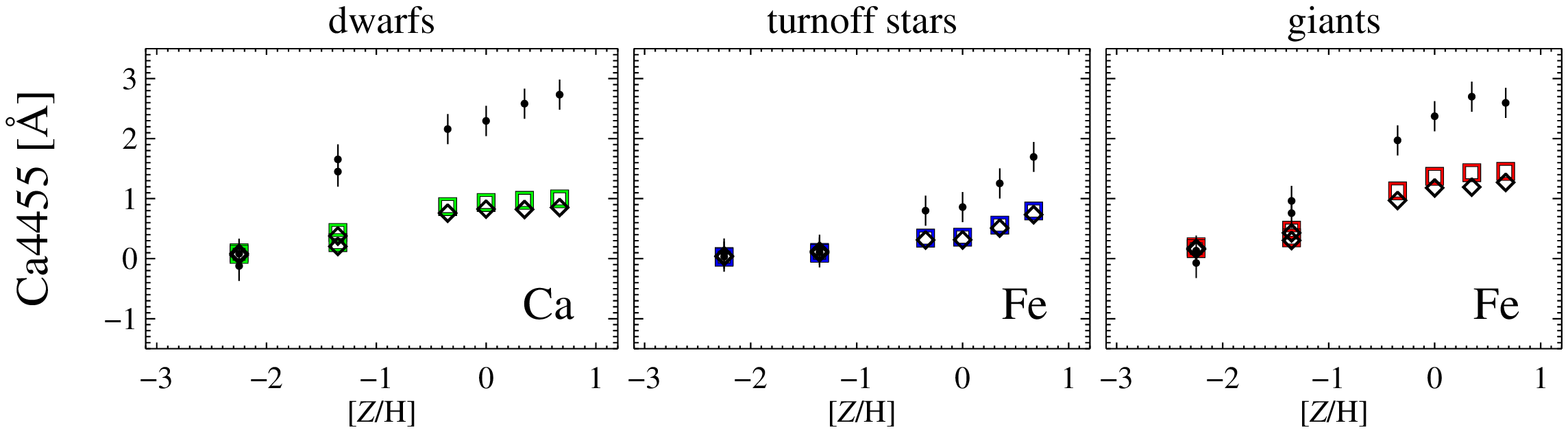}
   \centering
   \includegraphics[bb=388 53 534 738,angle=90,width=\textwidth,clip]{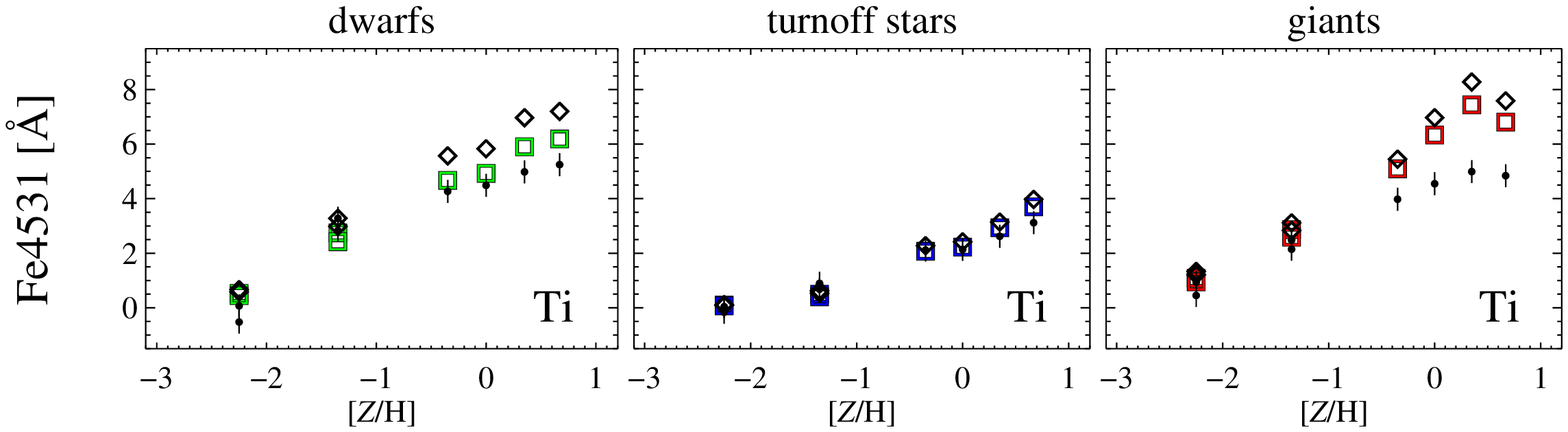}
   \centering
   \includegraphics[bb=388 53 534 738,angle=90,width=\textwidth,clip]{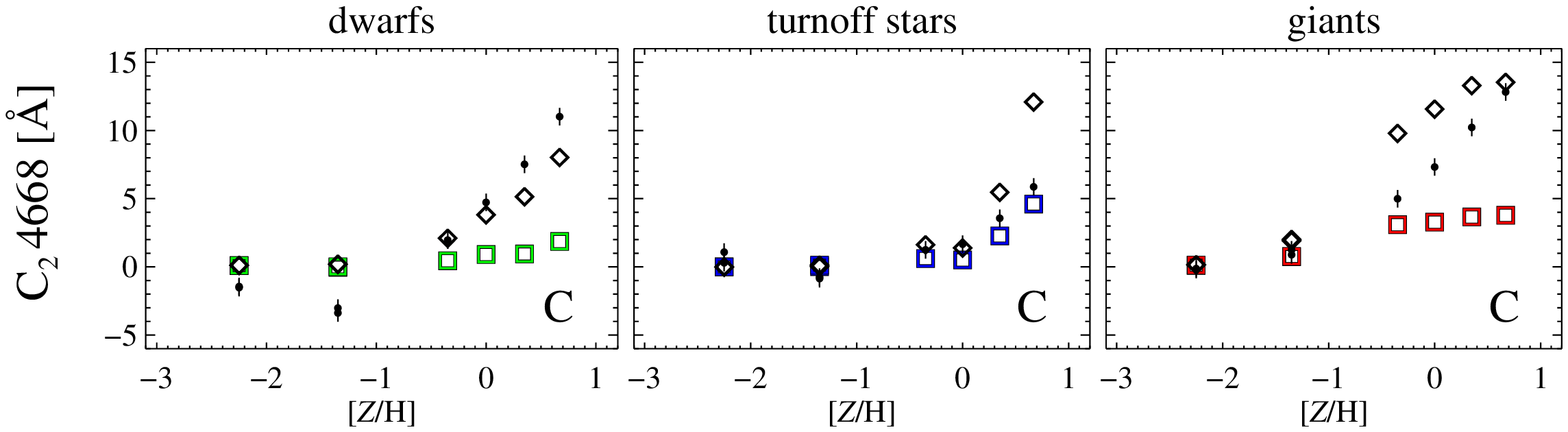}
   \centering
   \includegraphics[bb=388 53 534 738,angle=90,width=\textwidth,clip]{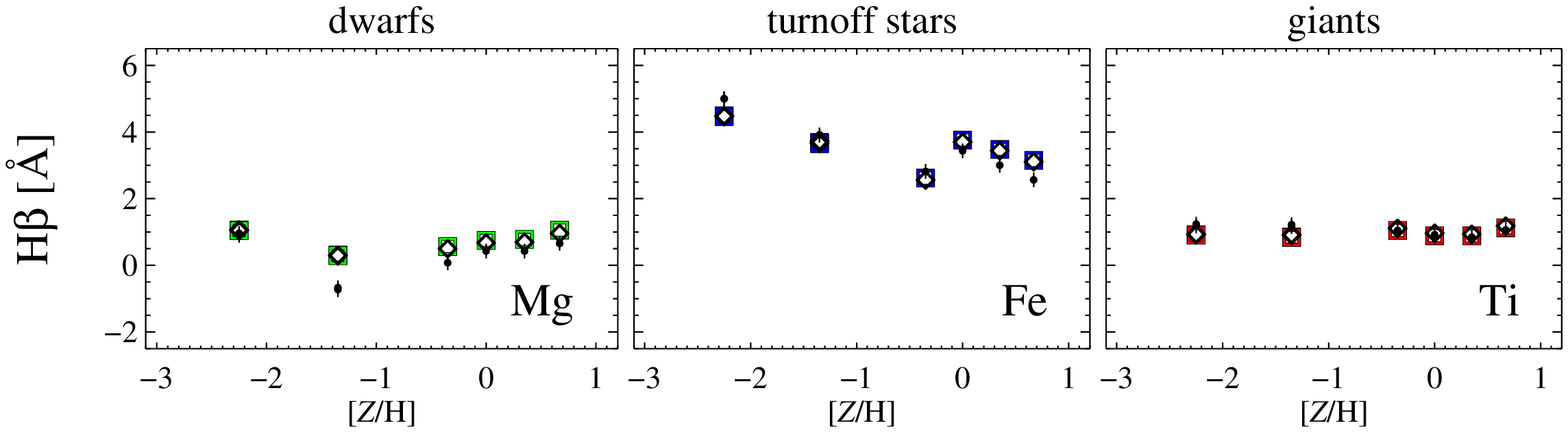}
   \centering
   \includegraphics[bb=350 53 534 738,angle=90,width=\textwidth,clip]{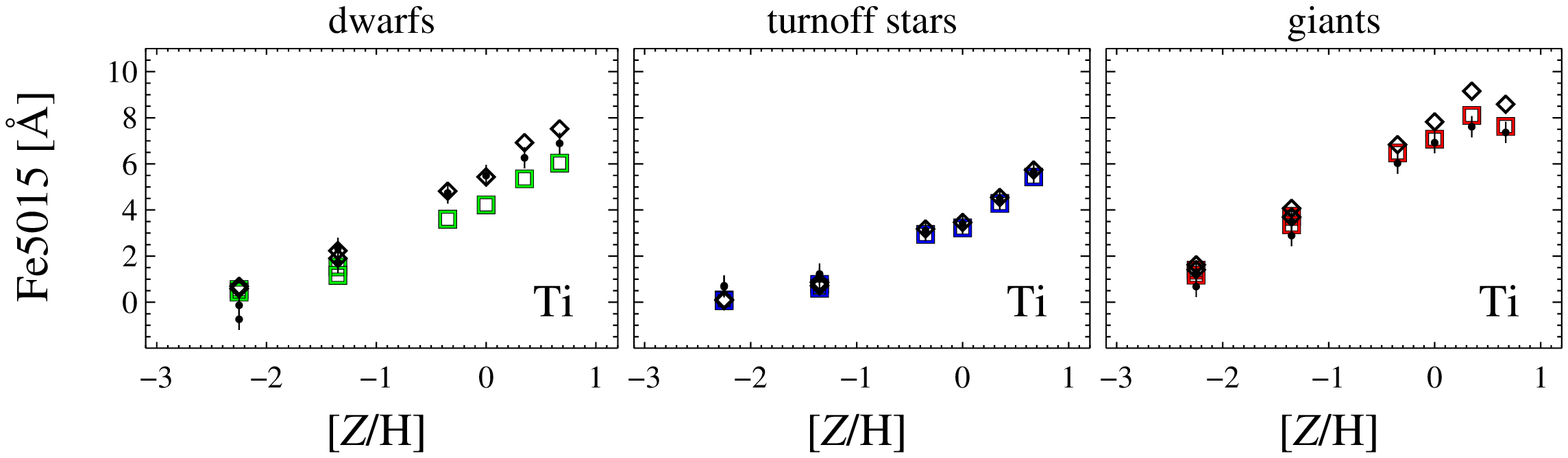}
   \caption{Same as Fig.~\ref{Fe4383}, here for Ca4455, Fe4531, C$_2\,$4668, H$\beta$ and Fe5015}
         \label{Fe5015}\label{Ca4455}\label{Fe4531}\label{Fe4668}\label{Hbeta}
\end{figure*}

\clearpage

\begin{figure*}
   \centering
   \includegraphics[bb=388 53 570 738,angle=90,width=\textwidth,clip]{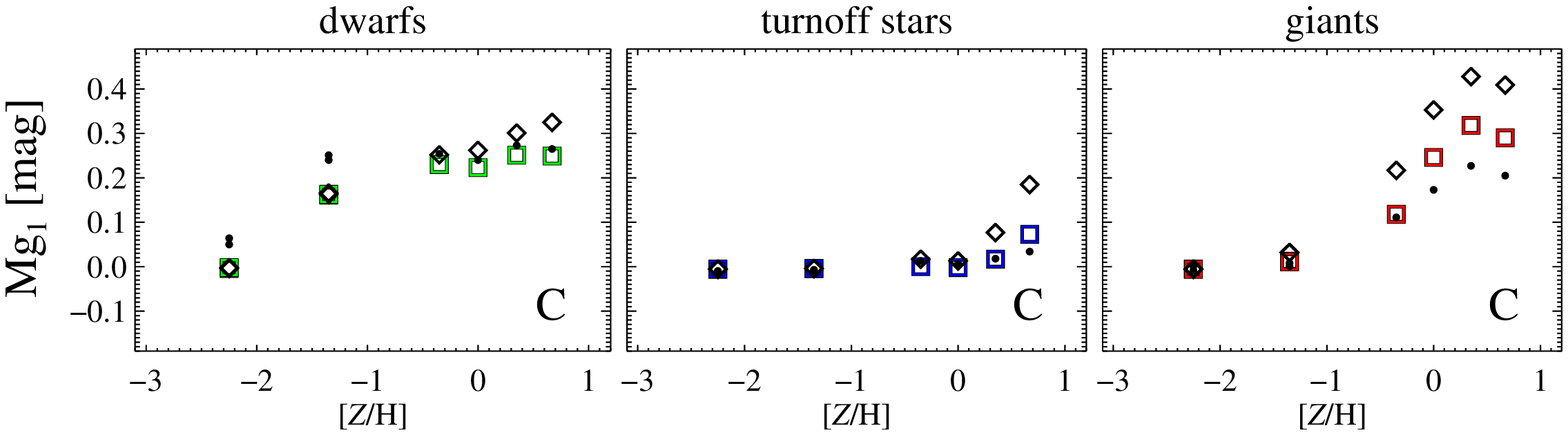}
   \centering
   \includegraphics[bb=388 53 534 738,angle=90,width=\textwidth,clip]{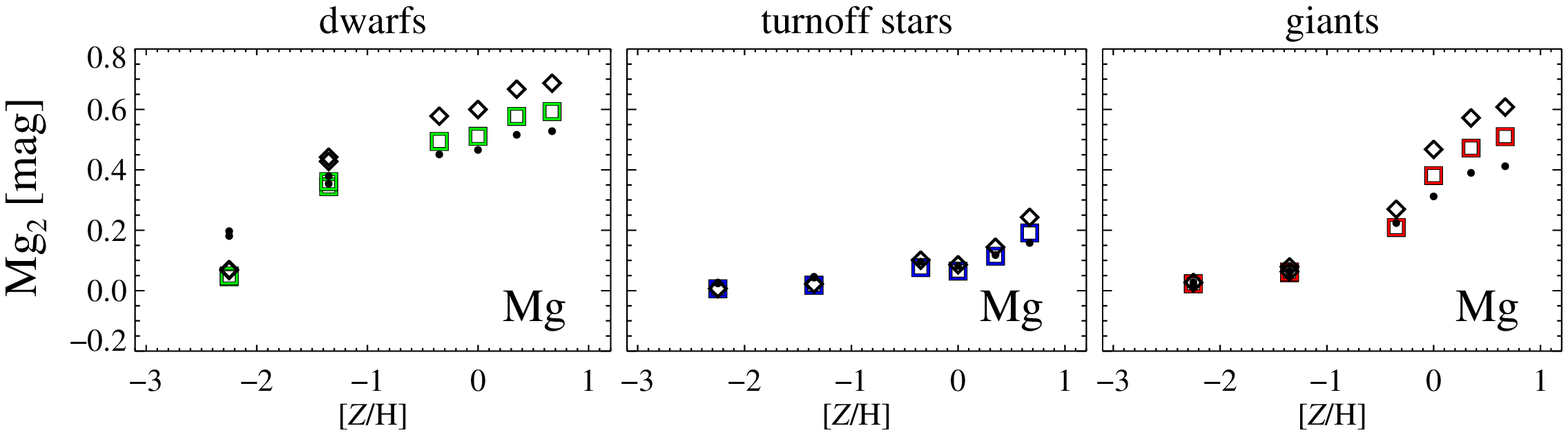}
   \centering
   \includegraphics[bb=388 53 534 738,angle=90,width=\textwidth,clip]{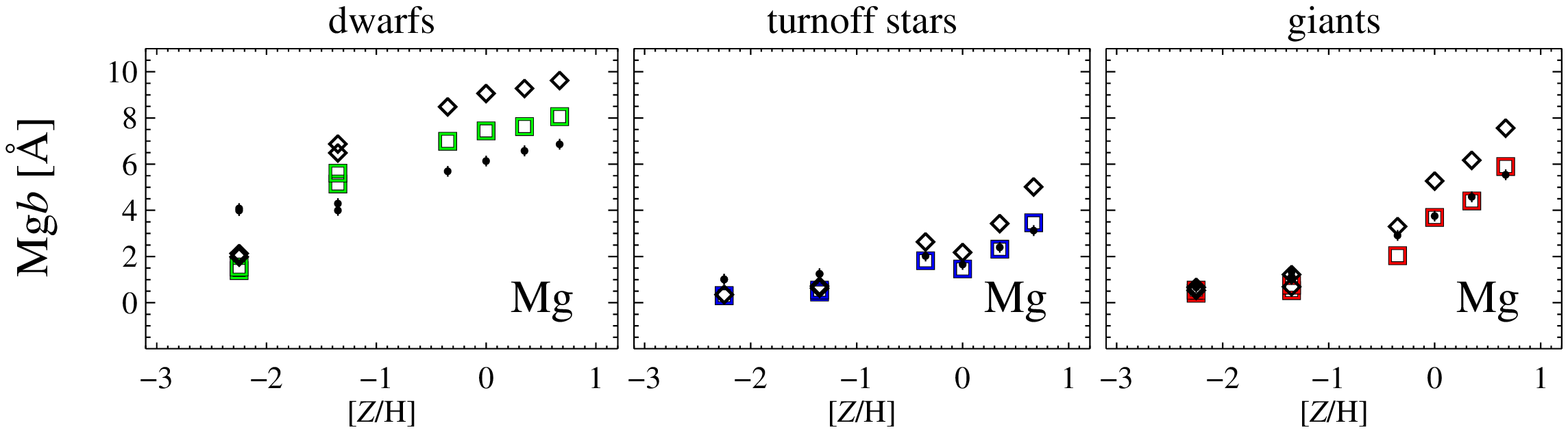}
   \centering
   \includegraphics[bb=388 53 534 738,angle=90,width=\textwidth,clip]{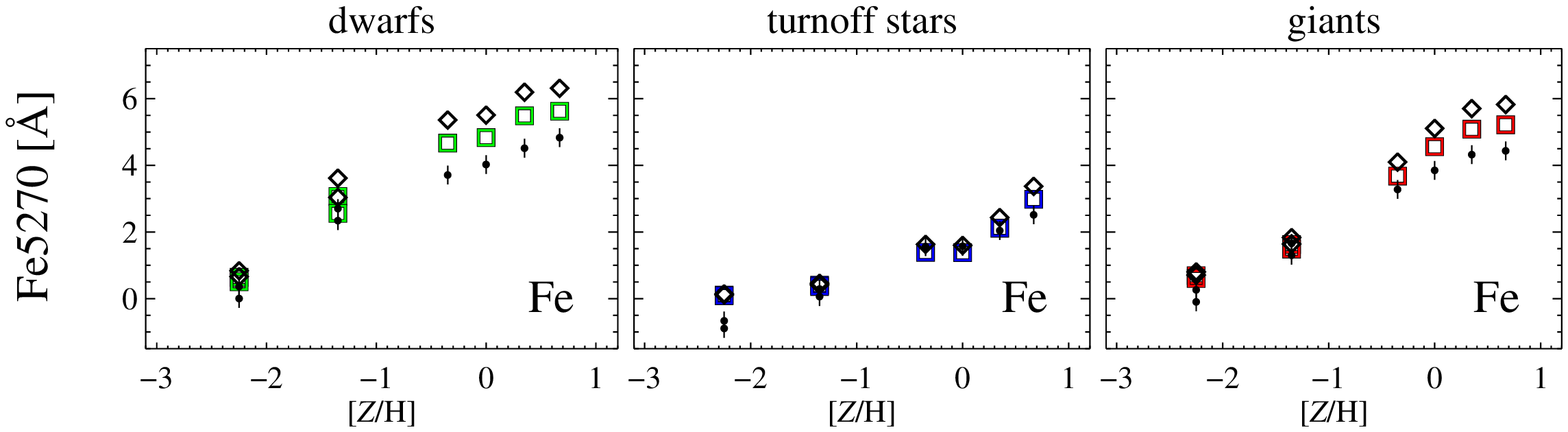}
   \centering
   \includegraphics[bb=350 53 534 738,angle=90,width=\textwidth,clip]{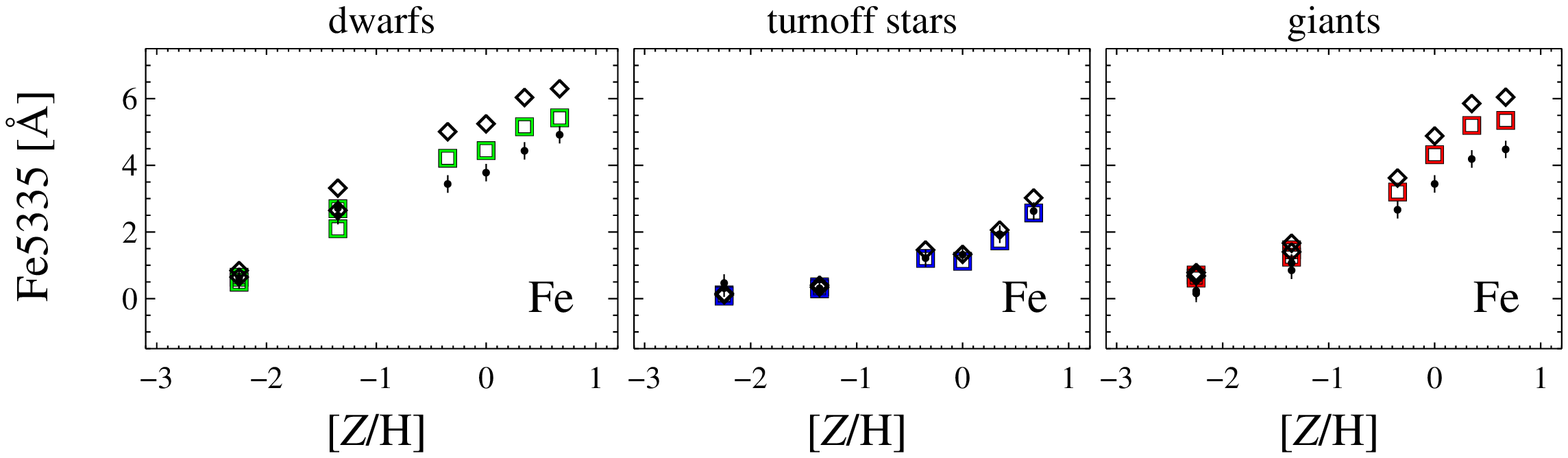}
   \caption{Same as Fig.~\ref{Fe4383}, here for Mg$_1$, Mg$_2$, Mg$_b$, Fe5270 and Fe5335}
         \label{Fe5335}\label{Mg1}\label{Mg2}\label{Mgb}\label{Fe5270}
\end{figure*}

\clearpage

\begin{figure*}
   \centering
   \includegraphics[bb=388 53 570 738,angle=90,width=\textwidth,clip]{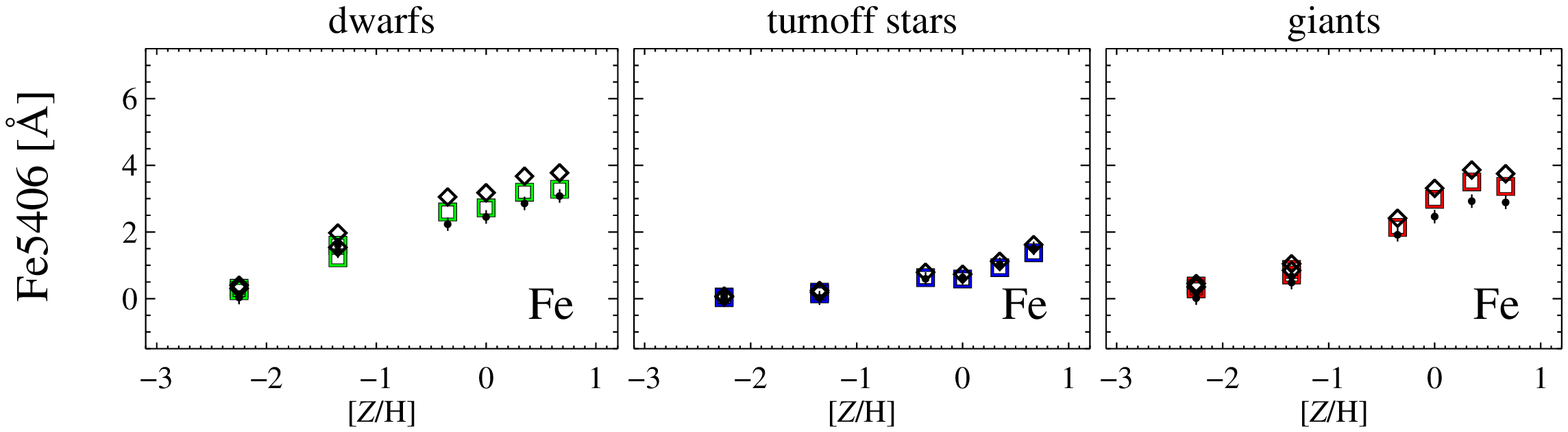}
   \centering
   \includegraphics[bb=388 53 534 738,angle=90,width=\textwidth,clip]{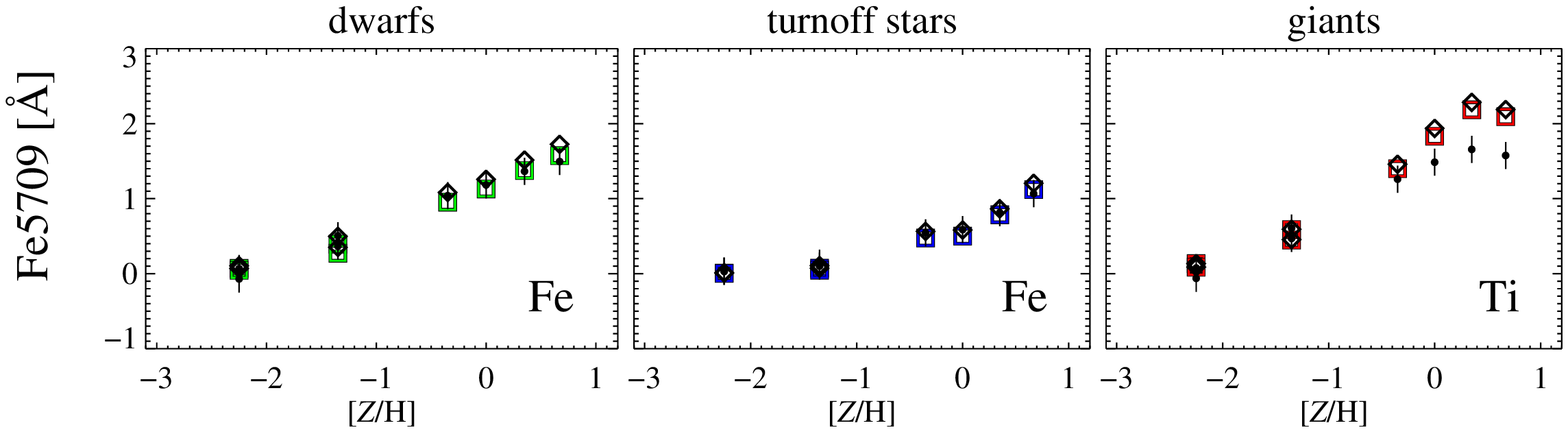}
   \centering
   \includegraphics[bb=388 53 534 738,angle=90,width=\textwidth,clip]{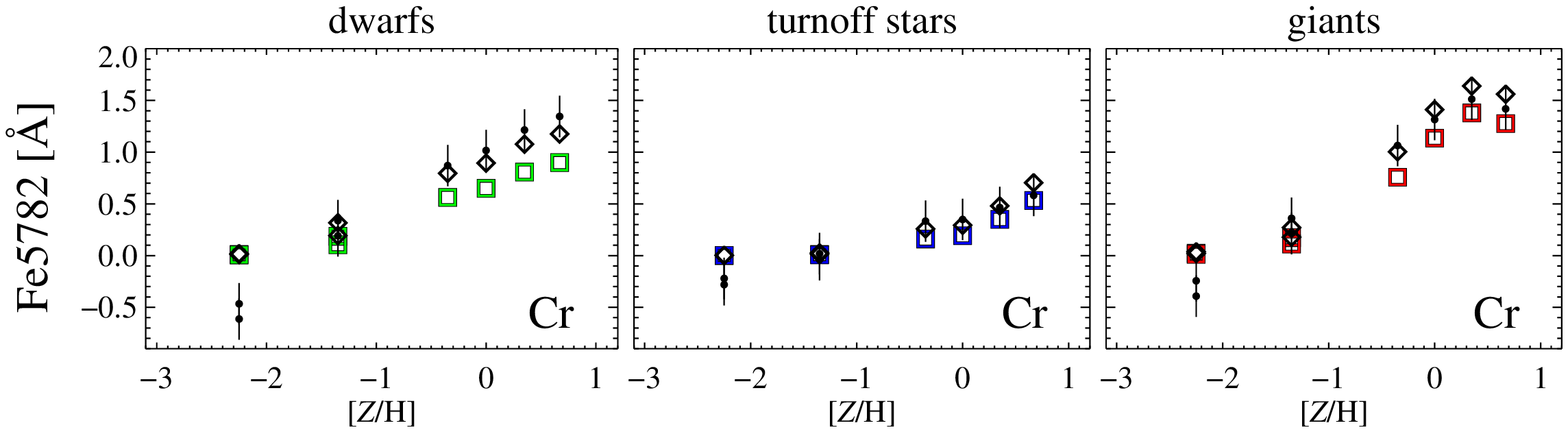}
   \centering
   \includegraphics[bb=388 53 534 738,angle=90,width=\textwidth,clip]{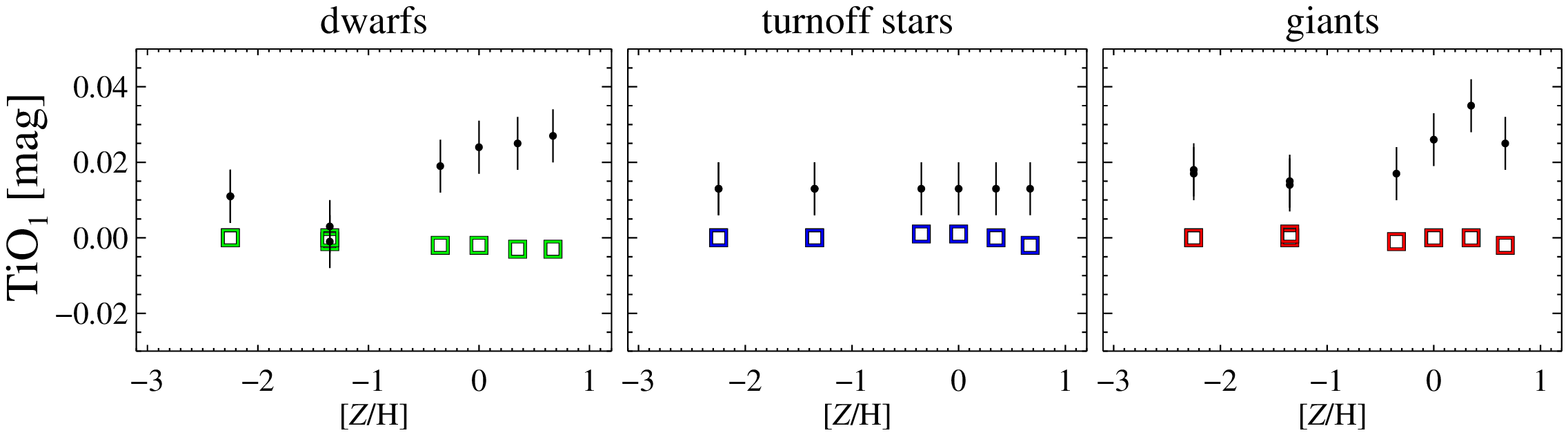}
   \centering
   \includegraphics[bb=350 53 534 738,angle=90,width=\textwidth,clip]{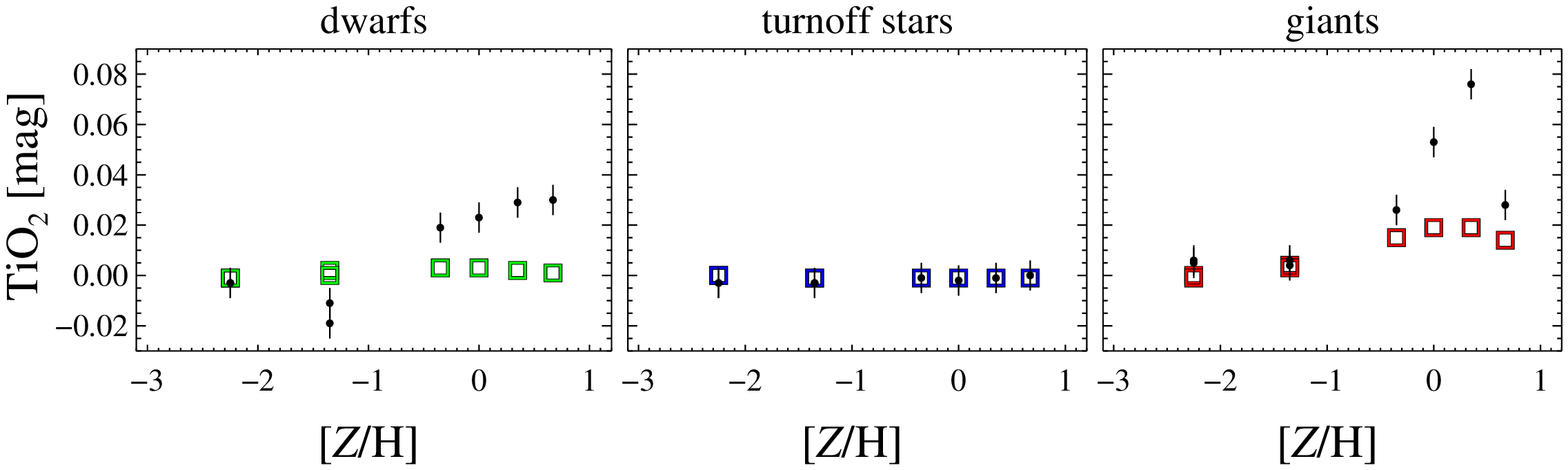}
   \caption{Same as Fig.~\ref{Fe4383}, here for Fe5406, Fe5709, Fe5782, TiO$_1$ and TiO$_2$}
         \label{TiO2}\label{Fe5406}\label{Fe5709}\label{Fe5782}\label{TiO1}
\end{figure*}

\clearpage

\begin{figure*}
   \centering
   \includegraphics[bb=388 53 570 738,angle=90,width=\textwidth,clip]{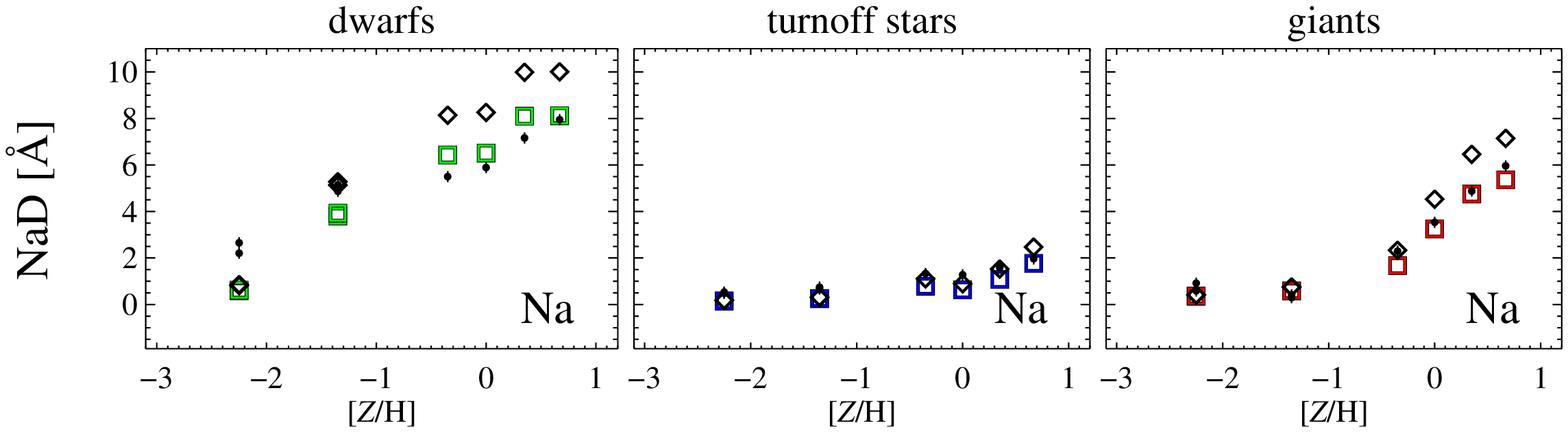}
   \centering
   \includegraphics[bb=388 53 534 738,angle=90,width=\textwidth,clip]{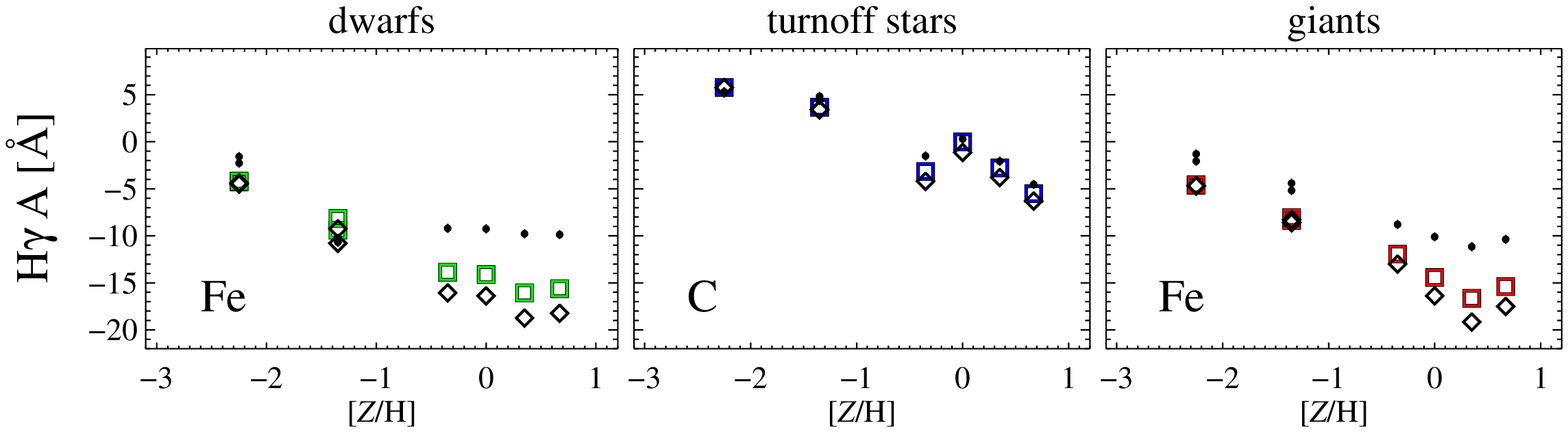}
   \centering
   \includegraphics[bb=388 53 534 738,angle=90,width=\textwidth,clip]{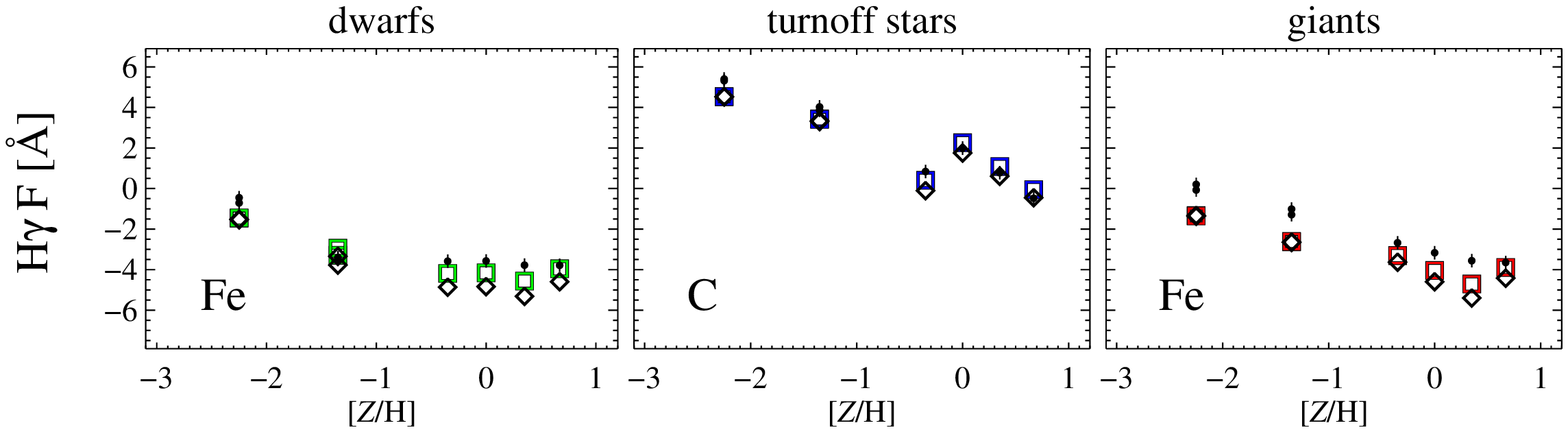}
   \centering
   \includegraphics[bb=388 53 534 738,angle=90,width=\textwidth,clip]{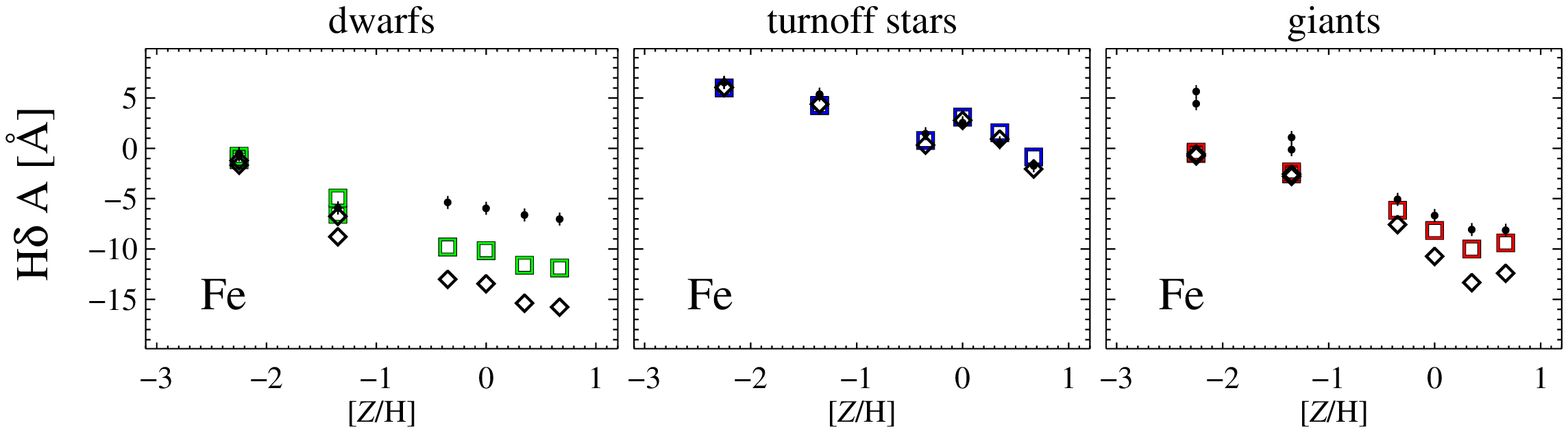}
   \centering
   \includegraphics[bb=350 53 534 738,angle=90,width=\textwidth,clip]{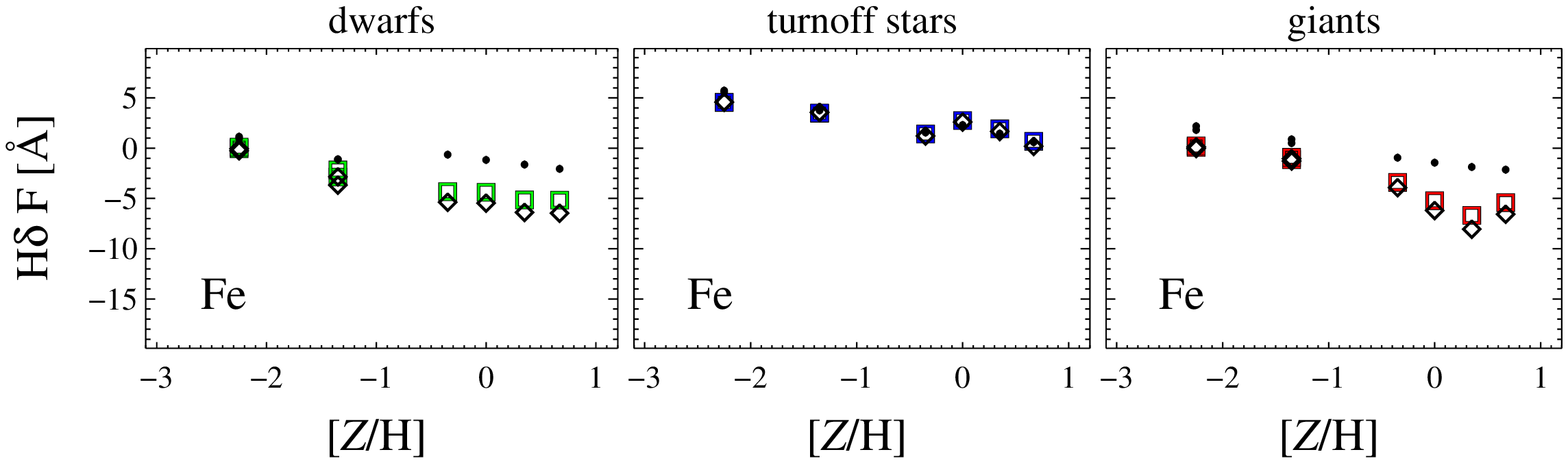}
   \caption{Same as Fig.~\ref{Fe4383}, here for NaD, H$\gamma$\,A, H$\gamma$\,F, H$\delta$\,A and H$\delta$\,F}
         \label{HdeltaF}\label{NaD}\label{HgammaA}\label{HgammaF}\label{HdeltaA}
\end{figure*}

\clearpage


\clearpage
\thispagestyle{empty}
\hspace*{-11cm}
\includegraphics[width=2.3\linewidth]{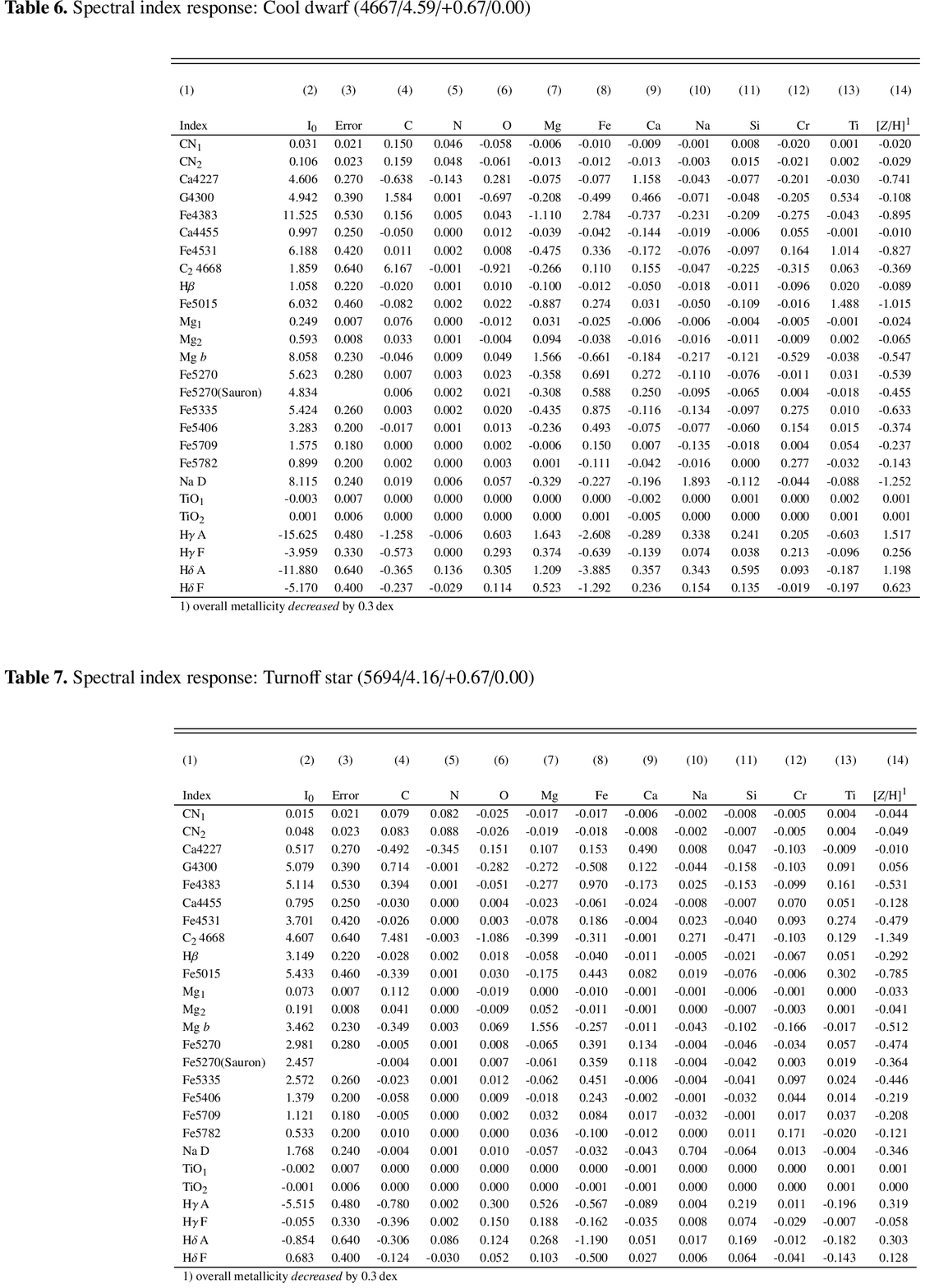}

\clearpage
\thispagestyle{empty}
\hspace*{-11cm}
\includegraphics[width=2.3\linewidth]{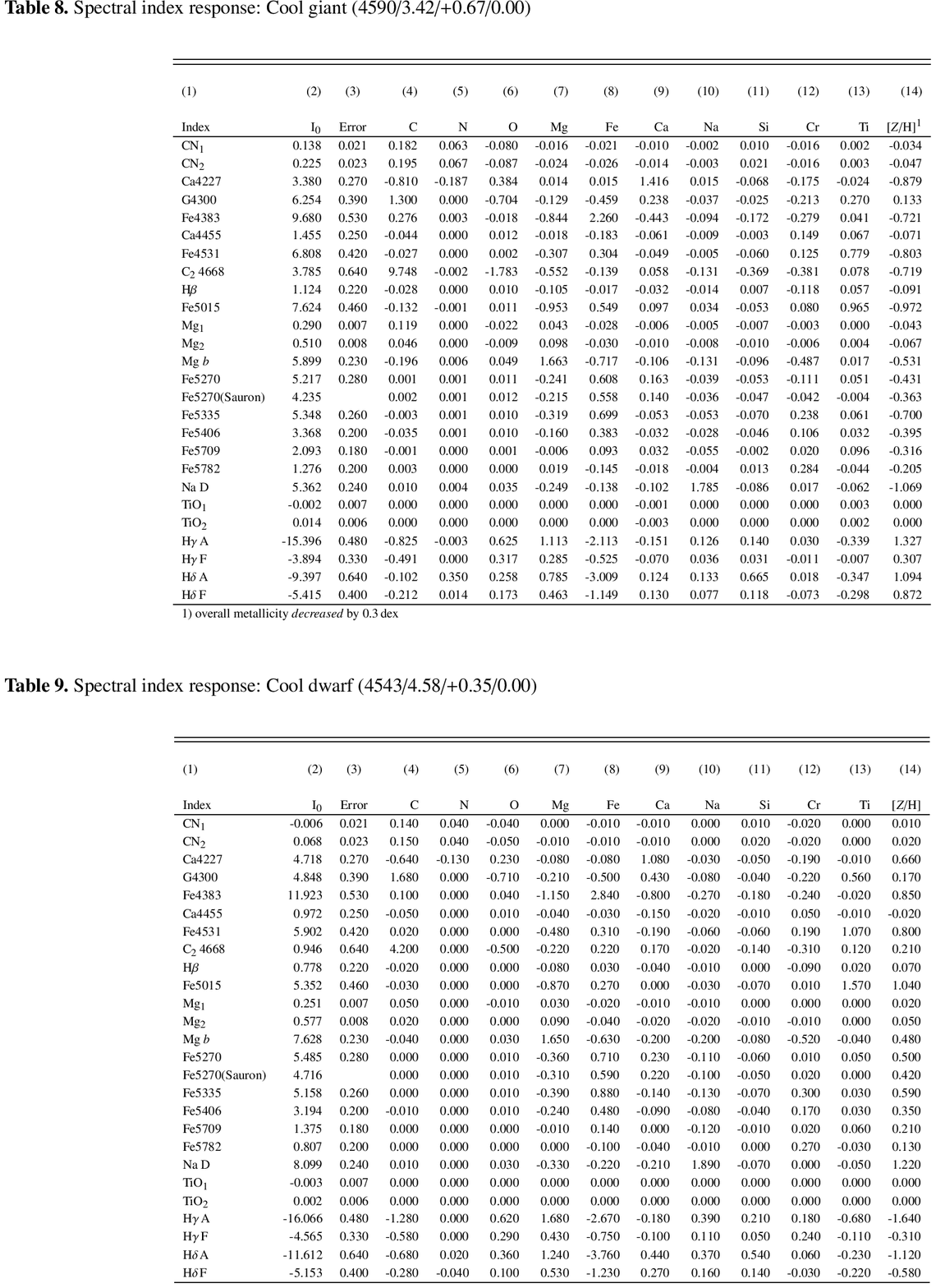}

\clearpage
\thispagestyle{empty}
\hspace*{-11cm}
\includegraphics[width=2.3\linewidth]{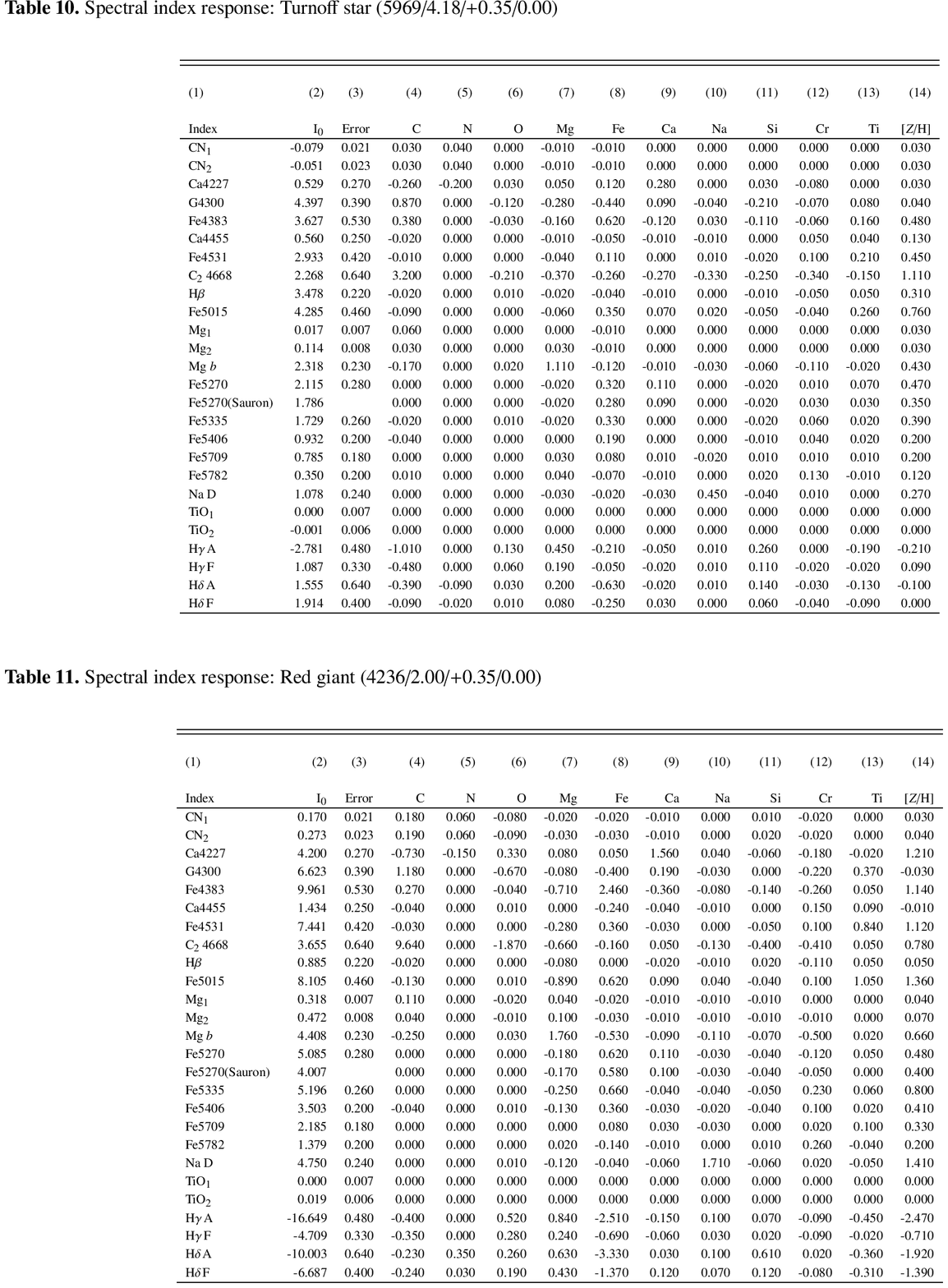}

\clearpage
\thispagestyle{empty}
\hspace*{-11cm}
\includegraphics[width=2.3\linewidth]{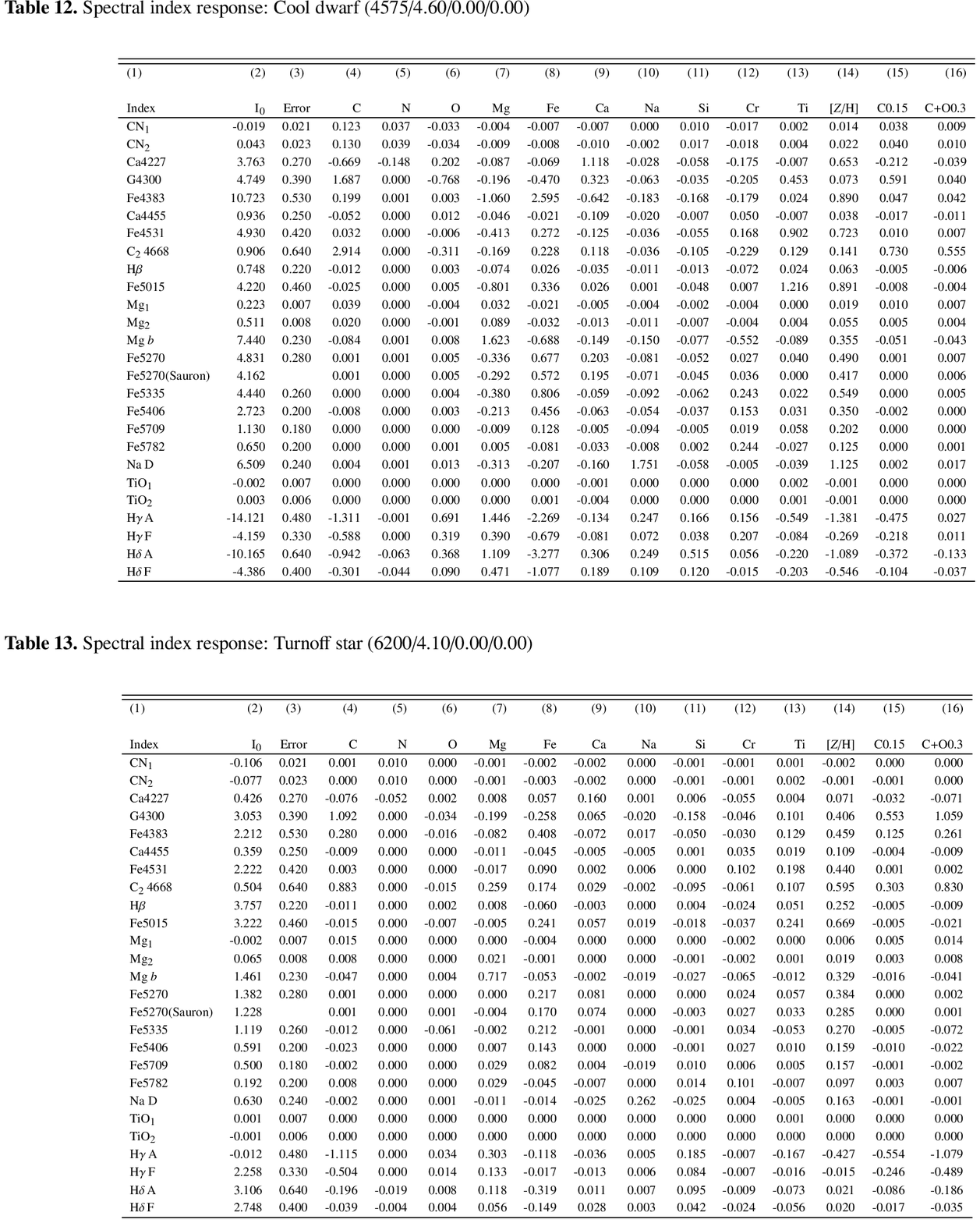}

\clearpage
\thispagestyle{empty}
\hspace*{-11cm}
\includegraphics[width=2.3\linewidth]{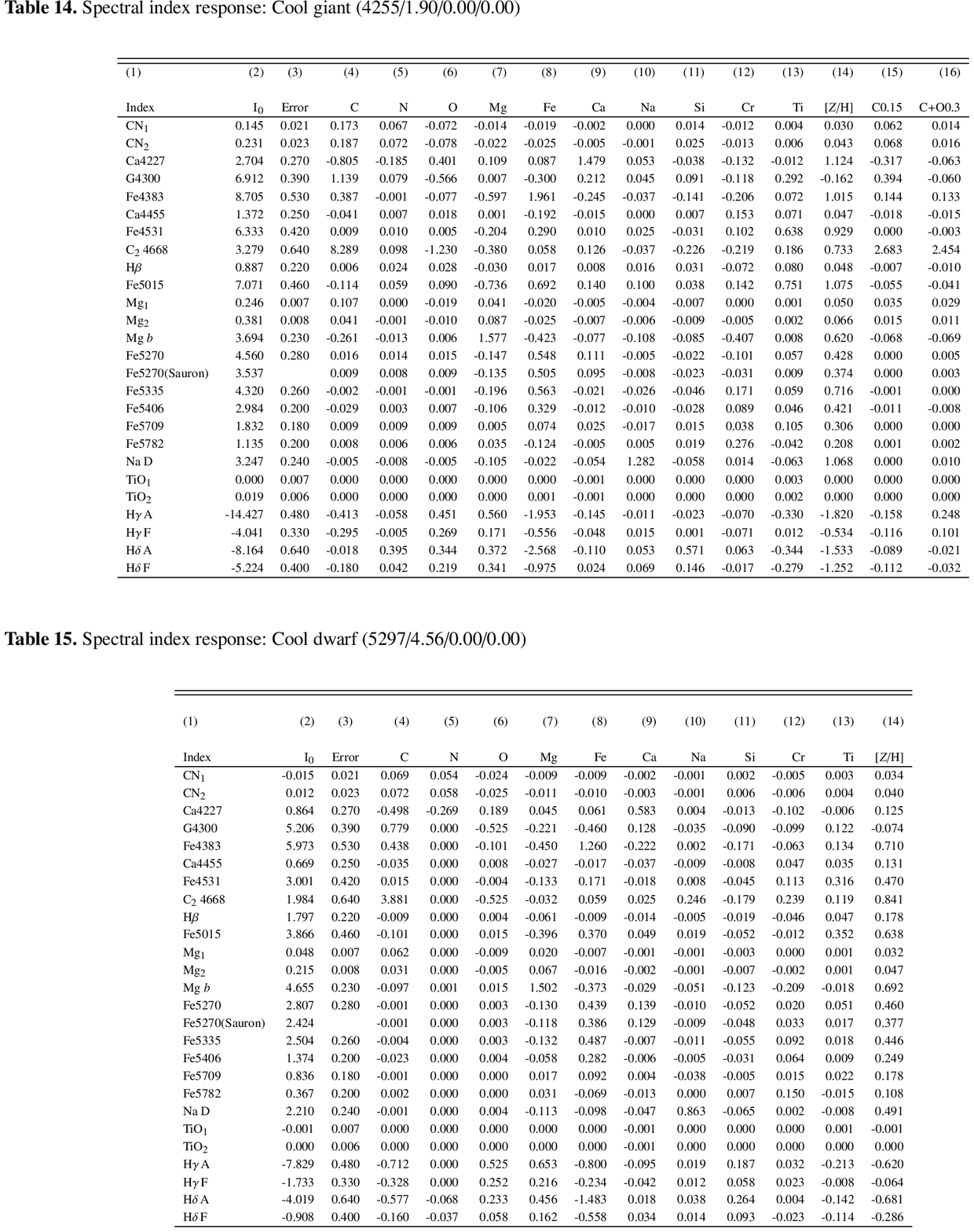}

\clearpage
\thispagestyle{empty}
\hspace*{-11cm}
\includegraphics[width=2.3\linewidth]{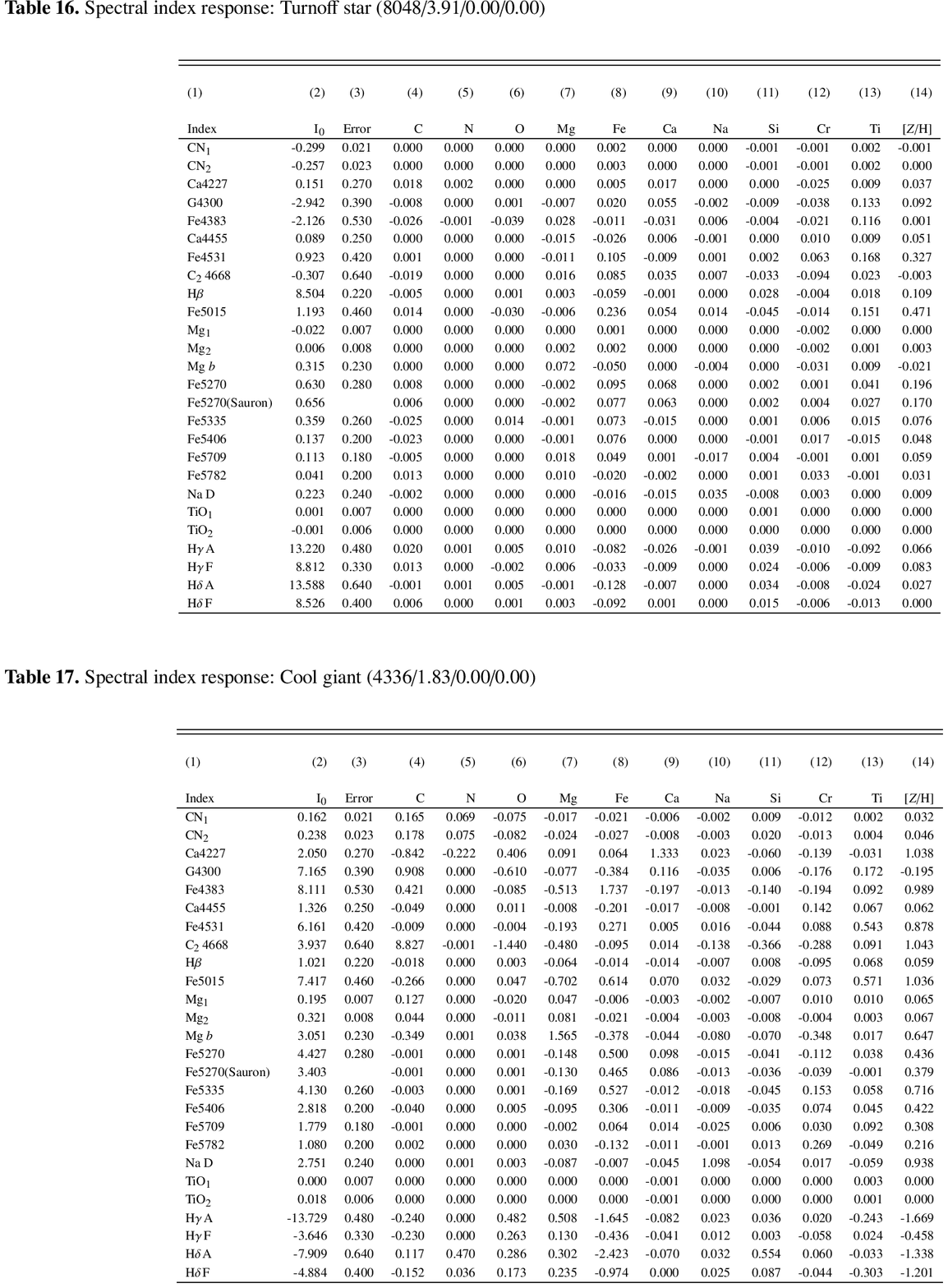}

\clearpage
\thispagestyle{empty}
\hspace*{-11cm}
\includegraphics[width=2.3\linewidth]{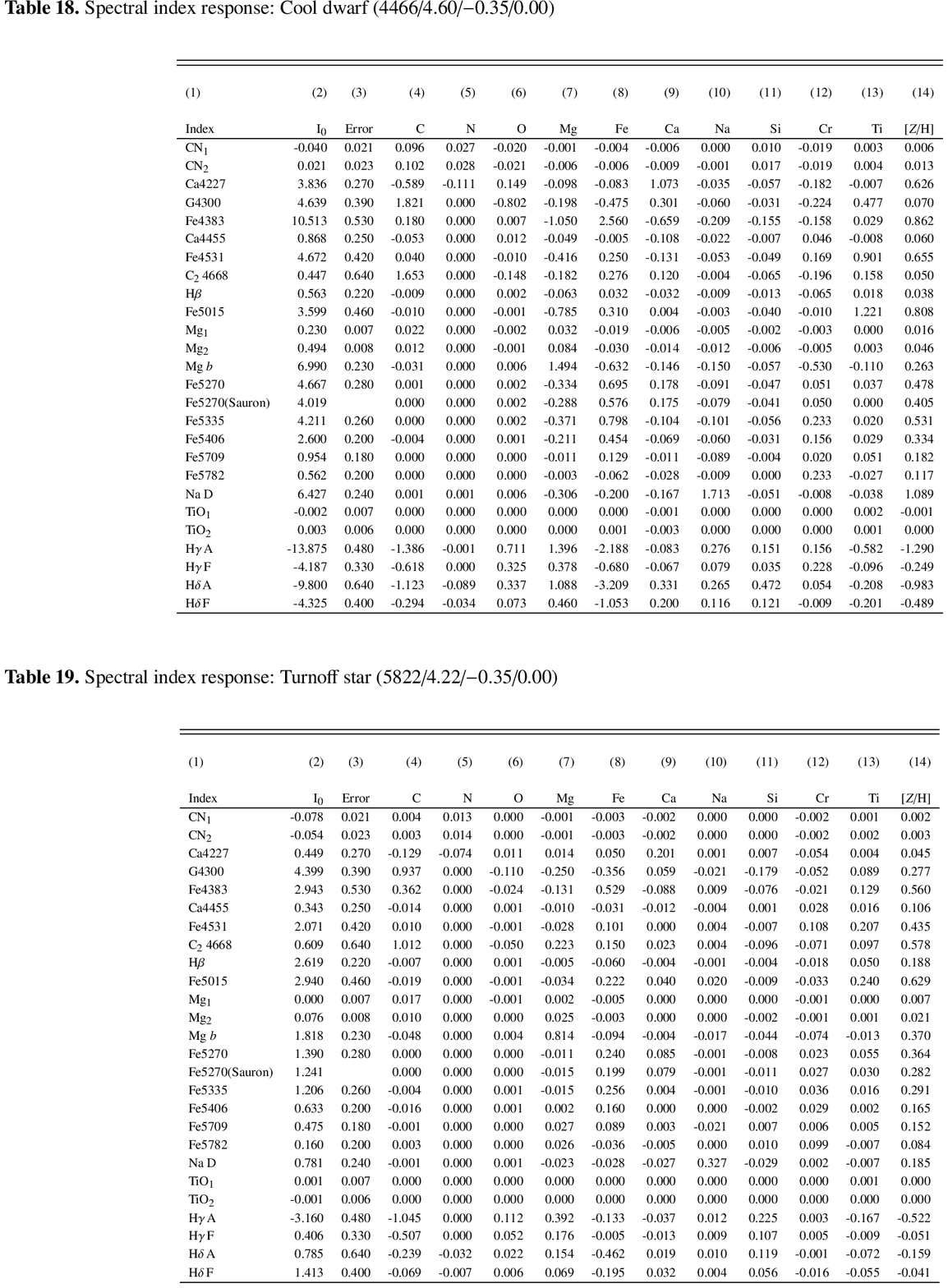}

\clearpage
\thispagestyle{empty}
\hspace*{-11cm}
\includegraphics[width=2.3\linewidth]{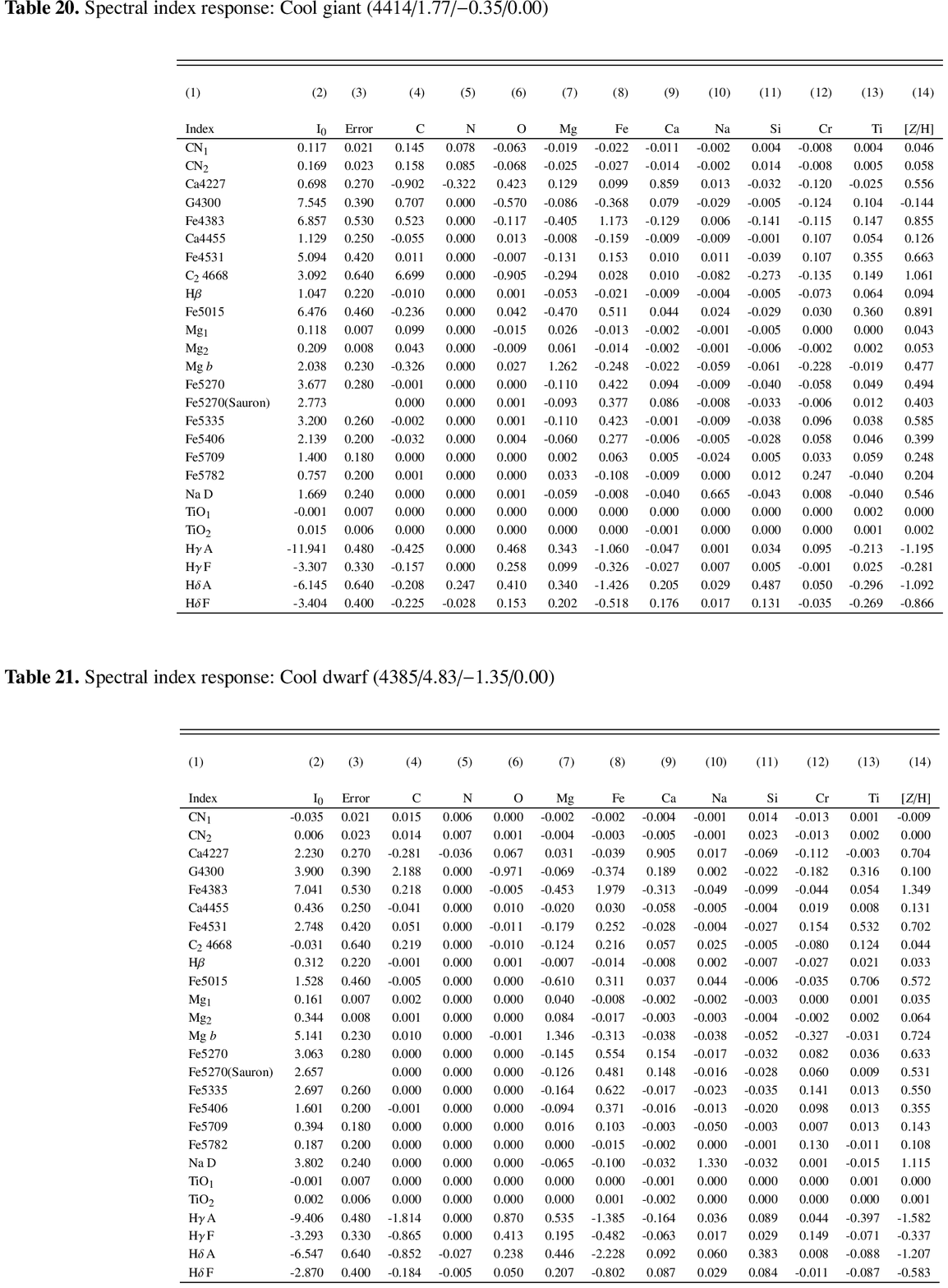}

\clearpage
\thispagestyle{empty}
\hspace*{-11cm}
\includegraphics[width=2.3\linewidth]{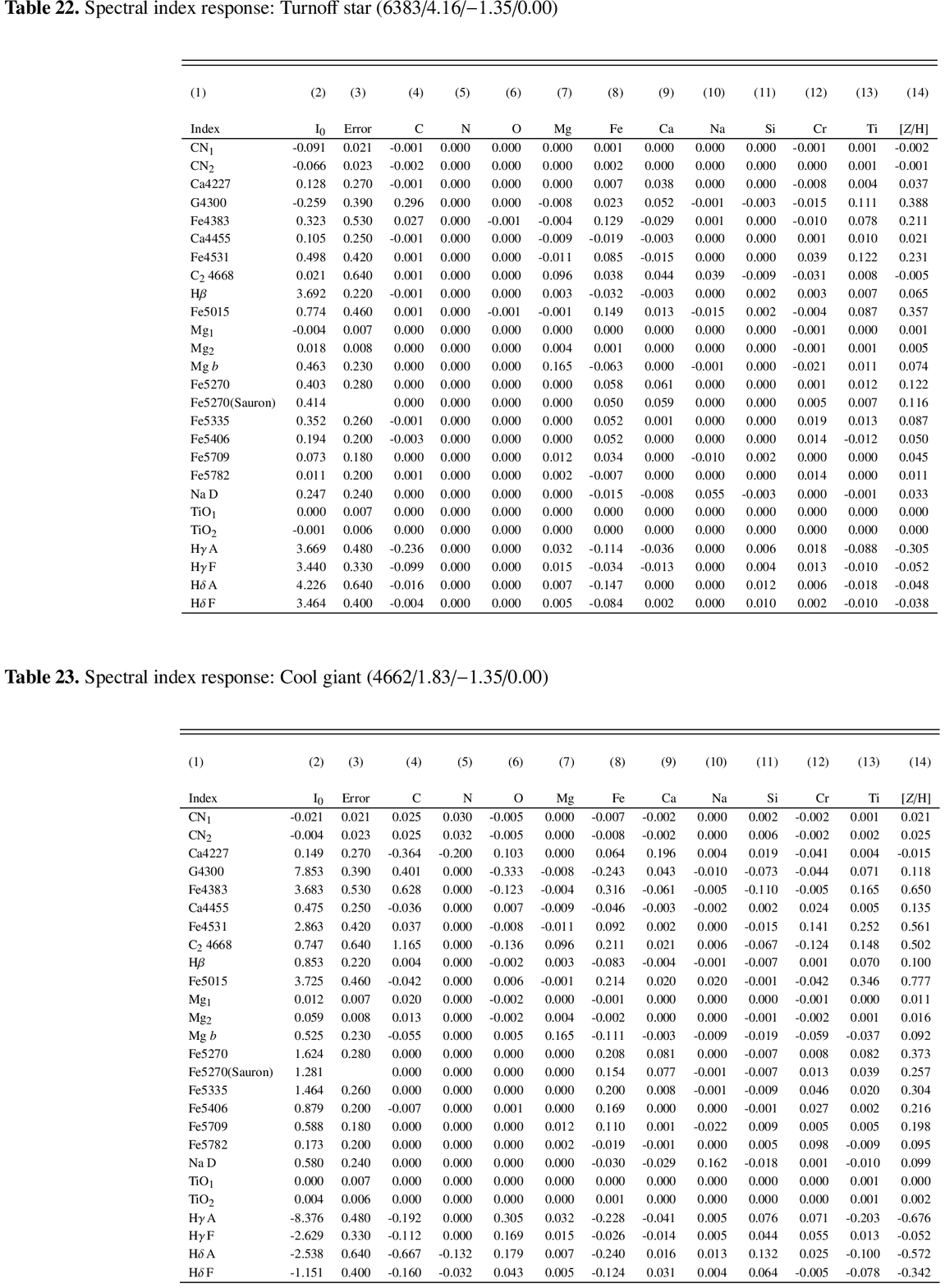}

\clearpage
\thispagestyle{empty}
\hspace*{-11cm}
\includegraphics[width=2.3\linewidth]{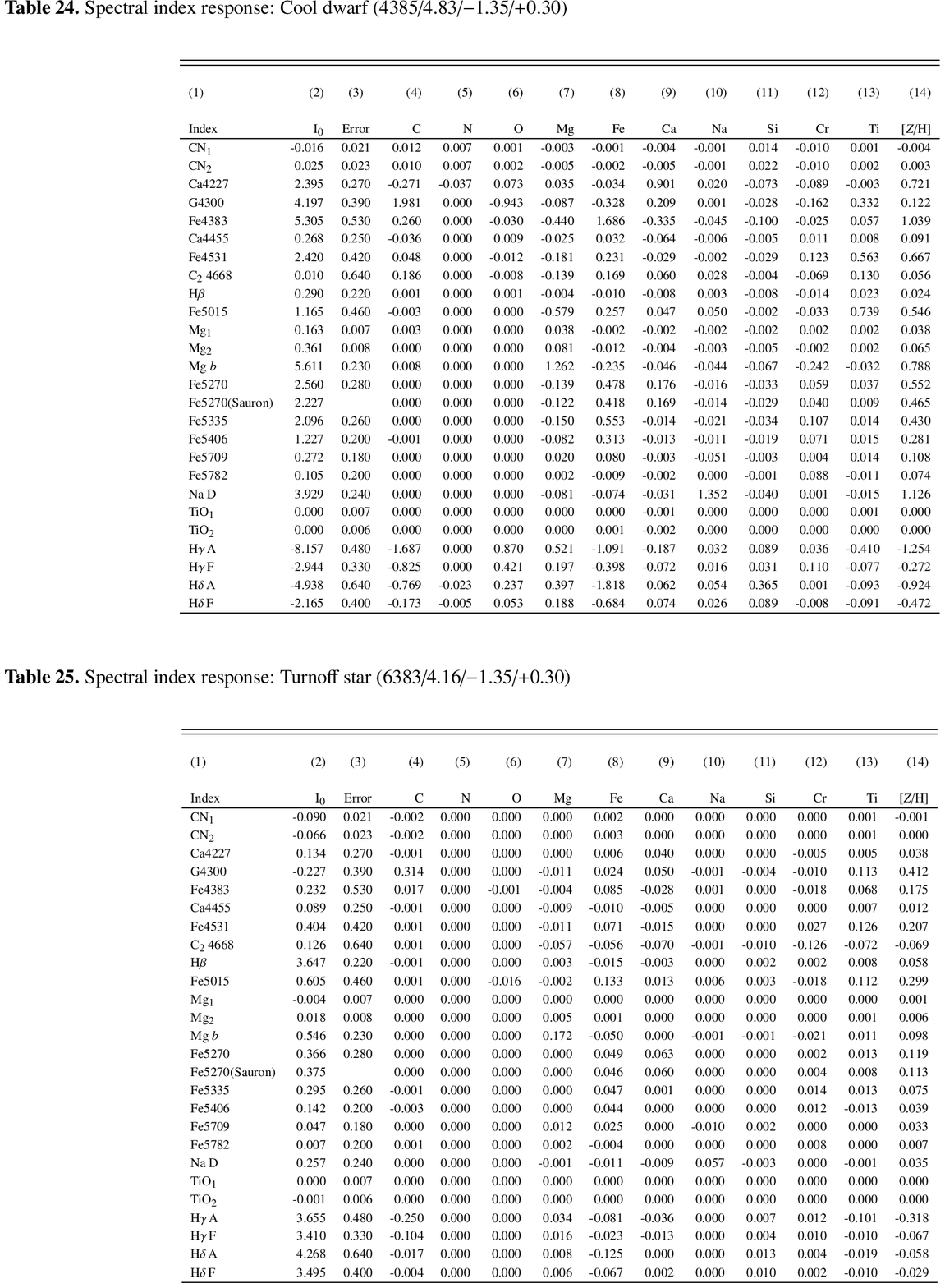}

\clearpage
\thispagestyle{empty}
\hspace*{-11cm}
\includegraphics[width=2.3\linewidth]{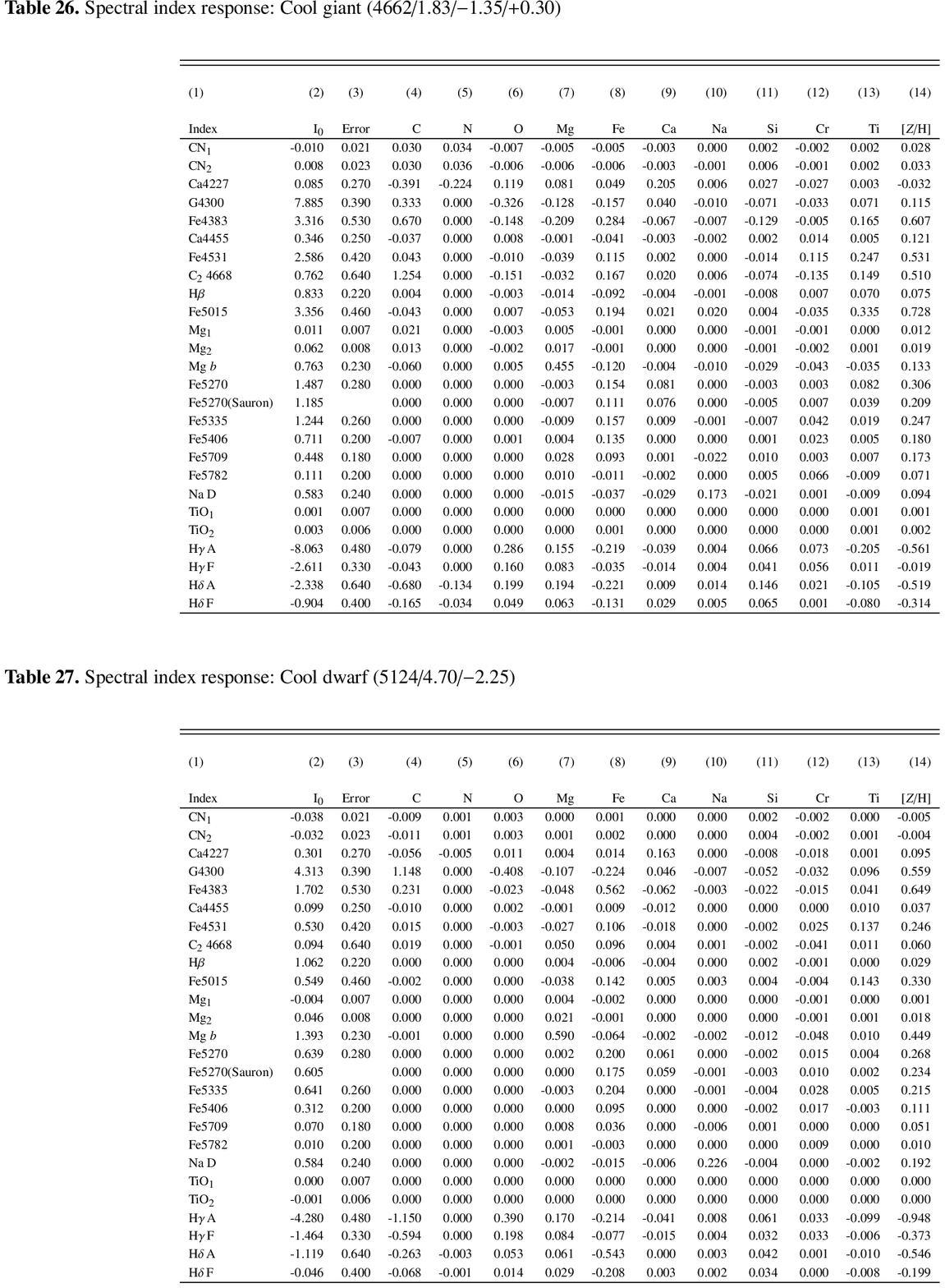}

\clearpage
\thispagestyle{empty}
\hspace*{-11cm}
\includegraphics[width=2.3\linewidth]{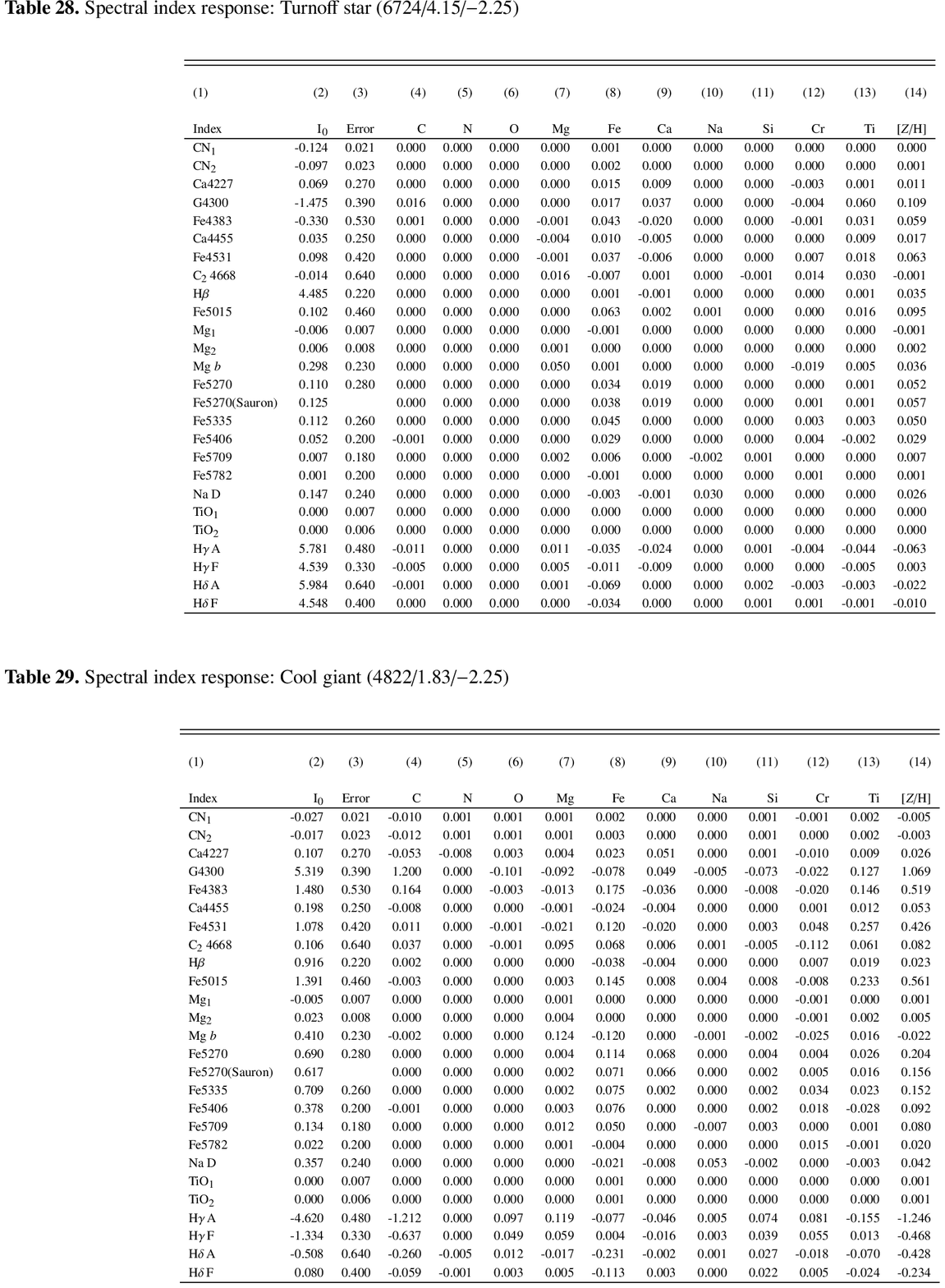}

\clearpage
\thispagestyle{empty}
\hspace*{-11cm}
\includegraphics[width=2.3\linewidth]{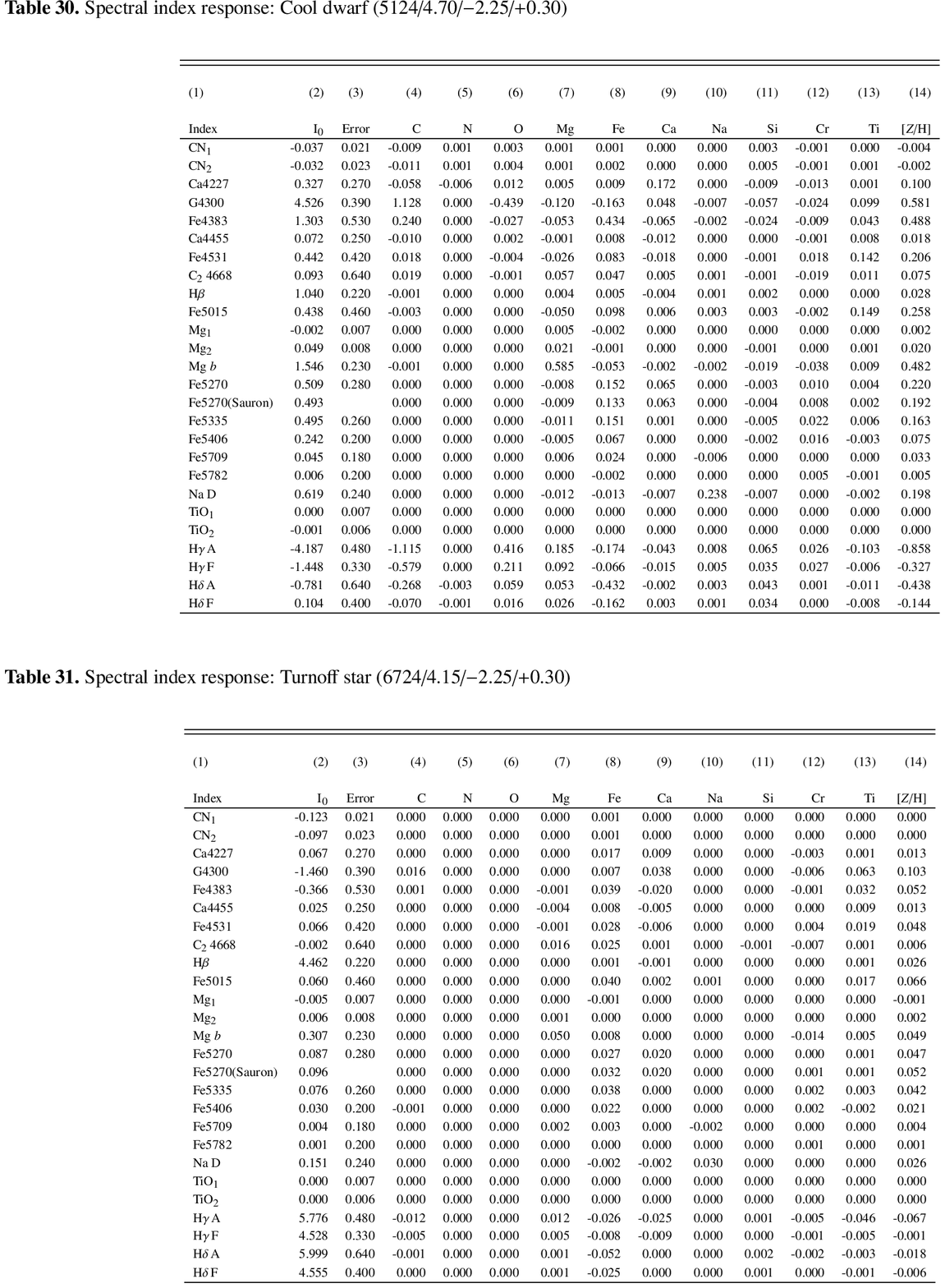}

\clearpage
\thispagestyle{empty}
\hspace*{-11cm}
\includegraphics[width=2.3\linewidth]{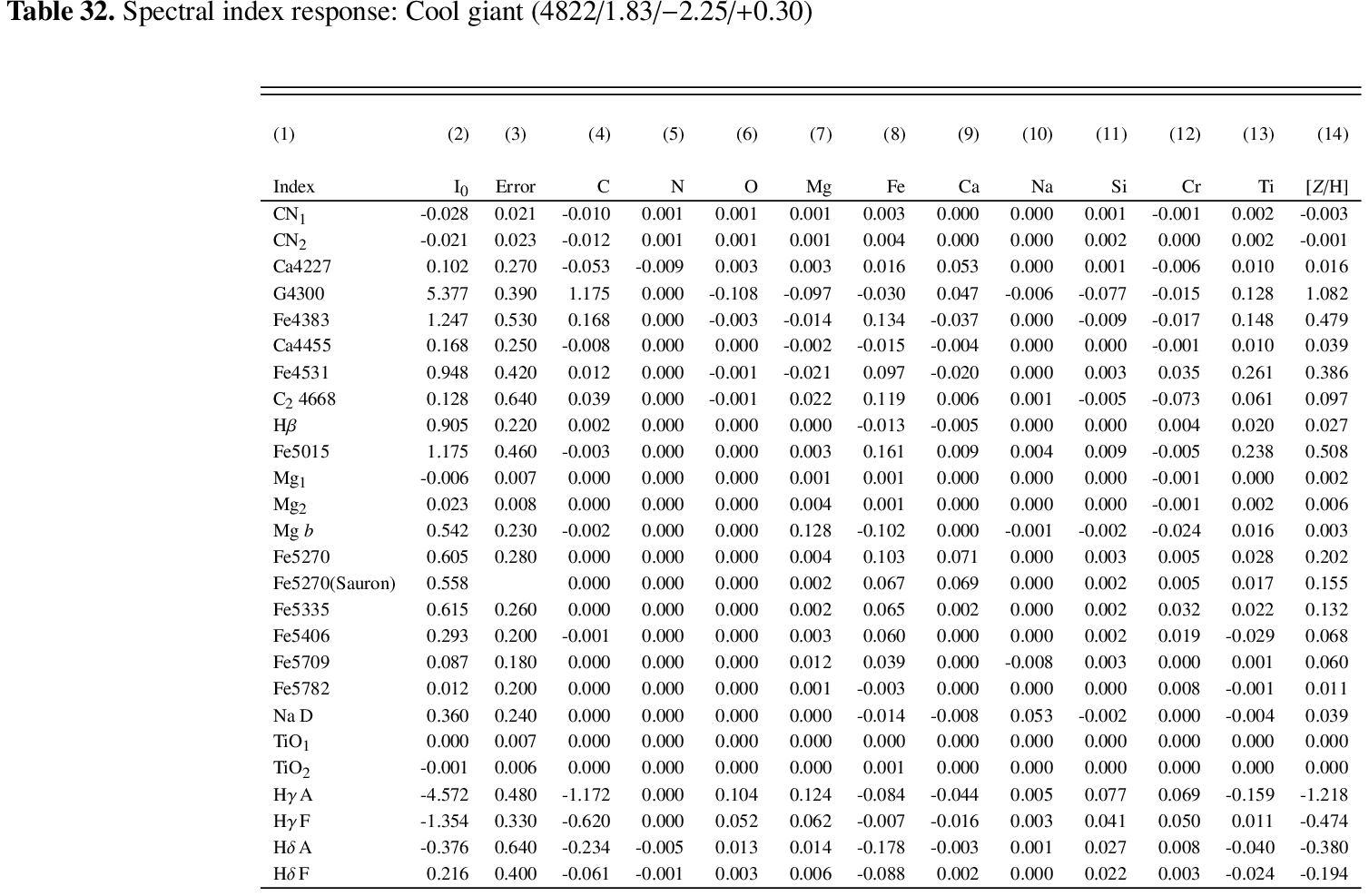}


\begin{thebibliography}{}
\bibitem[1995]{AOM95} Anstee, S.D., \& O'Mara, B.J. 1995, MNRAS 276, 859
\bibitem[2002]{barklem} Barklem, P.S., Stempels, H.C., Allende Prieto, C., Kochukhov, O.P., Piskunov, N., O'Mara, B.J. 2002, A\&A 385, 951
\bibitem[1998]{bernkopf} Bernkopf, J. 1998, A\&A 332, 127
\bibitem[1958]{bohm-vitense} B{\"o}hm-Vitense, E. 1958, Z. f. Astrophys. 46, 108
\bibitem[1995]{Boretal95} Borges, A.C., Idiart, T.P., de Freitas Pacheco, J.A., Th\'{e}venin, F. 1995, AJ, 110, 2408
\bibitem[1984]{BFGK84} Burstein, D., Faber, S.M., Gaskell, C.M., Krumm, N. 1984, ApJ 287, 586
\bibitem[1992]{canutomazzitelli} Canuto, V.M., \& Mazzitelli, I. 1992, ApJ 389, 724
\bibitem[1994]{CD94} Carollo, C.M., Danziger, I.J. 1994, MNRAS, 270, 523
\bibitem[1997]{CCC97} Cassisi, S., Castellani, M., Castellani, V. 1997, A\&A, 317, 10
\bibitem[2001]{Davetal01} Davies, R.L., Kuntschner, H., et al. 2001, ApJ, 548, L33
\bibitem[1993]{DSP93} Davies, R.L., Sadler, E.M., Peletier, R.F. 1993, 262, 650
\bibitem[1997]{D97} Decin, L., Cohen, M., Eriksson, K. et al. 1997, {\em in:} Proceedings of the first ISO workshop on Analytical Spectroscopy, A.M. Heras, K. Leech, N. R. Trams and Michael Perry (eds.), ESA (Noordwijk, The Netherlands), 185
\bibitem[2004]{F-Betal04} Falcon-Barroso, J., Peletier, R.F., Emsellem, E. et al. 2004, MNRAS, 350, 35
\bibitem[1993]{FAG93} Fuhrmann, K., Axer, M., Gehren, T. 1993, A\&A 271, 451
\bibitem[1975a]{G75a} Gehren, T. 1975a, LTE-Sternatmosph{\"a}renmodelle (I), University of Kiel, Germany
\bibitem[1975b]{G75b} Gehren, T. 1975b, LTE-Sternatmosph{\"a}renmodelle (II), University Kiel, Germany
\bibitem[2001]{GKS02} Gehren, T., Korn A.J., Shi J. 2001, A\&A 380, 645
\bibitem[1977]{Gray77} Gray, D.F. 1977, ApJ 218, 530
\bibitem[1998]{GS98} Grevesse, N. \& Sauval, A.J. 1998, Space Science Reviews 85, 161
\bibitem[1993]{Goretal93} Gorgas, J., Faber, S.M., Burstein, D., Gonz\'{a}lez, J.J., Courteau, S., Prosser, C. 1993, ApJS, 86, 153
\bibitem[1975]{GBEN75} Gustafsson, B., Bell, R.A., Eriksson, K., Nordlund, \AA. 1975, A\&A 42, 407
\bibitem[2002]{Hetal02} Houdashelt, M.L., Trager, S.C., Worthey, G., Bell, R A. 2002, BAAS, 34, 1118
\bibitem[1997]{ITF97} Idiart, T.P., Th\'{e}venin, F., de Freitas Pacheco, J.A. 1997, AJ, 113, 1066
\bibitem[1992]{K92} Kurucz, R.L. 1992, Rev. Mex. Astron. Astrof. 23, 45
\bibitem[1993a]{ATLAS} Kurucz, R.L. 1993a, CD-ROM No. 13 ``ATLAS9'', Smithonian Astrophysical Observatory, Cambridge, USA
\bibitem[1993b]{ODF} Kurucz, R.L., 1993b, CD-ROMs ``Opacities for stellar atmospheres'', Smithonian Astrophysical Observatory, Cambridge, USA
\bibitem[2002]{K02} Kurucz, R.L. 2002, {\em in:} Atomic and Molecular Data and their Applications, D.R. Schultz, P.S. Krstic and F. Ownby (eds.), AIP Conf. Proc. 636, 134
\bibitem[1995]{kubell}  Kurucz, R.L. \& Bell, B., 1995, CD-ROM No. 23 ``Atomic Line List'', Smithonian Astrophysical Observatory, Cambridge, USA
\bibitem[1984]{KFBT84} Kurucz, R.L., Furenlid, I., Brault, J., Testerman, L. 1984, Solar Flux Atlas from 296 to 1300 nm, Kitt Peak National Solar Observatory
\bibitem[1975]{kupey} Kurucz, R.L. \& Peytremann, E., 1975, SAO Spec. Rep. 362, 1219
\bibitem[2004]{LRPLFS04} LeBorgne, D., Rocca-Volmerange, B., Prugniel, P., Lan\c{c}on, A., Fioc, M., Soubiran, C. 2004, A\&A 425, 881
\bibitem[1997]{LCB97} Lejeune, T., Cuisinier, F., Buser, R. 1997, A\&AS 125, 229
\bibitem[2003]{MCCetal03} Madgwick, D.S., Coil, A.L., Conselice, C.J. et al. 2003, ApJ 599, 997
\bibitem[1998]{Maraston98} Maraston, C. 1998, MNRAS, 300, 872
\bibitem[2003]{Maretal03} Maraston, C., Greggio, L., Renzini, A., Ortolani, S., Saglia, R.P., Puzia, T., Kissler-Patig, M. 2003, A\&A, 400, 823
\bibitem[2004]{Maraston04} Maraston, C. 2004, MNRAS, submitted
\bibitem[1993]{PDK93} Peterson, R.C., Dalle Ore, C.M., \& Kurucz, R.L. 1993, ApJ 404, 333
\bibitem[1998]{pfeiffer} Pfeiffer, M.J., Frank, C., Baum{\"u}ller, D., Fuhrmann, K., Gehren, T. 1998, A\&AS 130, 381
\bibitem[2004]{nob04} Przybilla, N. \& Butler, K. 2004, ApJ 610, 61
\bibitem[2000]{Saletal00} Salasnich, B., Girardi, L., Weiss, A., Chiosi, C. 2000, A\&A, 361, 1023
\bibitem[2004]{tc04} Tantalo, R. \& Chiosi C. 2004, MNRAS 353, 917
\bibitem[2003a]{TMB03} Thomas, D., Maraston, C., Bender R. 2003a, MNRAS 339, 897 (TMB03)
\bibitem[2003b]{TMB03b} Thomas, D., Maraston, C., Bender, R. 2003b, MNRAS 343, 279
\bibitem[2004]{TMK04} Thomas, D., Maraston, C., Korn, A. 2004, MNRAS 351, L19
\bibitem[2004]{Thoetal04} Thomas, D., Maraston, C., Bender, R., Mendes de Oliveira, C. 2005, ApJ, 621, 673
\bibitem[1995]{TB95} Tripicco, M.J. \& Bell, R.A. 1995, ApJ 110, 3035
\bibitem[1968]{unsoeld} Uns{\"o}ld, A. 1968, {\em in}: Physik der Sternatmosph{\"a}ren, Springer (Berlin)
\bibitem[1996]{vant} van't Veer-Menneret, C., \& M\'{e}gessier, C. 1996, A\&A 309, 879
\bibitem[2005]{W04} Worthey, G. 2004, AJ, 128, 2826
\bibitem[1994]{WFG92} Worthey, G., Faber, S.M., Gonz\'{a}lez, J.J. 1992, ApJ, 398, 69
\bibitem[1994]{WFGB94} Worthey, G., Faber, S.M., Gonz\'{a}lez, J.J., \& Burstein D. 1994, ApJS 94, 687
\bibitem[1997]{WO97} Worthey, G. \& Ottaviani, D.L. 1997, ApJS 111, 377
\end{thebibliography}
\end{document}